\documentclass[a4paper,12pt]{article}
\pdfoutput=1
\usepackage{jheppub}
\usepackage[english]{babel}
\usepackage[utf8]{inputenc}
\usepackage{dcolumn}
\usepackage{soul}

\usepackage{graphicx}
\usepackage{bm,amsmath,amssymb}
\usepackage[mathscr]{eucal}
\usepackage{makeidx}
\usepackage{subfig}
\usepackage{amsmath}
\usepackage{amssymb}
\usepackage{wasysym}
\usepackage{amsthm}
\usepackage{mathrsfs}
\usepackage{graphicx}
\usepackage{fancyhdr}
\usepackage{array}
\usepackage{simplewick}
\usepackage{latexsym}
\usepackage[all]{xy}
\usepackage{enumerate}
\usepackage{dsfont}
\usepackage{slashed}
\usepackage{float}
\usepackage[titletoc]{appendix}
\usepackage{ulem}

\title{Criticality  from Einstein-Maxwell-dilaton holography at finite temperature and density}

\author[a]{Alfonso Ballon-Bayona}
\author[a]{Henrique Boschi-Filho}
\author[a,b]{Eduardo Folco Capossoli}
\author[a,c]{Diego M. Rodrigues} 

\affiliation[a]{Instituto de Física,
Universidade Federal do Rio de Janeiro, Caixa Postal 68528, RJ
21941-972 -- Brazil}
\affiliation[b]{Departamento de F\'\i sica / Mestrado Profissional em Práticas de Educação Básica (MPPEB), Col\'egio Pedro II, 20.921-903 - Rio de Janeiro-RJ - Brazil}
\affiliation[c]{Centro de Ciências Naturais e Humanas, 
Universidade Federal do ABC,\\
Rua Santa Ad\'elia 166, 09210-170, Santo Andr\'e, SP, Brazil}
\emailAdd{aballonb@if.ufrj.br}
\emailAdd{boschi@if.ufrj.br}
\emailAdd{eduardo\_capossoli@cp2.g12.br} 
\emailAdd{diegomhrod@gmail.com}

\abstract{
We investigate consistent charged black hole solutions to the Einstein-Maxwell-Dilaton (EMD) equations that are asymptotically AdS. The solutions are gravity duals to phases of a non-conformal plasma at  finite temperature and density. For the dilaton we take a quadratic ansatz leading to linear confinement at zero temperature and density. We consider a grand canonical ensemble, where the chemical potential is fixed, and find a rich phase diagram  involving the competition of small and large black holes. The phase diagram contains a critical line and a critical point similar to the van der Waals-Maxwell liquid-gas  transition. As the critical point is approached, we show that the trace anomaly in the plasma phases vanishes signifying the restoration of conformal symmetry in the fluid. We find that the heat capacity and charge susceptibility diverge as $C_V \propto (T-T^c)^{-\alpha}$ and $\chi \propto (T-T^c)^{-\gamma}$ at the critical point with universal critical exponents $\alpha=\gamma=2/3$. Our results suggest a description of the thermodynamics near the critical point in terms of catastrophe theories. In the limit $\mu \to 0$ we compare our results with lattice results for $SU(N_c)$ Yang-Mills theories.}


\begin{document}
	\maketitle
	
	\newcommand{\limit}[3]
	{\ensuremath{\lim_{#1 \rightarrow #2} #3}}
	
\section{Introduction} 

Matter under extreme conditions is an exciting and challenging research field in high energy physics. In the case of Quantum Chromodynamics (QCD) exploring hadronic matter at extreme conditions is an excellent opportunity for understanding  non-perturbative aspects of QCD and gain new insights into strongly coupled phenomena such as confinement and chiral symmetry breaking.  A good example is the celebrated quark-gluon plasma (QGP) \cite{Shuryak:2014zxa,Cabibbo:1975ig, Collins:1974ky}  observed in high energetic heavy ion collisions at the  Relativistic Heavy Ion Collider (RHIC) in Brookhaven. The QGP is a new state of matter that behaves as an ideal fluid and whose constituents are deconfined quarks and gluons. Lattice QCD has correctly predicted a crossover transition to the QGP \cite{Aoki:2006we}  to occur at approximately $T \approx 155$ MeV \cite{Bazavov:2011nk}.

Just like ordinary matter, e.g. water, hadronic matter in QCD  is expected to have different phases. It is therefore natural to consider the so called QCD phase diagram, which is usually described in a temperature ($T$) vs. chemical potential ($\mu$) plane. A finite chemical potential is associated with non-zero baryonic/quark density.
At $\mu=0$, as described above,  increasing the temperature we expect a transition from a hadronic gas to the QGP. At $T=0$, on the other hand, for increasing $\mu$ we expect a transition  to nuclear matter and then for sufficiently large $\mu$ a transition to another state of matter named color superconductor. Currently, there is an experimental program devoted to investigating the QCD phase diagram from relativistic heavy ion collisions, see e.g. \cite{STAR}. Neutron stars are also a very interesting framework for investigating the transition to nuclear matter at low temperatures, see e.g. \cite{Piekarewicz:2018zbi}.

One important aspect of the QCD phase diagram concerns the possible existence of a Critical End Point (CEP) in the $T - \mu$ plane \cite{Stephanov:2004wx,Gupta:2011wh}. For  large $\mu$  and small $T$ it is expected a first order transition from the hadronic gas to the QGP, at $\mu=0$, lattice QCD predicts a crossover transition to the QGP. One therefore concludes that a CEP, represented by the coordinates ($\mu^c, T^c$), should be located at the interface between the crossover and first order transition. Together with the experimental achievements, theoretical efforts have been made in order to understand QCD matter under extreme conditions. Perturbative QCD methods leads to reliable results in the regime of large temperatures or large chemical potentials \cite{Fraga:2001id, Haque:2014rua, Schmidt:2017bjt}, Nambu-Jona-Lasinio models \cite{Nambu:1961tp, Nambu:1961fr} also provide good results related to the QCD phase diagram (for recent work see \cite{Ferreira:2018sun}).   Lattice QCD, works very well at $\mu=0$ but faces a difficult challenge at finite $\mu$ due to the sign problem \cite{Aarts:2015tyj}; there have been, however, recent promising developments \cite{Ratti:2018ksb}.

In this work we rather follow the gauge/gravity duality approach to investigate criticality in non-conformal plasmas at finite temperature and density. Non-conformal plasmas arising from the gauge/gravity duality are far from the real QGP found in QCD but they provide very useful insights regarding the breaking and restoration of conformal symmetry. They are also useful for investigating the real time dynamics associated with perturbations and the response of the fluid. 
 The gauge/gravity approach applied to QCD is usually called holographic QCD and it is based on the AdS/CFT 
correspondence. According to gauge/gravity duality the physics of four dimensional strongly coupled non-conformal plasmas maps to five dimensional black hole solutions that are asymptotically AdS. Non-conformal plasmas described in the gauge/gravity approach satisfy the property of conformal symmetry restoration at very  high temperatures, which is associated with the vanishing of the trace anomaly $E-3p$, where $E$ and $p$ are the energy density  and pressure  respectively. This property is in agreement with  the QGP equation of state obtained from lattice QCD \cite{Borsanyi:2010cj}.

Since the advent of AdS/CFT correspondence, the study of black hole physics has gained a lot of interest due to its holographic connection with thermal phase transitions in strongly coupled field theories. In particular, the study of the thermodynamic phase structure of charged AdS black holes in global coordinates has stood out in the pioneering papers \cite{Chamblin:1999tk,Chamblin:1999hg} and, subsequently, in \cite{Wu:2000id}. In these works it was shown that these charged AdS black holes present a variety of fascinating features and critical phenomena, including a rich phase structure with first-, second-order and continuous phase transitions. Moreover, one of the most interesting features of these charged AdS black holes was the striking similarities with the van der Waals-Maxwell liquid-gas phase transition. The analogy with the van der Waals-Maxwell liquid-gas system and the study of criticality of charged AdS black holes was further explored and analysed in \cite{Niu:2011tb,Kubiznak:2012wp,Zeng:2015wtt} (for a review see \cite{Kubiznak:2016qmn}). Furthermore, holographic van der Waals-Maxwell-like systems were extended to the case of Born-Infeld AdS black holes in \cite{Banerjee:2011cz} and Gauss-Bonet AdS black holes \cite{Cai:2013qga,Zeng:2016aly}.

The rich thermodynamic phase structure of charged AdS black holes has also been applied to investigate the phase diagram of holographic QCD in the context of Einstein-Maxwell-Dilaton (EMD) holography by Gubser {\it{et. al}} in \cite{Gubser:2008ny,Gubser:2008yx}. The criticality of holographic QCD plasmas in the $T-\mu$ plane was first studied in  \cite{DeWolfe:2010he,DeWolfe:2011ts} within the context of EMD holography. Several other studies followed dealing with different aspects of the thermodynamics, including phase transitions between different black hole branches and a quantitative  comparison with lattice QCD results for some holographic models \cite{Li:2011hp,Cai:2012xh,He:2013qq,Li:2014hja,Rougemont:2015wca,Rougemont:2017tlu,Critelli:2017oub,Li:2017ple, Chen:2018vty, Chen:2019rez}. In  these works, however, universal aspects of charged black holes in EMD holography were not investigated. In particular, the relation between charged black hole transitions in EMD holography and the liquid-gas transition, described by the van der Walls model, was not explored.  

Holographic models based on Einstein-Maxwell theory has also been useful for describing non-conformal plasmas in the presence of a magnetic field \cite{DHoker:2009mmn,DHoker:2009ixq,Rougemont:2015oea,Critelli:2016cvq,Rodrigues:2017cha,Rodrigues:2017iqi,Braga:2018zlu,Rodrigues:2018pep,Rodrigues:2018chh,Braga:2019yeh} or an electric field \cite{Costa:2015gol,Costa:2017tug}. Still in the context of EMD holography, another important ingredient considered was the effect of anisotropy in strongly coupled gauge theories \cite{Giataganas:2017koz}, and its interplay with an applied magnetic field in holographic QCD \cite{Gursoy:2018ydr,Bohra:2019ebj}.  
Criticality has also been investigated in gauge/gravity models applied to condensed matter physics. This approach is called AdS/CMT and has led to new insights for strongly coupled systems in condensed matter near the  quantum  critical  point  \cite{Charmousis:2010zz, Sachdev:2010ch, Hartnoll:2009sz}.

In this work we describe non-conformal plasmas  that admit  a CEP in the $T- \mu$ phase diagram using EMD holography. We consider analytic background solutions generated by a quadratic dilaton profile, to investigate the phase structure and  criticality of asymptotically charged AdS black holes in the grand canonical ensemble \footnote{The background solutions considered here are similar to the ones considered recently in \cite{Mahapatra:2020wym} for investigating asymptotically AdS charged hairy black hole solutions, and in \cite{He:2020fdi}, for studying  analytically magnetic catalysis in holographic QCD.}. 
We find a rich phase structure due to the presence of three black hole solutions. One of them will always be unstable whereas the other two will compete leading to a first order transition.  We describe the thermodynamics and phase transitions between the different black hole branches using the thermodynamic observables, which are obtained from the grand canonical potential which, in turn, is reconstructed from the first law of thermodynamics in a consistent way. Interestingly, we find that the thermodynamics of charged black hole solutions in EMD holography display strong similarities with the thermodynamics of charged black holes in global AdS, the latter described by Chamblin et al in the pioneer works \cite{Chamblin:1999tk,Chamblin:1999hg}. In particular, we find that EMD holography lead to a relation between the temperature and the horizon radius for charged black holes qualitatively similar to the corresponding relation for charged black holes in global AdS.   

We find in this work a critical line in the $T-\mu$ phase diagram, associated with the first order transition, that ends on a CEP. The critical point in our work will correspond  to the situation where the unstable black hole disappears and the two other black holes merge.
We develop an analogy between the phase transitions found in this work and the van der Waals-Maxwell liquid-gas transition. In this analogy the temperature $T$ and horizon radius $z_h$ are mapped to the pressure $P$ and volume $V$ of the van der Waals model. The curves at fixed $\mu$ in our work will play the role of  the isothermal curves in the van der Waals model. We present a systematic description of all the relevant thermodynamic quantities in our model. In particular, we show explicitly the following results: i) Conformal symmetry is explicitly broken at $\mu=0$ due to the deformation of the 4d theory, dual to the dilaton backreaction to the AdS black brane. ii) Conformal symmetry breaking due to the dilaton backreaction  persists at finite $\mu$ and the trace anomaly $E-3p$ is in general non-zero. iii) As we reach the critical point in the $T-\mu$ phase diagram, the trace anomaly goes to zero. We interpret this result as the emergence of a non-trivial CFT at the critical point.

We also perform a careful and universal analysis of the thermodynamics in the neighborhood of the critical point. From that analysis we find universal critical exponents for the specific heat and charge susceptibility. These results hold for any holographic model, based on EMD theory with a minimal coupling, displaying a critical line that ends on a CEP as the chemical potential increases. For the specific heat we find a  critical exponent $\alpha=2/3$, which is in agreement with the result found in  \cite{Chamblin:1999tk,Chamblin:1999hg} for charged AdS black holes in global coordinates.  For the charge susceptibility we find a critical exponent $\gamma=2/3$, which is identical to the result for the specific heat. We are not aware of any previous calculation in holography leading to a similar result for the charge susceptibility.  From our analysis for the thermodynamics near the critical point, we suggest a connection between criticality of non-conformal plasmas in EMD holography and catastrophe theories of type  $A_3$. Lastly,  in the limit $\mu \to 0$ we compare our results against the lattice results for  $SU(N_c)$ Yang-Mills theories and conclude that our model is compatible with the thermodynamics of $SU(N_c)$ Yang-Mills theories in the large $N_c$ limit.

This work is organised as follows. In section \ref{Sec:HQCD} we review the EMD holographic theory, and construct analytically the charged black brane solutions dual to non-conformal plasmas. In section \ref{Sec:Thermodynamics} we study the black holes temperature and derive the grand canonical potential along with its related thermodynamic quantities. In section \ref{Sec:Results} we present our results for the thermodynamic observables and the phase diagram in the $T-\mu$ plane. In section \ref{Sec:CriticalThermod} we analyse the thermodynamics of our holographic model near the critical point and from this analysis we find universal critical exponents for the heat capacity and charge susceptibility. In section \ref{Sec:Lattice} we compare our results in the limit $\mu \to 0$ against the lattice results for $SU(N_c)$ Yang-Mills theories. Finally, in section \ref{Sec:Conclusions} we discuss the results and present our concluding remarks. Appendix \ref{App:RNAdS} briefly reviews the thermodynamics of pure Reissner-Nordstr\"om AdS black brane. Appendix \ref{App:AltMethod} describes an alternative method for reconstructing the grand canonical potential and appendix \ref{App:Susceptibility} describes the charge susceptibility in our model. 

\medskip 

 Note added in v2: After the first version of our paper appeared on arXiv, the paper \cite{Mamani:2020pks}  appeared on arXiv proposing a holographic QCD model that also leads to criticality in the $T-\mu$ phase diagram.  Although the critical point in \cite{Mamani:2020pks} is located in a different position, our universal analysis near the critical point should also apply to the model of \cite{Mamani:2020pks}.

\section{Non-conformal plasmas at finite temperature and density from EMD holography}
\label{Sec:HQCD}

In this section we present our framework for describing non-conformal plasmas  at finite temperature and density arising from 5d  Einstein-Maxwell-Dilaton (EMD) theory. First we describe the general EMD equations and then we describe the ansatz that leads to black hole solutions. The latter will be interpreted as 5d gravity duals of non-conformal fluids in the dual 4d theory.

\subsection{Einstein-Maxwell-Dilaton theory}

In this subsection we describe the five dimensional Einstein-Maxwell-Dilaton (EMD) theory in the Einstein frame and obtain the corresponding field equations.

The EMD action is given by:
\begin{equation} \label{EMD action}
S = \dfrac{1}{16\pi G_5}\int d^{5}x \sqrt{-g}\left(R - \dfrac{4}{3}g^{\mu\nu}\partial_{\mu}\phi\partial_{\nu}\phi + V(\phi)-\dfrac{1}{4}F_{\mu\nu}F^{\mu\nu} \right),
\end{equation}
where $ G_5 $ is the 5-dimensional Newton's constant, $g=\mathrm{det}(g_{\mu\nu})$, $ R $ is the Ricci scalar, $\phi$ is the dilaton field, $V(\phi)$ is the dilaton potential and $F_{\mu \nu}$ is the usual electromagnetic field strength defined as $F_{\mu \nu}  = \partial_{\mu} A_{\nu} - \partial_{\nu} A_{\mu}$.

The field equations derived from the  action \eqref{EMD action} are given by
\begin{eqnarray}
G_{\mu\nu} -\dfrac{4}{3}\left(\partial_{\mu}\phi\partial_{\nu}\phi - \dfrac{1}{2}g_{\mu\nu}(\partial \phi)^2\right) - \dfrac{1}{2}g_{\mu\nu}V(\phi) - \frac{1}{2}\left(F_{\mu\alpha}F_{\nu}^{\,\alpha} - \dfrac{1}{4}g_{\mu\nu}F^2\right) &=& 0, \label{EinsteinEqn} \label{eom1}\qquad \\ 
\nabla^{2}\phi + \dfrac{3}{8}\frac{\partial\,V(\phi )}{\partial\,\phi}&=& 0, \label{DilatonEqn} \label{eom2}\\
\nabla_{\mu}F^{\mu\nu} &=& 0, \label{eom3}
\end{eqnarray}
with the Einstein tensor $ G_{\mu\nu} $ defined as
\begin{equation}
G_{\mu\nu} = R_{\mu\nu}- \dfrac{1}{2}g_{\mu\nu}R.
\end{equation}
The field equations \eqref{EinsteinEqn} correspond to the Einstein equations in the presence of an energy momentum tensor due to the scalar and gauge fields. The equation \eqref{eom2} describes the dynamics of the scalar field $\phi$ (the dilaton) in a curved space whereas the equations \eqref{eom3} correspond to the Maxwell equations for the gauge field $A_\mu$ in curved space.

\subsection{Ansatz for finite $T$ and $\mu$}

We will consider the following ansatz for a charged black brane solution coupled to a scalar field:\footnote{We work in a convenient frame where the metric is written in terms of a function  $\zeta(z)$ and the blackening factor $f(z)$ \cite{Ballon-Bayona:2017sxa}.} 
\begin{eqnarray}
ds^2 &=& \dfrac{1}{\zeta(z)^2}\left(\dfrac{dz^2}{f(z)} - f(z)dt^2 + d\vec{x}^2\right), \label{Metriczeta} \\
\phi &=& \phi(z), \label{phi} \\ 
A &=& A_{t}(z)\,dt,  \label{At}
\end{eqnarray}
\noindent 
where $f(z)$ is the horizon function and $A_t$ is the time component of the $U(1)$ gauge field $A_\mu(z)$. The former will be related to the temperature $T$ whereas the latter will be related to the chemical potential of the dual theory, as we will discuss later in this work. 

The ansatz in \eqref{Metriczeta}-\eqref{At} leads to to static charged black holes enjoying $SO(3)$ symmetry. It is a very general ansatz for gravity backgrounds describing 4d non-conformal fluids at finite temperature and density. Conformal symmetry breaking in the 4d fluid, characterised by the trace anomaly, will be a consequence of a deformation due to a 4d scalar operator ${\cal O}$ (dual to $\phi$). 

Plugging the ansatz  \eqref{Metriczeta}-\eqref{At} into the field equations  \eqref{eom1}-\eqref{eom3} lead to the following independent equations: 
\begin{eqnarray}
\frac{\zeta''(z)}{\zeta(z)} - \frac{4}{9}\phi'(z)^2 &=& 0,\label{breqn3} \\
\frac{\zeta'(z)}{\zeta(z)} - \frac{A_{t}''(z)}{A_{t}'(z)} &=& 0,\label{Atreqn3} \\
\frac{d}{dz}\left(\zeta(z)^{-3}f'(z)\right) - \frac{A_{t}'(z)^2}{\zeta(z)} &=& 0,\label{freqn3} 
\end{eqnarray}
where $'$ denotes derivative w.r.t $z$, and there is an extra field equation involving the dilaton potential:
\begin{eqnarray}  \label{potential}
V(\phi) = 12\zeta'(z)^2 f(z) - 3\zeta'(z)f'(z)\zeta(z) - \frac{4}{3}f(z)\zeta (z)^2\phi'(z)^2 + \frac{1}{2}\zeta(z)^{4} A_{t}'(z)^2.  \;\; 
\end{eqnarray}
We will solve the  equations for the functions $\zeta(z)$, $ f(z) $, $ A_{t}(z) $ and $ V(\phi) $ for a given dilaton profile $\phi(z)$.  In particular, solving the differential equation \eqref{breqn3} one finds the scale factor $\zeta(z)$ for a given dilaton profile $\phi(z)$. This differential equation is linear in $\zeta$ and admits analytic solutions when the dilaton is given by a power-law ansatz. In this work we consider the quadratic ansatz: 
\begin{equation}\label{QuadraticDilaton}
\phi(z) = k z^2,
\end{equation}
with $k$ a positive constant to be  fixed latter.  It is important to mention that the  dilaton in \eqref{QuadraticDilaton} fulfills the IR criterion established in \cite{Gursoy:2007cb, Gursoy:2007er} for a good dilaton profile leading,  for instance, to confinement in the dual gauge theory as well as linear Regge trajectories for scalar and tensor glueballs \cite{Gursoy:2007er,Ballon-Bayona:2017sxa}. Note that in this work we take into account the dilaton backreaction in the geometry \footnote{This is in contrast with phenomenological soft wall models where backreaction is not taken into account.}. 

The quadratic ansatz for the dilaton \eqref{QuadraticDilaton} corresponds to a dual relevant operator of conformal dimension two.
For the case of zero temperature and zero chemical potential, corresponding to $f=1$ and $A_t=0$, one finds that the dilaton potential in \eqref{potential} can be expanded\footnote{Note that the dilaton mass saturates the Breitenlohner-Freedman bound.} at small $\phi$ as $V(\phi)= 12 - (4/3) m^2 \phi^2$, with $m^2=-4$. This is  compatible with the AdS/CFT dictionary $m^2=\Delta (\Delta -4)$ with $\Delta=2$ the conformal dimension of the 4d operator ${\cal O}$. Near the boundary, the source and the VEV coefficients  have non-zero values, both equal to $k$. This corresponds to a Dirichlet condition in the UV that fixes the source coefficient and an IR condition that fixes the VEV coefficient.

An alternative procedure in EMD holography consists of providing an ansatz for the dilaton potential $V(\phi)$ and then solving the EMD equations in order to find  $\phi(z)$, $\zeta(z)$ $A_t(z)$ and $f(z)$. Considering an ansatz for the dilaton potential that enjoys the same IR and UV asymptotics chosen in this work would lead to the same thermodynamics and phase diagram \footnote{The approach of providing an ansatz for the dilaton potential is useful for investigating holographic renormalisation but in general the background solutions are not analytic. }.

It is by now well understood that the deformation of AdS space due to a dilaton field $\phi$ in 5d is dual to the deformation of a 4d CFT due to a scalar operator ${\cal O}$. In the 4d theory the deformation can be written as $\int d^4 x \, \phi_0 \langle {\cal O} \rangle$, where $\phi_0$ and $\langle {\cal O} \rangle$ are the coupling and VEV of the scalar operator respectively. Conformal symmetry breaking due to the scalar operator is described by the Ward identity   $T^{\mu}_{\, \, \mu} = \phi_0 (4 - \Delta) \langle {\cal O} \rangle $ for the trace of the stress-energy tensor \cite{Skenderis:2002wp}. This Ward identity is reminiscent of the QCD trace anomaly  \cite{Gubser:2008yx,Ballon-Bayona:2017sxa}; see also \cite{Ballon-Bayona:2018ddm}. The parameter  $\Delta$ is the conformal dimension of the scalar operator ${\cal O}$. For the quadratic ansatz in \eqref{QuadraticDilaton} we easily find that $\Delta=2$ (relevant operator) whereas $\phi_0$ and $\langle {\cal O} \rangle$ are both proportional to $k$. We therefore expect a non-vanishing trace anomaly and therefore the explicit breaking of conformal symmetry.  In this work we will show that this is the case even at finite temperature and density. However, we will find a very special point in the phase diagram where the trace anomaly vanishes and conformal symmetry is restored. This is the critical point (the point where the critical line ends in the phase diagram). The vanishing of the trace anomaly at this point indicates the presence of a non-trivial CFT. Later in this work we will describe the thermodynamics near this special point. 

\subsection{Solving the EMD Equations} 

\medskip 

In principle, for any given dilaton profile, $ \phi(z) $, one can solve the  equations \eqref{breqn3}, \eqref{Atreqn3} and \eqref{freqn3} for the unknown functions $\zeta(z)$, $ f(z) $ and $ A_{t}(z)$, either analytically or numerically, with the constraint that the background must asymptotically approach a charged AdS black hole in the UV ($z\rightarrow0$), which corresponds on the boundary to a gauge theory at finite temperature and density. 

In this subsection we solve the equations \eqref{breqn3}, \eqref{Atreqn3} and \eqref{freqn3} for the particular case of a quadratic dilaton profile \eqref{QuadraticDilaton}.

\subsubsection{Solution for $\zeta(z)$}

The differential equation in \eqref{breqn3} is second order and linear in $\zeta(z)$. For the quadratic dilaton profile $\phi(z)=kz^2$ the equation can easily be solved and we impose the condition $\zeta(z)=z$ at small $z$ to obtain an asymptotically AdS space. The solution reads 
\begin{equation}
\zeta(z) = z\,\,_0F_{1}\left(\dfrac{5}{4};\dfrac{k^2\,z^4}{4}\right) \label{zetasol},
\end{equation}
where $_0F_{1}\left(a;z\right) $ is the confluent hypergeometric function. This solution can also be written as
\begin{equation}\label{zetaBessel}
\zeta(z) =  \Gamma\left(\frac{5}{4}\right) \left ( \frac{3}{k} \right )^{1/4} \sqrt{z} \;  I_{\frac{1}{4}} \left ( \frac23 k z^2 \right ), 
\end{equation}
\noindent where $\Gamma (x)$ is the usual Gamma function and $I_{\alpha}(x)$ is the modified or hyperbolic Bessel function of the first kind. 

Note that the function $\zeta(z)$ admits an expansion in the regime $k z^2 \ll 1$
\begin{equation} \label{zetaExp}
\zeta(z) = z \Big [ 1  + \frac{4}{45} (k z^2)^2 + \frac{8}{3465}(k z^2)^4  + \dots \Big ]\, .
\end{equation}

\begin{figure}[ht]
	\centering
	\includegraphics[scale = 0.3]{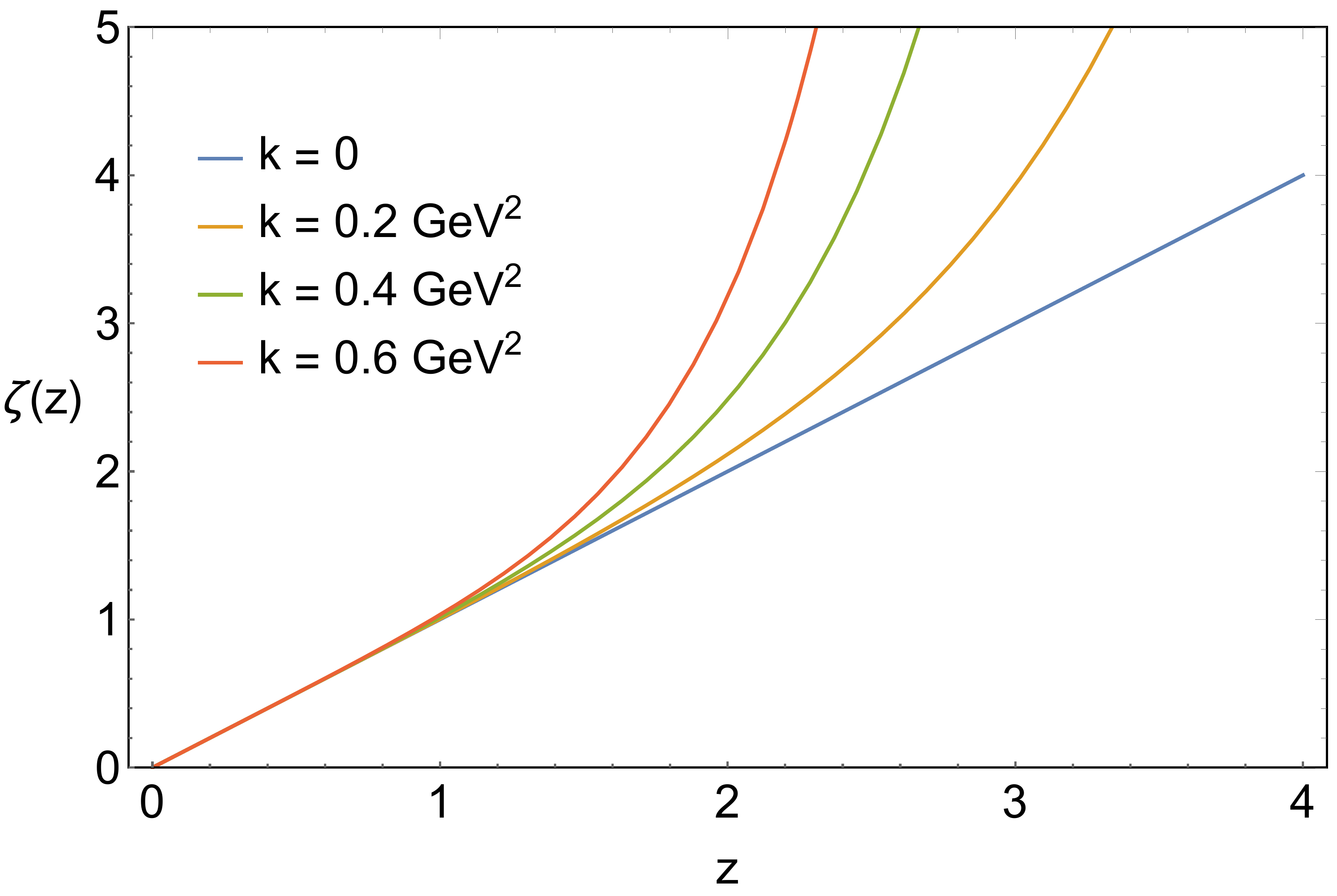}
	\caption{Behaviour of the warp factor $ \zeta(z) $ from Eq. \eqref{zetasol}, or eq. \eqref{zetaBessel}, as a function of $z$ for some values of $k$, expressed in GeV$^2$. For $k=0$ one recovers the AdS solution. The greater the value of $k$ the greater the departure of $\zeta(z)$ from the pure AdS solution for large $z$. For small $z$ all solutions coalesce to the AdS solution.}
	\label{fig:zetaz}
\end{figure}

In figure \ref{fig:zetaz} we display the behaviour of $ \zeta(z) $ as a function of $ z $, for some choices of the parameter $k$. Note that for the choice $k=0$ one recovers the AdS solution since in this particular case $\zeta(z)=z$.  

\subsubsection{Solution for the Gauge Field Component $ A_{t}(z) $}

Now, we proceed to solve the differential equation \eqref{Atreqn3} for the time component $A_t$ of the Maxwell field $A_{\mu}$. Its general solution can be written as 
\begin{equation}
A_{t}(z) = c_2 - c_{3}\int_0^{z} \zeta(y)\,dy,
\end{equation}
where $ c_2 $ and $ c_3 $ are integration constants, and $ \zeta(y) $ is the scale factor found in \eqref{zetasol}. One can fix $ c_2 $ and $ c_3 $ by imposing the regularity condition at the horizon $ A_{t}(z=z_h) = 0$, so that the norm $||A_{t}(z)|| = g^{tt}|A_{t}|^2$ is well defined, and the AdS/CFT dictionary $A_{t}(z=0) = \mu$, where $ \mu $ is the chemical potential of the dual gauge theory.  
Using these conditions the  solution for $ A_{t}(z) $ takes the form
\begin{eqnarray}
A_{t}(z) &=& \mu\, \dfrac{\int_{z}^{z_h} \zeta(y)\,dy}{\int_0^{z_h} \zeta(y)\,dy}
= \mu \Big [ 1 - \frac{C_2 (z)}{C_2(z_h)} \Big ], \label{Asol}
\end{eqnarray}
where we have introduced the function 
\begin{equation} \label{C2}
 C_2 (z) = \int_0^{z} \zeta (y) dy \,.    
\end{equation}
\noindent 

In the UV ($z\rightarrow0$)  $ A_{t}(z) $ should reduce to
\begin{eqnarray} \label{AtUV}
A_{t}(z) = \mu\Big [ 1-\left(\dfrac{z}{z_h}\right)^2 \Big ] \, ,
\end{eqnarray}
which is the solution corresponding to the Reissner-Nordstr\"om (RN) AdS black brane \cite{Hartnoll:2009sz}. Indeed, if we expand the integrand $\zeta(y)$ around $ y=0 $ up to linear order in \eqref{Asol} and integrate the result we find \eqref{AtUV}.

The potential $A_{t} (z)$  couples to the charge density operator $j^t= \bar q \gamma^t q$  at the boundary. 
The interaction term  $ \langle j^t \rangle A_t (0)$ is identified with the term $\rho \, \mu$ in the grand canonical ensemble. Therefore the quark density $\rho$ can be obtained from the holographic dictionary
\begin{equation} \label{chargedict}
\rho = \langle j^t \rangle = \frac{\delta S}{ \delta A_t} = - \sigma \Big [ \frac{1}{\zeta(z)} \partial_z A_t  \Big ]_{z= \epsilon} \, 
\end{equation}
where $\sigma = (16 \pi G_5)^{-1}$. We have defined the boundary at $z=\epsilon$ and we take in the end the limit $\epsilon \to 0$. 
Plugging the solution \eqref{Asol} into \eqref{chargedict} we find 
\begin{equation} \label{chargedensity}
\rho(z_h, \mu) = \frac{\sigma}{C_2 (z_h)} \mu \,.     
\end{equation}
The quark density $\rho$ depends linearly in $\mu$ and has a non-trivial dependence in $z_h$. It is finite already so it does not require any renormalisation procedure.

\subsubsection{Solution for the Horizon Function $ f(z) $}

\begin{figure}[ht]
	\centering
	\includegraphics[scale = 0.3]{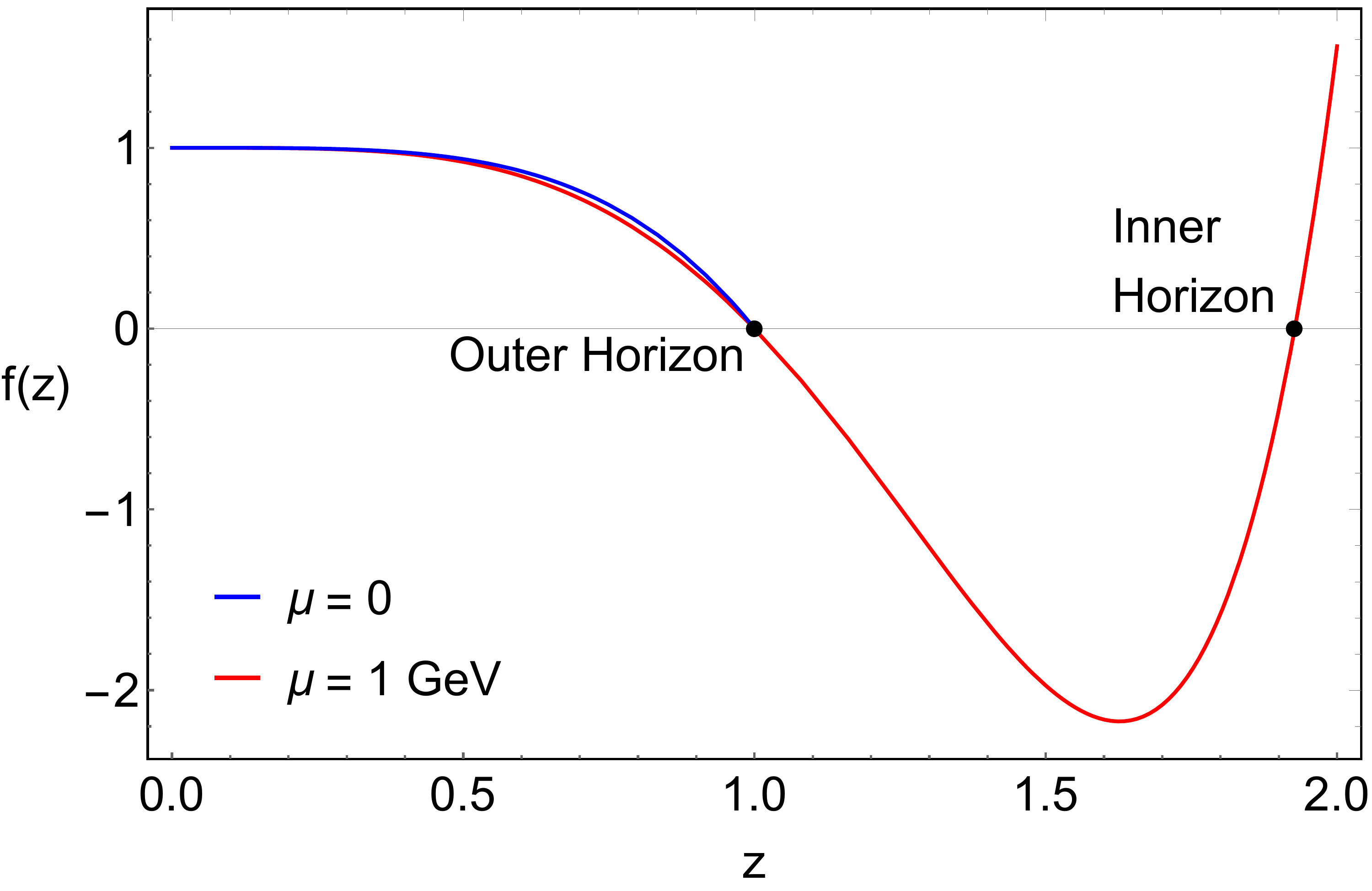}
	\caption{Behaviour of $ f(z) $ from Eq. \eqref{horfun} as a function of $z$ for zero and finite chemical potential $\mu$. We set $z_h = 1$ GeV$^{-1}$. The black dots represent the outer and the inner horizons.}
	\label{fig:f(z)}
\end{figure}

Now we perform the last and most important step which is solving the equation \eqref{freqn3} for the horizon function $ f(z)$. Plugging the solution for $A_t(z)$, cf.\eqref{Asol}, one can write the solution for $ f(z) $ in a formal way as
\begin{equation} \label{horfun}
f(z) = 1 - \dfrac{C_4(z)}{C_4(z_h)} + \dfrac{\mu^2}{C_4(z_h)\,C_2(z_h)^2}\Big [C_4(z_h)\,C_6(z) - C_4(z)\,C_6(z_h)\Big ],
\end{equation}
where the $C_4$ and $C_6$ functions are given by 
\begin{eqnarray} 
C_4(z) &=& \int_0^{z} \zeta(y)^3 \, dy,\label{C4}\\
C_6(z) &=& \int_0^{z}C_2(y)\,\zeta(y)^3\,dy\label{C6}\,,
\end{eqnarray}
\noindent and $C_2(z)$ is given by Eq. \eqref{C2}. The integration constants were chosen in order to satisfy the conditions $f(0) = 1$ (AdS asymptotics) and $f(z_h) = 0$ (horizon property). In figure \ref{fig:f(z)} we show the profile of the horizon function $f(z)$ as a function of $z$ for $\mu=0$ and $\mu\neq0$. Note that the functions $C_n(z)$ have dimension $[{\rm Length}]^n$.

Note that in the regime $k z^2 \ll 1$ the functions above admit the following expansions:
\begin{align}
C_2 (z) &=  \frac{z^2}{2} \Big [ 1 + \frac{4}{135} (k z^2)^2 + {\cal O} (k z^2)^4  \Big ] \,,  \nonumber \\
C_4(z) &= \frac{z^4}{4} \Big [ 1 + \frac{2}{15} (k z^2)^2 + {\cal O} (k z^2)^4 \Big ]    \, , \nonumber \\
C_6 (z)  &=  \frac{z^6}{12} \Big [ 1 + \frac{8}{45} (k z^2)^2  + {\cal O} (k z^2)^4\Big ] \,.
\end{align}
These expansions will be useful later.

From this point we will set the value of constant $k = 0.18$ GeV$^2$. This choice leads at $T=0$ to the appropriate value for the $\rho$ meson mass \cite{Li:2013oda,Chelabi:2015cwn}.

In figure \ref{fig:f(z)} one can see, in the case of finite chemical potential, the presence of two horizons: the inner and the outer. This is characteristic of charged black holes. The first one is a non-physical and the second one is the physical horizon, which satisfies
the inequality
\begin{equation}\label{fconstraint}
f'(z_h)<0,
\end{equation}
which implies a constraint between the $C$ functions to be described in the next section.

Finally, as a consistency check, in the UV we should obtain the corresponding $ f(z) $ for a charged AdS black brane solution. Indeed, one can check that \footnote{This expression is equivalent to eq. $(60)$ of  \cite{Hartnoll:2009sz} for the case $\gamma^2=3$ that corresponds to taking the Maxwell coupling  equal to the gravitational coupling as we did.}
\begin{equation}
f(z\rightarrow0) = 1-\left(1+\frac{\bar{\mu}^2}{3}\right)\left(\dfrac{z}{z_h}\right)^4 +\frac{\bar{\mu}^2}{3}\left( \dfrac{z}{z_h}\right)^6; \quad \bar{\mu}:= \mu\,z_h.
\end{equation}
Therefore, in the UV we recover the horizon function $ f(z) $ associated with the RN AdS$_5$ black brane.

\section{Thermodynamics}
\label{Sec:Thermodynamics}


In this section we study the thermodynamics of our holographic model consisting of 5d asymptotically AdS charged black branes coupled to a scalar field in the grand canonical ensemble.

\subsection{The black hole entropy}

The black hole entropy $ S $ is related to the horizon area by the Bekenstein-Hawking formula
\begin{equation}\label{bhentropy}
S = \dfrac{\mathcal{A}}{4\,G_5} = \dfrac{4\pi\sigma}{\zeta^{3}(z_h)},
\end{equation}
where we used the relation $G_5=1/(16\pi \sigma)$. The constant $\sigma$  will later be fixed by the Stefan-Boltzmann law at high temperatures.

\begin{figure}[ht]
	\centering
	\includegraphics[scale = 0.45]{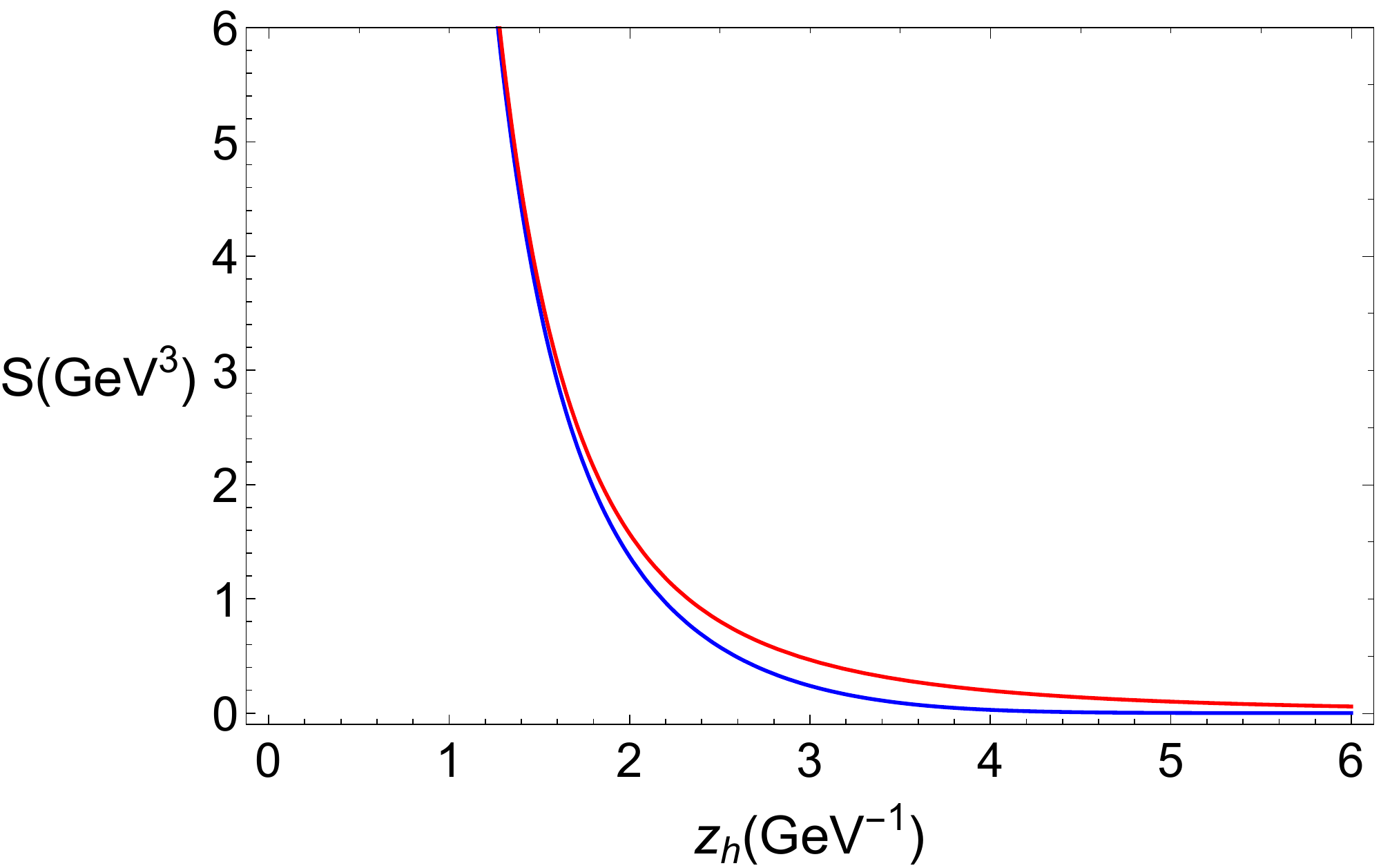}
	\caption{Entropy $ S  ({\rm GeV}^3)$ as a function of $z_h$ (${\rm GeV}^{-1}$), from eq. \eqref{bhentropy}  for $\sigma=1$. The blue  curve represents the result for $k$ = 0.18 GeV$^2$ whilst the red curve represents the result for $k=0$, corresponding to the RN AdS$_5$ black brane.  }
	\label{fig:Entropyvszh}
\end{figure}

Since the function $\zeta(z_h)$ monotonically increases with $z_h$ the entropy is a decreasing function of $z_h$ We plot in figure \ref{fig:Entropyvszh} the entropy in our model (blue curve) compared with the entropy of the pure charged AdS black brane (red curve). 

\subsection{The black holes temperature} 
\label{subsec:Temperature}

In this section we present a systematic description of the temperature of charged black branes  dual to non-conformal plasmas. We will obtain general expressions in terms of a general scale factor $\zeta(z)$, related to a general dilaton profile $\phi(z)$ and then we present numerical results for the profile in \eqref{zetasol}, obtained as a solution of the EMD equations in the case of the quadratic dilaton profile $\phi(z)= k z^2$. 

The analysis starts with the Hawking formula \cite{Hawking:1982dh} for the black hole (BH) temperature:
\begin{equation} \label{Tdef}
T = \dfrac{|f'(z_h)|}{4\pi}.
\end{equation}
This formula arises from the requirement of smoothness of the metric in $d\tau$  and $dz$  near the horizon radius $z=z_h$, where $\tau$ is the imaginary time with period $\beta =1/T$. 

Below we describe the temperature $T$ as a function of the horizon radius $z_h$ and the chemical potential $\mu$. First we describe the simpler case $\mu=0$ and then the full result for finite $\mu$.

\subsubsection{Zero Density $ \mu=0 $}
\indent
For zero chemical potential the horizon function in \eqref{horfun}  reduces to
\begin{equation}
 f(z) = 1 - \frac{C_4(z)}{C_4(z_h)} \, ,
\end{equation}
with $C_4(z)$ defined in \eqref{C4}. The temperature becomes
\begin{equation}\label{Tmueq0}
T_{\mu=0}(z_h) = \dfrac{\zeta(z_h)^3}{4\,\pi\int_{0}^{z_h}\zeta(y)^3\,dy},
\end{equation}
and we will be interested in the solution \eqref{zetasol} for $ \zeta(z) $.  As explained in the previous section, in the conformal case we have $\zeta(z) = z$ and, for $\mu=0$, the horizon function reduces to $f(z) = 1 - z^4/z_h^4$ which corresponds to the usual AdS BH with flat horizon (Poincar\'e coordinates). In the non-conformal case we have and AdS deformation due to the dilaton $\phi(z) = k z^2$ controlled by an IR mass scale $\sqrt{k}$.

\begin{figure}[ht]
	\centering
	\includegraphics[scale = 0.4]{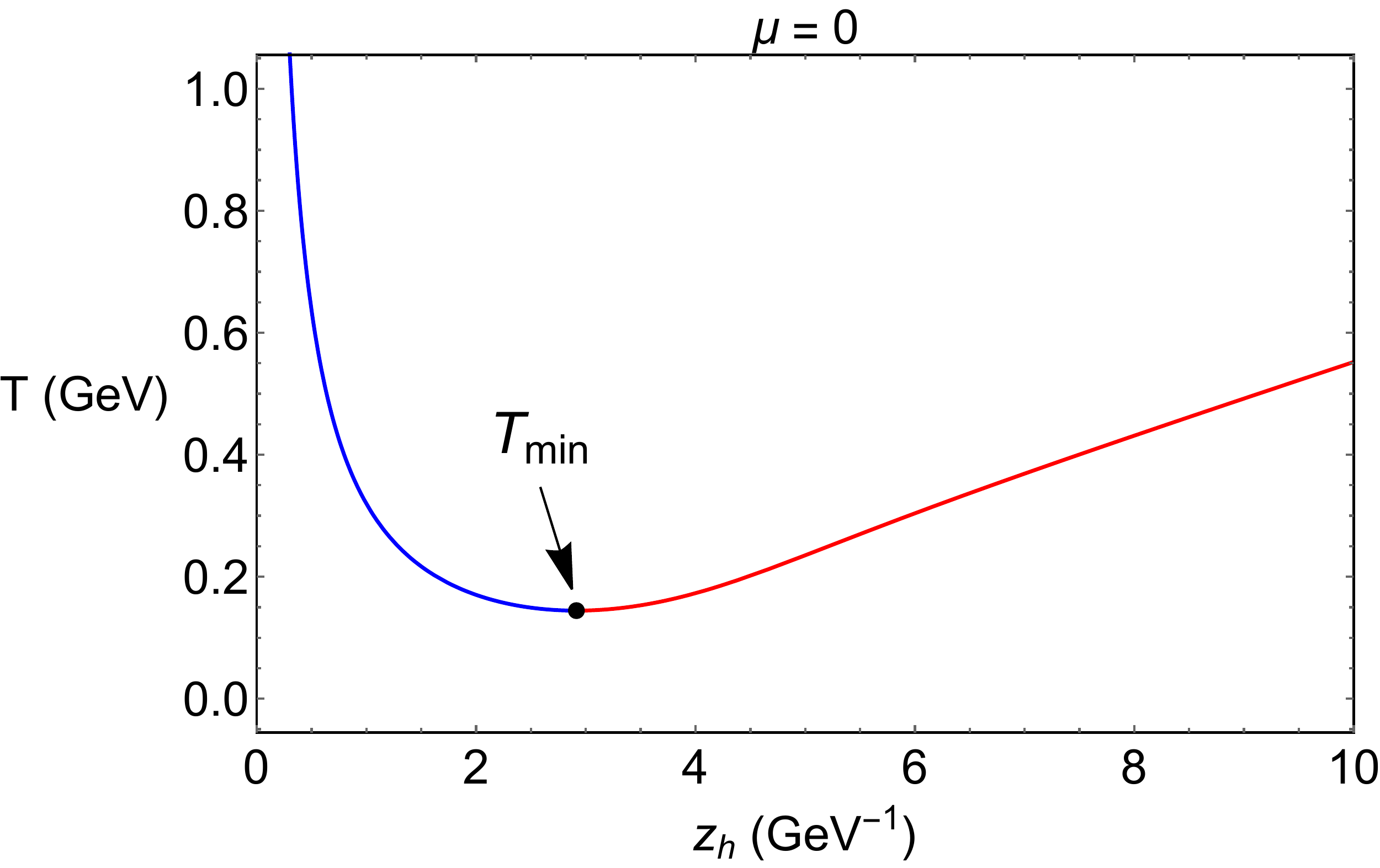}
	\caption{Temperature $ T $ (GeV) as a function of $z_h$ (GeV$^{-1}$) for $\mu=0$ and $k$ = 0.18 GeV$^2$ from Eq.\eqref{Tmueq0}. The blue and red curves represent the physical large BH and the non-physical small BH respectively. The black dot represents the minimum temperature.}
	\label{fig:T0(zh)}
\end{figure}

In figure \ref{fig:T0(zh)} , using Eq. \eqref{Tmueq0}, we show the behaviour of $T_{\mu=0}(z_h)$ (temperature vs horizon radius at $\mu=0$) for the numerical solution. Note that the temperature $T_{\mu=0}(z_h)$ presents a minimum value, $ T_{min} \simeq 144$ MeV, above which one can distinguish two black hole phases \cite{Bohra:2019ebj,Gursoy:2008za}:
\begin{enumerate}
	\item \textbf{Large BH}: $T_{\mu=0}(z_h)$ decreases as $ z_h $ increases (\textbf{stable phase}).
	\item \textbf{Small BH}: $T_{\mu=0}(z_h)$ increases as $ z_h $ increases (\textbf{unstable phase}).
\end{enumerate}
The two branches for $T(z_h)$ are depicted by blue and red curves. The blue curve represents the large BH that was present already in the case $k=0$ whereas the red curve represents the emergence of an non-physical small BH due to conformal symmetry breaking. The BH solutions exist only above $T_{min}$, which is related to the IR mass scale $\sqrt{k}$.

The stability/instability of the BH phases is based on the free energy for each case. One can show that the free energy is smaller for the large BH than the small BH, as will be clear later on.

\medskip 

{\bf Analytic approximation for the temperature}

\medskip 

In the regime $k z_h^2 \ll 1$ we can expand \eqref{zetasol} in powers of $k z^2$ so that \eqref{Tmueq0} becomes 
\begin{equation} \label{ApproxTmueq0}
T_{\mu=0}(z_h) = \frac{1}{\pi z_h} \Big [ 1 + \frac{2}{15} (k z_h^2)^2 + {\cal O} (k z_h^2)^4 \Big ]  \, ,  
\end{equation}

We can truncate \eqref{ApproxTmueq0} to estimate the minimum temperature. From the condition $T_{\mu=0}'(z_h^*)=0$ we find
\begin{equation}
z_h^* \approx \frac{5^{1/4}}{2^{1/4} \sqrt{k}} \quad, \quad 
T_{min} = T(z_h^*) \approx \frac{4}{3 \pi} \frac{2^{1/4}}{5^{1/4}} \sqrt{k} \,. \end{equation}
For $k=0.18 \, {\rm GeV}^2$ we obtain $z_h^*=2.96 \, {\rm GeV}^{-1}$ and $T_{min}=0.143 \, {\rm GeV}$. These values are quite close to the numerical results $z_h^*=2.92 \, {\rm GeV}^{-1}$ and $T_{min}=0.144 \, {\rm GeV}$ found from the exact  solution. 
The inverse of the temperature in \eqref{ApproxTmueq0} takes the form 
\begin{equation} \label{Approxbetamueq0}
\beta_{\mu=0}(z_h) = \frac{1}{T_{\mu=0}(z_h)} = \frac{ \pi z_h}{1 + \frac{2}{15} (k z_h^2)^2 + {\cal O} (k z_h^2)^4}    \,. \end{equation}
This expression looks similar to the analytic expression found in the case of an AdS BH in global coordinates \cite{Hawking:1982dh}.

\subsubsection{Finite Density: $\mu\neq0$}

Now we consider the effect of finite density which corresponds to turning on the chemical potential $\mu$. We will find that having non-zero $\mu$ allows for a new small BH solution that is stable or metastable and therefore competes with the large BH.

In  the case of finite chemical potential $\mu$, we use the general expression in \eqref{horfun} for the horizon function $f(z)$ and the temperature $ T $ in \eqref{Tdef} becomes
\begin{align} \label{T}
T(z_h,\mu) 
&= \dfrac{\zeta^{3}(z_h)}{4 \pi \,C_4(z_h)} 
\left\{ 1 - \mu^2\,\left [\dfrac{C_4(z_h)}{C_2(z_h)} - \dfrac{C_6(z_h)}{C_2(z_h)^2} \right ] \right\}  \nonumber \\
&\equiv a(z_h) \Big [ 1 - \mu^2 \, b(z_h) \Big ] \, .
\end{align}
From \eqref{bhentropy} and \eqref{T} we find the relation
\begin{equation}\label{Tcompact2}
TS = \sigma\left[ \dfrac{1}{C_4} + \dfrac{\mu^{2}}{C_4}\,\left( \dfrac{C_6}{C^2_2} -\dfrac{C_4}{C_2}\right) \right] \,,
\end{equation}
where the $C's$ depend only on the horizon radius $z_h$. 

Note that the constraint on the horizon function \eqref{fconstraint} is equivalent to the physical fact that the temperature is non-negative, i.e., $ T\geq0 $ or, similarly, it implies that the BPS bound, given by 
\begin{equation} \label{BPSbound}
1 - \mu^2\,b(z_h)  \geq 0,
\end{equation}
must be satisfied. For extremal BHs or, BPS BHs ($ T=0 $), we have the saturation of the BPS bound above, i.e,
\begin{equation}
1 - \mu^2\,b(z_h) = 0.
\end{equation}

In the UV one can show, by expanding the temperature \eqref{T} around $z_h\rightarrow0$, that it reduces to the RN AdS$_{5}$ temperature,  given by 
\begin{equation}
T = \dfrac{1}{\pi\,z_h}\left(1 - \dfrac{\bar \mu^2}{6} \right )\, .
\end{equation}
where $ \bar{\mu} = \mu\,z_h$. Note that the condition $T \geq 0$ implies that $0 \leq \mu\,z_h \leq \sqrt{6}$.

\begin{figure}[ht]
	\centering
	\includegraphics[scale = 0.3]{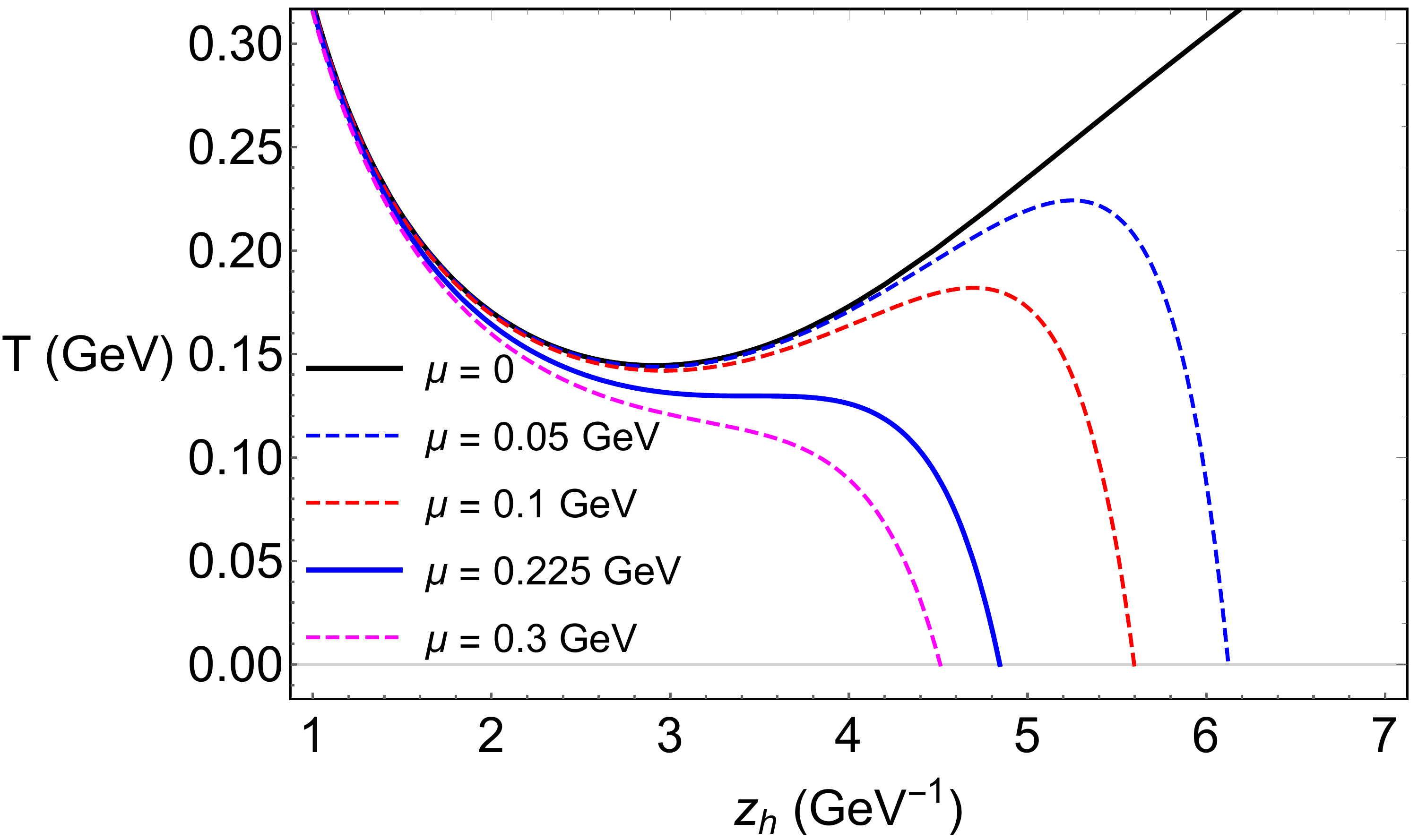}
	\hfill
	\caption{Temperature $ T $ (GeV) as a function of $z_h$ (GeV$^{-1}$), from Eq. \eqref{T},  for $k$ = 0.18 GeV$^2$  and different values of $\mu$.}
	\label{fig:Tfinitemu}
\end{figure}

In Fig.\ref{fig:Tfinitemu}  we display the temperature $ T $ as a function of the horizon radius $z_h$ for different values of the chemical potential $\mu$ using the formula \eqref{T}.

From the figure we see that the effect of a finite chemical potential on the curves $T(z_h)$ is to bring a new small BH solution (large $z_h$). 
Later in this work we will conclude that this new BH solution is physical (stable or metastable) and will compete with the large BH solution leading to a first order thermodynamic transition. As the chemical potential increases, the non-physical small BH will disappear and the physical small BH will end up merging with the large BH at sufficiently large $\mu$. We will see that this transition is very similar to the {\it van der Waals-Maxwell liquid-gas transition}.

One can distinguish four different scenarios for $T(z_h)$: (i) $ \mu=0$; (ii) $ \mu<\mu^c $; (iii) $ \mu=\mu^c $; (iv) $ \mu>\mu^c $, where $\mu^c$ is the critical chemical potential where the non-physical small BH disappears and the physical small BHs merges with the large BH.

\begin{figure}[ht]
	\centering
	\includegraphics[scale = 0.29]{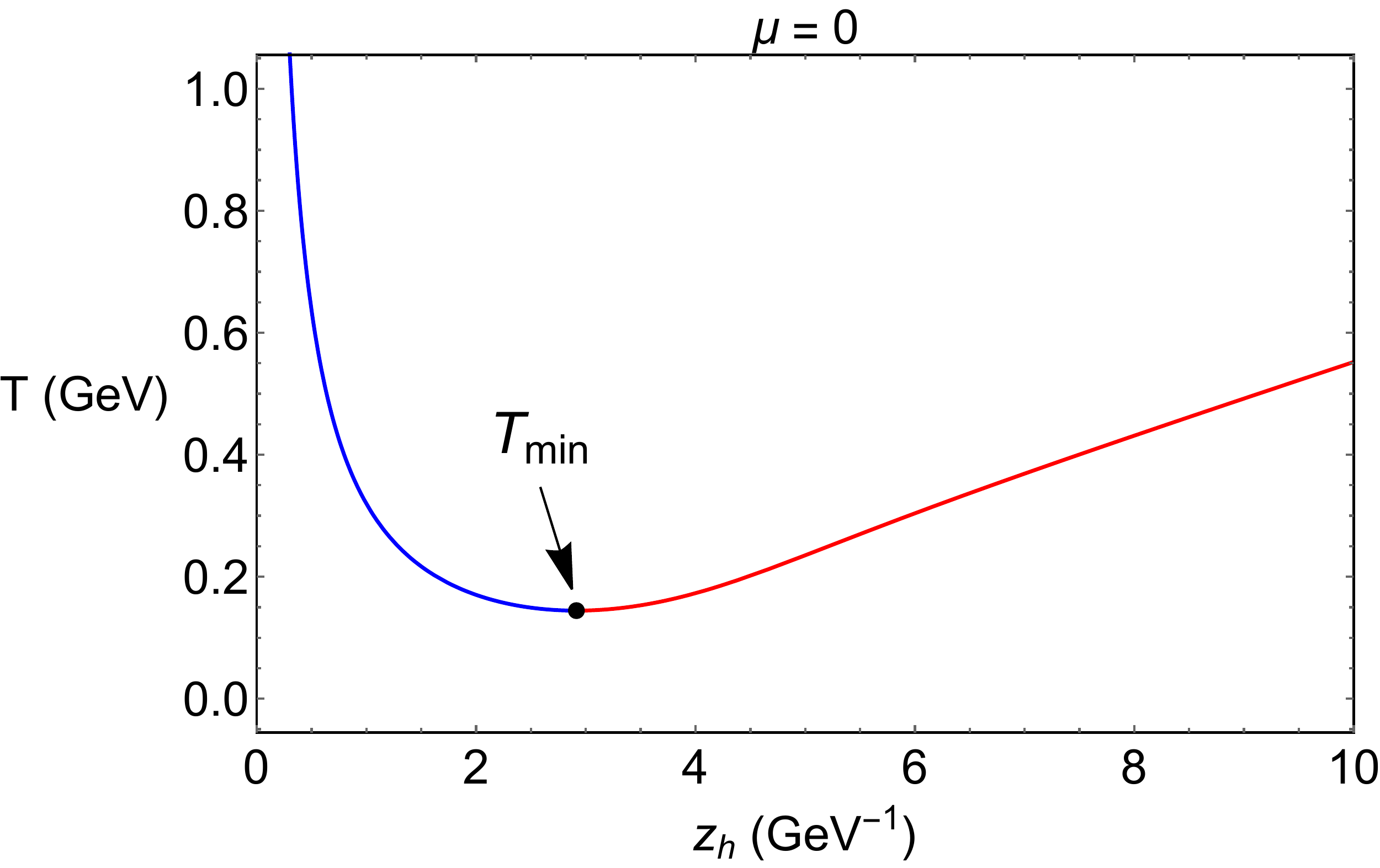}
	\hfill
	\includegraphics[scale = 0.24]{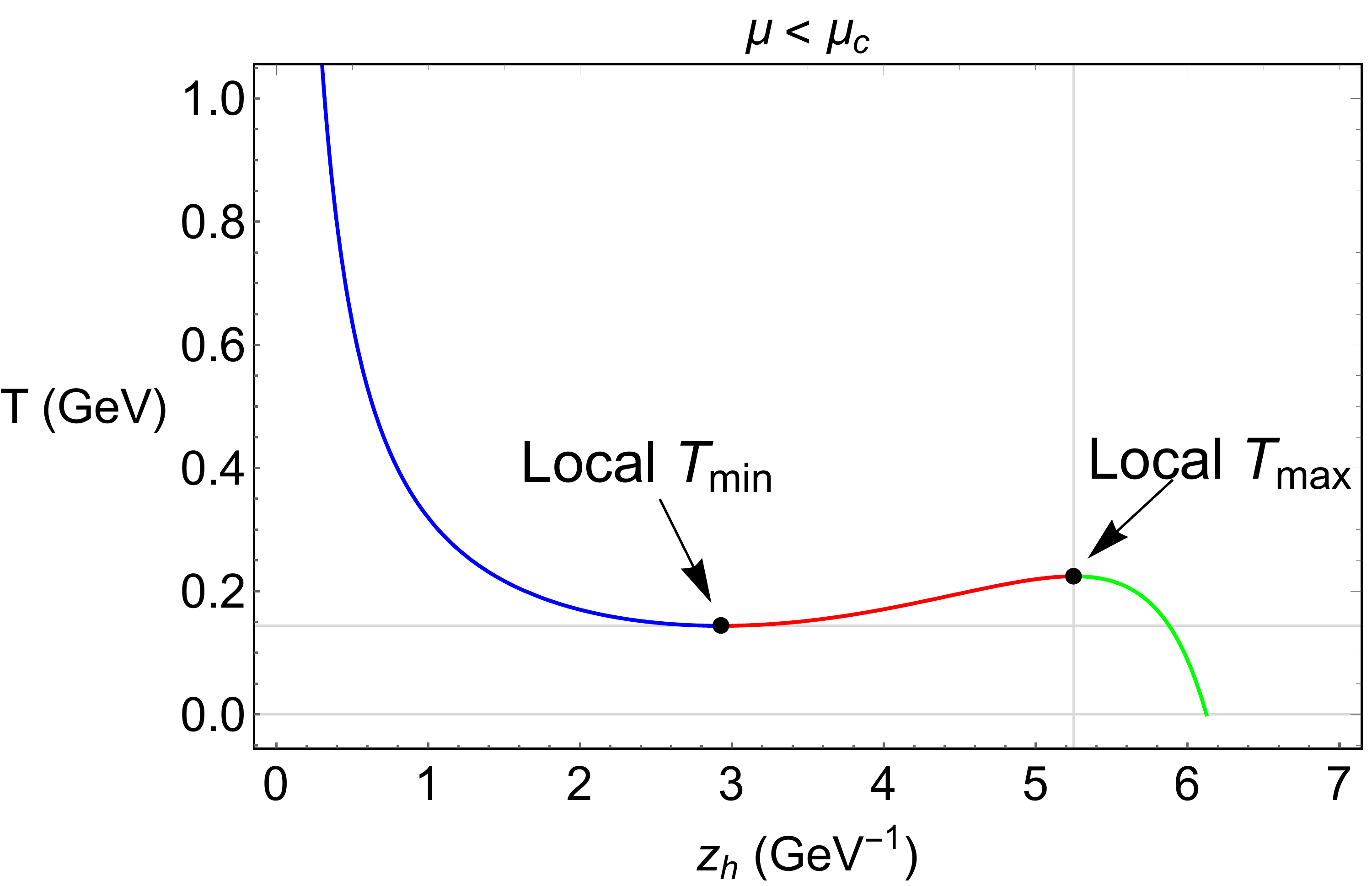}
	\vfill
	\includegraphics[scale = 0.24]{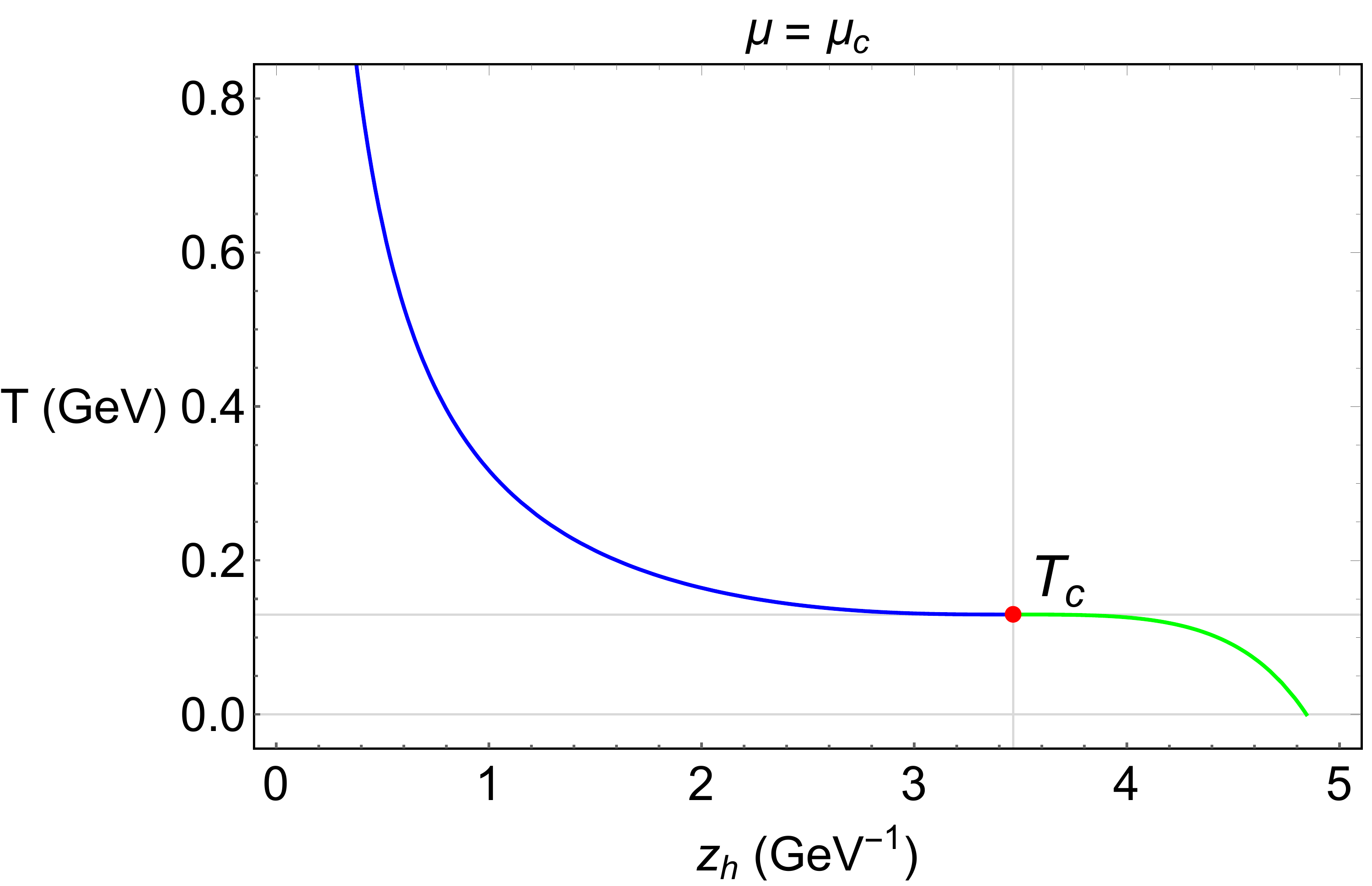}
	\hfill
	\includegraphics[scale = 0.24]{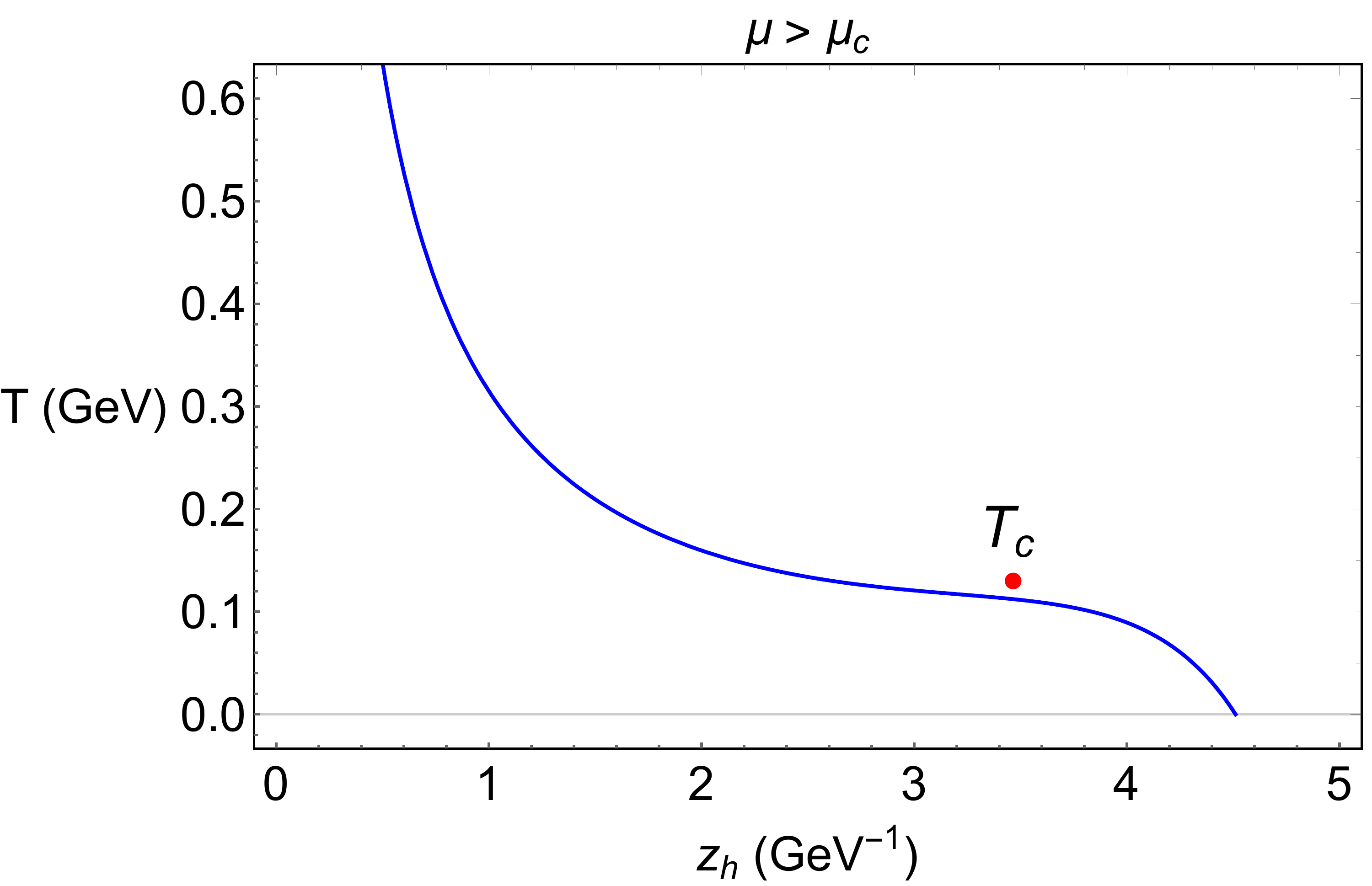}
		\caption{BH temperature $T$ as a function of the horizon radius $z_h$ for the scenarios $\mu=0$ (upper left panel), $\mu<\mu^c$ (upper right panel), $\mu=\mu^c = 0.225 \, {\rm GeV}$ (lower left panel) and $\mu>\mu^c$ (lower right panel). The blue, red and green curves represent the physical large BH, unphysical small BH and physical small BH respectively.}
	\label{fig:Tfinitemuv2}
\end{figure}

These scenarios are displayed in Fig. \ref{fig:Tfinitemuv2}. In all the plots the large BH is depicted by a blue curve, whereas the non-physical and physical small BHs are depicted by red and green curves respectively. The second plot in Fig. \ref{fig:Tfinitemuv2} shows the appearance of a local minimum and maximum for the temperature, $T_{min}$ and $T_{max}$ respectively. The third plot corresponds to the critical case  $\mu=\mu^c \approx 0.225 \, {\rm GeV}$ where the unphysical small BHs disappears. This corresponds to the situation where $T_{min}=T_{max}=T^c \approx 0.13 \, {\rm GeV}$ and $z_h=z_h^c \approx 3.47 \, {\rm GeV}^{-1}$. 
Later in the paper we will analyse the grand canonical potential and many other thermodynamic quantities in order to describe the phase diagram $T$ vs $\mu$. 

\medskip 

{\bf Analytic approximation for the temperature}

\medskip

In the  regime $k z_h^2 \ll 1$ the functions $a$ and $b$, defined in \eqref{T}, can be expanded as
\begin{align} \label{abExp}
a(z_h) &= \frac{1}{\pi z_h} \Big [ 
1 + \frac{2}{15} (k z_h^2)^2 + {\cal O} (k z_h^2)^4 \Big ] \nonumber \\
b(z_h) &=  \frac{z_h^2}{6} \Big [ 1 + \frac{2}{27} (k z_h^2)^2 + {\cal O} (k
z_h^2)^4 \Big ] 
\end{align}

Plugging  the expansions \eqref{abExp} in \eqref{T} we find the following expansion for the temperature:
\begin{equation} \label{TExp}
  T(z_h, \mu) = \frac{1}{\pi z_h} \Big [ 1 + \frac{2}{15} (k z_h^2)^2 + {\cal O} (k z_h^2)^4 \Big ]  \Big \{ 1 - \frac16 \mu^2 z_h^2 \Big [ 1 + \frac{2}{27} (k z_h^2)^2 + {\cal O} (k z_h^2)^4  \Big] \Big \} \, .   
\end{equation}
This expansion suggests the following analytic approximation 
\begin{equation} \label{AnalyticTemp}
T_{{\rm an}}(z_h , \mu) =  \frac{1}{\pi z_h} \Big [ 1 + \frac{2}{15} k^2 z_h^4 \Big ]  \Big [ 1 - \frac16 \mu^2 z_h^2 \Big ( 1 + \frac{2}{27} k^2 z_h^4 \Big ) \Big ]    \,.
\end{equation}
Imposing simultaneously the conditions $\partial_{z_h}T_{{\rm an}}(z_h^c,\mu^c)=0$ and $\partial_{z_h}^2 T_{{\rm an}}(z_h^c,\mu^c)=0$ we find 
\begin{equation}
z_h^c=   \frac{1.57}{\sqrt{k}} \quad , \quad 
\mu^c= 0.573 \sqrt{k} \quad , \quad 
T_{{\rm an}}^c = 0.295 \sqrt{k} \, . 
\end{equation} 
For $k=0.18 \, {\rm GeV}^2$ we obtain
\begin{equation}
z_h^c=   3.69 \, {\rm GeV}^{-1} \quad , \quad 
\mu^c=  0.244 \, {\rm GeV} \quad , \quad 
T_{{\rm an}}^c= 0.125 \, {\rm GeV} \, . 
\end{equation}
These values are close to the numerical results 
\begin{equation}
z_h^c=   3.47 \, {\rm GeV}^{-1} \quad , \quad 
\mu^c=  0.225 \, {\rm GeV} \quad , \quad 
T^c= 0.13 \, {\rm GeV} \, ,
\end{equation}
obtained from the exact solution. The inverse of the temperature in \eqref{AnalyticTemp} takes the form
\begin{equation} \label{Analyticbeta}
\beta_{{\rm an}} (z_h , \mu) = \frac{1}{T_{{\rm an}}(z_h , \mu)} =  \frac{\pi z_h}{ \Big [ 1 + \frac{2}{15} k^2 z_h^4 \Big ]  \Big [ 1 - \frac16 \mu^2 z_h^2 \Big ( 1 + \frac{2}{27} k^2 z_h^4 \Big ) \Big ] }    \,.    
\end{equation}
Both formulas \eqref{AnalyticTemp} and \eqref{Analyticbeta} provide good approximations for the full numerical solution in the regime $z_h < z_h^c $. In that regime those formulas  lead to curves very similar to those of figures \ref{fig:Tfinitemu} and \ref{fig:betafinitemu}. Interestingly, the formula in \eqref{Analyticbeta} is similar to the expression found in \cite{Chamblin:1999tk} for the RN AdS$_5$ BH in global coordinates. 

\subsection{Analogy between the non-conformal plasma transition and the van der Waals-Maxwell liquid-gas transition}

\begin{figure}[ht]
	\centering
	\includegraphics[scale = 0.3]{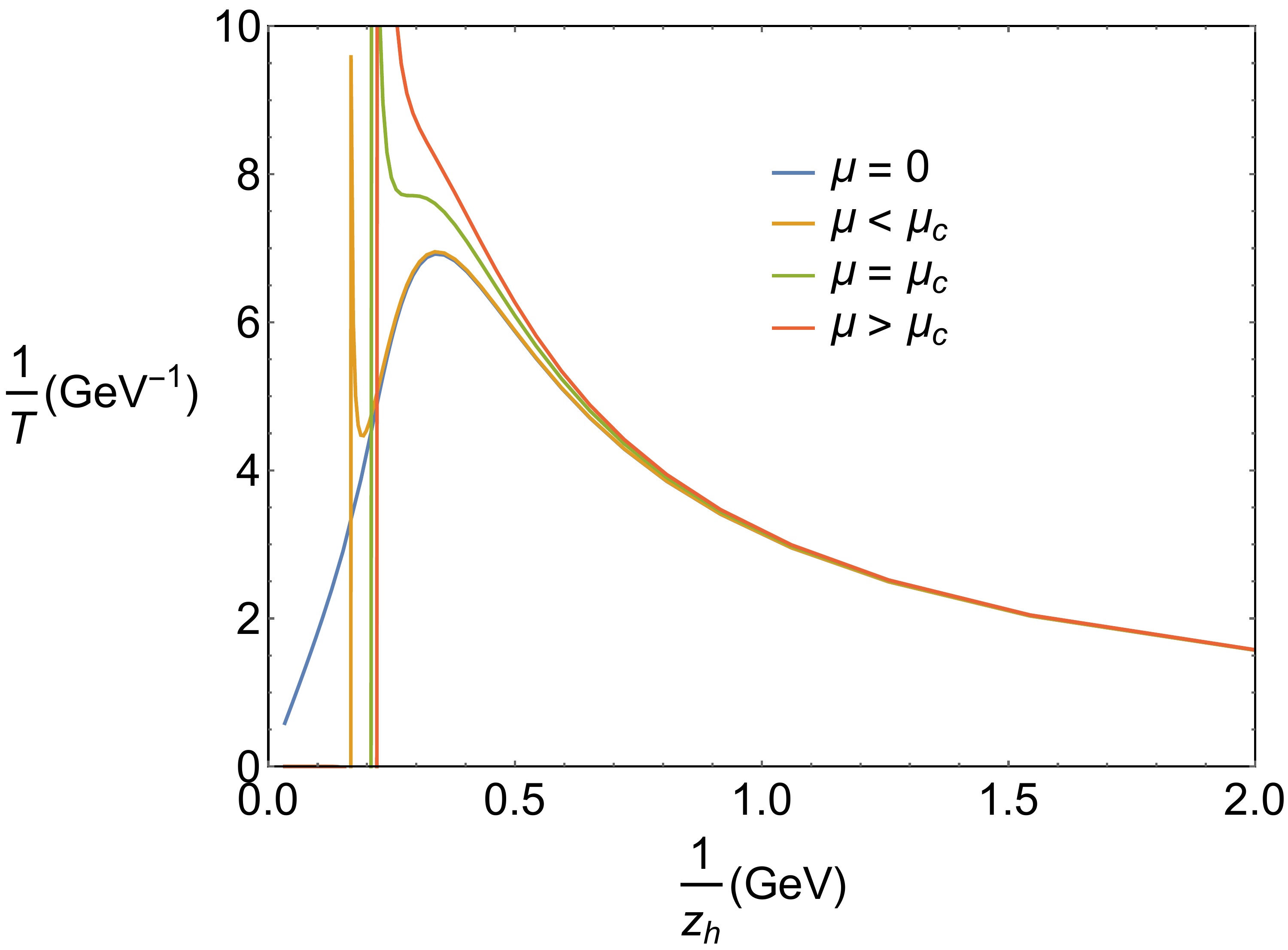}
	\hfill
	\caption{Plot of $\beta=1/T$ as a function of $1/z_h$ for the different scenarios described in Fig. \ref{fig:Tfinitemuv2}. The figure is analogous to the $P$ vs $V$ diagram in the liquid-gas transition described by the van der Waals model. The curves at fixed $\mu$ correspond to isothermal curves in the liquid-gas transition. The vertical lines represent the BPS bounds, obtained in \eqref{BPSbound}. }
	\label{fig:betafinitemu}
\end{figure}

We plot in Fig.  \ref{fig:betafinitemu} the inverse of the temperature $\beta = 1/T$ as a function of $z_h$ for different values of $\mu$. The vertical lines correspond to the physical BPS bounds for the horizon radius given in \eqref{BPSbound}. The transition from $\mu < \mu^c$ to $\mu> \mu^c$ is reminiscent of the liquid-gas transition described by the van der Waals model. This analogy was already observed in \cite{Chamblin:1999tk,Chamblin:1999hg} for the case of a charged AdS BH in global coordinates and works as follows. The quantities $\beta=1/T$ and $1/z_h$ in the non-conformal plasma transition play the role of the pressure $P$ and the volume $V$ in the liquid-gas transition. Furthermore, the chemical potential $\mu$ in the non-conformal plasma transition plays the role of the temperature $T$ in the liquid-gas transition. Then the curves at fixed $\mu$ in Fig. \ref{fig:betafinitemu} are the analog of isothermal curves in the liquid-gas transition.

We will confirm this analogy later in this work when we describe the final $T-\mu$ phase diagram. We will find a critical line, associated with a first order transition, ending on a critical point just as it happens in the $P-T$ diagram for the van der Waals-Maxwell liquid-gas transition. As suggested in \cite{Chamblin:1999tk,Chamblin:1999hg}, the similarity between the transition of non-conformal plasmas and the liquid-gas transition indicates the possibility of a description of the critical regime in terms of catastrophic theories \cite{catastrofe1, catastrofe2, catastrofe3}. Later in this work we will provide a universal description of the thermodynamics near the critical point that will allow us to establish a more concrete relation between the near criticality regime of non-conformal plasmas and the catastrophe theories of type $A_3$. 

\noindent

\subsection{Reconstructing the grand canonical potential}
\label{OmegaReconstruction}

Here, we consider the grand canonical ensemble where the chemical potential $\mu$ is kept fixed. In this case, in order to study the thermodynamics of the non-conformal plasmas, dual to the large and small BHs  in our holographic model, we need to compute the grand canonical potential $\Omega$. In holography, the grand canonical potential is usually obtained by evaluating the Euclidean on-shell action \eqref{EMD action} together with the appropriate boundary terms, namely, the Gibbons-Hawking term and counterterms. 

In this work we will reconstruct the grand canonical potential using the results for the Bekenstein-Hawking entropy \eqref{bhentropy}, the temperature \eqref{T} and the 1st law of thermodynamics which, in the grand canonical ensemble, reads
\begin{equation} \label{FirstLaw}
d\Omega = - p\,dV - S\,dT - Q\,d\mu . 
\end{equation}
In addition we have the thermodynamic identity $\Omega = E - TS - \mu\,Q$, where $E$ is the energy, $T$ is the temperature, given by \eqref{T}, $S$ is the entropy, given by \eqref{bhentropy}, $\mu$ the chemical potential and $Q$ the charge. Furthermore, we also have the relation $\Omega = -pV$, where $p$ is the pressure and $V\equiv V_3$ the volume over $\mathbb R^3$. 

From now on we will fix the volume $V$ to one so that the first law in \eqref{FirstLaw} reduces to 
\begin{eqnarray} \label{FirstLawv2}
d\Omega = - S\,dT - Q\,d\mu,
\end{eqnarray}
and we have $\Omega = -p$.  The quantities $\Omega$, $S$ and $Q$ now are interpreted as the  potential , entropy  and charge densities. Likewise, $E$ now represents the energy density.

We found in the previous section that the temperature was a function of the horizon radius $z_h$ and the chemical potential $\mu$. This function can be written as 
\begin{eqnarray}
T(z_h, \mu) = a(z_h) \Big [ 1 - \mu^2 b(z_h) \Big ] \,.
\end{eqnarray}
Then the differential $dT$ takes the form
\begin{eqnarray}
dT = \frac{ \partial T}{ \partial z_h} d z_h + \frac{\partial T}{\partial \mu} d \mu \, ,
\end{eqnarray}
with 
\begin{align}
 \frac{ \partial T}{ \partial z_h} &=  a'(z_h) - \mu^2 \Big [a'(z_h) b(z_h) + a(z_h) b'(z_h) \Big ] \, , \nonumber \\
 \frac{\partial T}{\partial \mu} &= - 2 \mu \, a(z_h) b(z_h) \,. 
\end{align}
From the first law in \eqref{FirstLawv2} we obtain
\begin{eqnarray} \label{dOmega}
d \Omega = - S \frac{ \partial T}{ \partial z_h} d z_h 
- \Big [  S \frac{\partial T}{\partial \mu} + Q \Big ] d \mu  \,.
\end{eqnarray}
This means that we can recast the grand canonical potential as a function of the variables $z_h$ and $\mu$. In the case of the RN AdS$_5$ black brane it was found that \cite{Hartnoll:2009sz}
\begin{align} \label{OmegaAdSRN}
\Omega(z_h , \mu) &= - \sigma  \Big [ z_h^{-4} + \frac13 \mu^2 z_h^{-2} \Big ]  \nonumber \\
&\equiv - A_{RN}(z_h) - B_{RN}(z_h) \mu^2  \,,   
\end{align}
where we have introduced the functions $A_{RN}(z_h)$ and $B_{RN}(z_h)$. 
We expect to recover the result in \eqref{OmegaAdSRN} in the absence of the dilaton field, which means in our model setting the parameter $k$ to zero. 

Now we will reconstruct the grand canonical potential.  The reconstruction method presented here should work for any 5d static charged black brane solution with  $\mathbb R^3$ symmetry arising from EMD theory. In that case one can choose coordinates where the metric is entirely described in terms of the scale factor $\zeta(z)$ and the horizon function $f(z)$, both functions of the radial coordinate $z$ solely. In turn, the temperature $T$ will be a function of $z_h$ and $\mu$ with the general form given in \eqref{T}. As a proof of principle, we will recover the  RN AdS$_5$  black brane solution in \eqref{OmegaAdSRN} for the case $k=0$, when the Einstein-Maxwell-Dilaton theory reduces to the Einstein-Maxwell theory (the dilaton term vanishes and the potential becomes the cosmological constant). 

The first step in the reconstruction method is taking the following ansatz for the grand canonical potential
\begin{eqnarray} \label{OmegaAnsatz}
\Omega(z_h, \mu) = - A(z_h) - \mu^2 B(z_h)   \,. 
\end{eqnarray}
Then the differential $d \Omega$ takes the form
\begin{eqnarray} \label{dOmega2}
d \Omega = - \Big [  A'(z_h) + \mu^2 B'(z_h)\Big ] d z_h  - 2 \mu  B(z_h ) d \mu \,. 
\end{eqnarray}
Comparing \eqref{dOmega} and \eqref{dOmega2} we arrive at the following identities
\begin{align} \label{ABEqs}
- A'(z_h) - \mu^2 B'(z_h) &= \frac{\partial \Omega}{ \partial z_h} 
 = - S \frac{ \partial T}{ \partial z_h}  \nonumber \\ 
&=  - S(z_h) a'(z_h) + \mu^2 S(z_h) \Big [a'(z_h) b(z_h) + a(z_h) b'(z_h) \Big ] \,,  
\end{align}
and 
\begin{align} \label{chargeEq}
- 2 \mu B(z_h) &= \frac{\partial \Omega}{ \partial \mu} =  - S \frac{\partial T}{\partial \mu} - Q  \nonumber \\ 
&= 2 \mu \, S(z_h)   \, a(z_h) b(z_h) - Q \,. 
\end{align}
From \eqref{ABEqs} we find trivial first order differential equations for the functions $A(z_h)$ and $B(z_h)$. Integrating these equations from infinity to $z_h$ we arrive at the relations
\begin{eqnarray} \label{ABsols}
A(z_h) &=& \int_{\infty}^{z_h} S(z) a'(z) dz + A_{\infty} \nonumber \\
B(z_h) &=& -\int_{\infty}^{z_h} S(z) \Big [a'(z) b(z) + a(z) b'(z) \Big ] dz + B_{\infty} \, ,
\end{eqnarray}
where $A_{\infty}= A(z_h \to \infty)$ and $B_{\infty}= B(z_h \to \infty)$ are integration constants. 
The procedure described above is equivalent to using directly the prescription
\begin{eqnarray} \label{OmegaFromT}
\Omega(z_h, \mu) = - \int_{\infty}^{z_h} S(z) \partial_z T(z, \mu) + \Omega_{\infty}(\mu) \, ,
\end{eqnarray}
with 
\begin{eqnarray} \label{OmegaInfty}
\Omega_{\infty}(\mu) = \Omega(z_h \to \infty , \mu) = - A_{\infty} - B_{\infty} \mu^2 \,. 
\end{eqnarray}
The integration constants $A_{\infty}$ and $B_{\infty}$ are associated with the renormalisation scheme dependence of the dual theory. In the limit $k \to 0$ we want to recover the potential in  \eqref{OmegaAdSRN} corresponding to the  RN AdS$_5$ black brane, up to a constant. We therefore fix $B_{\infty}=0$ and will fix $A_{\infty}$ by the requirement $\Omega \le 0$ at $\mu=0$. We will require that  the grand potential $\Omega$ vanishes at $z_h=z_{hmin}$ or, equivalently, at $T=T_{min}\simeq 144$ MeV as in \cite{Mamo:2016dew}. It turns out that $B(z_h)$ is non-negative so from \eqref{OmegaAnsatz} we see that negativity of $\Omega$ at $\mu=0$ guarantees negativity at finite $\mu$. In general, the integration constant $A_{\infty}$ is a function of the parameter $k$, responsible for conformal symmetry breaking. For the undeformed background, i.e, $k=0$, the criterion would correspond to fixing $A_{\infty}=0$, since for this case $T_{min} = 0$ and $z_{hmin}\to\infty$. Below we describe this limiting case in more detail. 

In  the case $k=0$ the EMD theory reduces to the EM theory (with a cosmological contant term).  The entropy \eqref{bhentropy} and temperature \eqref{T} of the charged black brane solutions reduce to \begin{equation}\label{bhentropyTRN}
S_{RN} =  \dfrac{4\pi\sigma}{z_h^3},
\end{equation}
\begin{equation} \label{TRN}
T_{RN}(z_h,\mu) 
 = \frac{1}{\pi z_h} \Big [ 1 - \frac16 \mu^2 z_h^2 \Big ]  
= a_{RN}(z_h) \Big [ 1 - \mu^2 \, b_{RN}(z_h) \Big ] \, ,
\end{equation}
where we used the expansions \eqref{zetaExp} and \eqref{TExp}. Plugging these results in our formula for the grand canonical potential \eqref{OmegaFromT}, performing the integral in $z$ and setting the integration constants $A_{\infty}$ and $B_{\infty}$ to zero one easily finds the result in \eqref{OmegaAdSRN} for the RN AdS$_5$ black brane, as promised. We remark that the result in \eqref{OmegaAdSRN} was found using holographic renormalisation and therefore it is an important check for our formula \eqref{OmegaFromT} in the case $k=0$. We remark, however, that our formula \eqref{OmegaFromT} should be valid in the more general case of non-zero $k$. For more details on the thermodynamics in the case $k=0$, see appendix \ref{App:RNAdS}. It is worth mentioning  that in the case $k=0$, the trace anomaly $E-3p$ vanishes even at finite temperature and density and therefore the 4d fluid enjoys conformal symmetry. This is also shown in appendix \ref{App:RNAdS}.

We have described in this subsection a procedure for reconstructing the grand canonical potential $\Omega$ from the knowledge of the Bekenstein-Hawking entropy and the first law of thermodynamics. This method is physically well-grounded and as a proof of concept we have recovered the RN AdS$_5$  black brane solution for the case $k=0$. In appendix \ref{App:AltMethod} we describe an alternative method  using an auxiliary potential that simplifies the thermodynamic relations.

\subsection{Extracting other thermodynamic quantities} 

The thermodynamic quantity $Q$ can be interpreted as the charge density in the grand canonical ensemble. From \eqref{chargeEq} we find the relation
\begin{eqnarray}
Q &=& 2 \mu \Big [  B(z_h) +  S(z_h) a(z_h) b(z_h) \Big ]  \cr 
&=& 2 \mu \int_{\infty}^{z_h} S'(z) a(z) b(z) 
\equiv d(z_h) \, \mu \,.  
\end{eqnarray}
It turns out that this charge density will be similar but not equal to the holographic charge density, obtained in \eqref{chargedensity}. This has to do with the presence of the small unstable BH in the regime of intermediate $z_h$, as described in the previous section. The small unstable BH leads to an unphysical increasing behaviour for the temperature when increasing $z_h$, which means decreasing the entropy. This is in contrast with the physical situation at small and large $z_h$ where the temperature decreases for increasing $z_h$ (decreasing entropy).

With the expressions for the grand canonical potential $\Omega$ and the charge density $Q$, one can compute the energy density $E$ using the thermodynamic relation
\begin{equation}
E = \Omega + TS + \mu\,Q.
\end{equation}
Another important thermodynamic quantity is the trace anomaly, which measures the breaking of conformal symmetry, and is given by
\begin{equation}
\left\langle T^{a}_{\;\;a}\right\rangle = E-3p = 4\,\Omega + TS + \mu\,Q \,. \end{equation}

We finish this section describing the global and local stability conditions. The  condition of thermodynamic stability corresponds to the minimum of the grand canonical potential $\Omega$, which is expressed by the following conditions 
\begin{equation}
(\delta\Omega)_{T,\,\mu} = 0, \quad (\delta^{2}\Omega)_{T,\,\mu} \leq 0.
\end{equation}
These are the global stability conditions \cite{DeWolfe:2010he}. Furthermore, we must also require the local stability of the grand canonical potential against small fluctuations, which is equivalent to demand the positivity of the determinant of the hessian matrix, given by the second derivatives of the grand canonical potential with respect to the temperature $T$ and chemical potential $\mu$ as
\begin{equation}
\mathcal{H} = \left( 
\begin{array}{cc}
-\dfrac{\partial^2\Omega}{\partial T^2} & - \dfrac{\partial^{2}\Omega}{\partial\mu\,\partial T}\\
- \dfrac{\partial^{2}\Omega}{\partial T\,\partial\mu}& -\dfrac{\partial^2\Omega}{\partial \mu^2} 
\end{array}
\right) = \left( 
\begin{array}{cc}
\dfrac{\partial S}{\partial T} & \dfrac{\partial S}{\partial\mu}\\
\dfrac{\partial S}{\partial\mu} & \dfrac{\partial Q}{\partial \mu} 
\end{array}\right)
\end{equation}
Therefore, det\,$\mathcal{H}\geq0$ implies
\begin{equation}
\left(\dfrac{\partial S}{\partial T}\right)_{V,\,\mu}\left(\dfrac{\partial Q}{\partial \mu}\right)_{V,\,T}-\left(\dfrac{\partial S}{\partial\mu}\right)^2\geq0.
\end{equation}
Since the entropy $S$ does not depend on $\mu$ explicitly, we have
\begin{equation}
\left(\dfrac{\partial S}{\partial T}\right)_{V,\,\mu}\left(\dfrac{\partial Q}{\partial \mu}\right)_{V,\,T}\geq0,
\end{equation}
which can be further simplified  to 
\begin{equation}
\frac{\chi\,C_{V}}{T}\geq 0,
\end{equation}
where $C_V$ is the specific heat at constant volume and $\chi$ is the charge susceptibility, and they are defined by
\begin{equation}
C_{V} = T\,\left(\dfrac{\partial S}{\partial T}\right)_{V,\,\mu}, \quad \chi = \left(\dfrac{\partial Q}{\partial \mu}\right)_{V,\,T}.
\end{equation}
Thus, one can conclude that the local thermodynamic stability condition, i.e., the positivity of the Jacobian of the hessian matrix $\mathcal{H}$, implies the positivity of the response functions $C_V$ and $\chi$.

Later in this work we will investigate the response functions $C_V$ and $\chi$ near the critical regime. We will find that these quantities diverge at the critical point and we will extract the corresponding critical exponents for a family of charged asymptotically AdS BHs arising from EMD equations.

\section{Numerical Results: Thermodynamic Observables}
\label{Sec:Results}

Here, we present our numerical results for the thermodynamic observables relevant for investigating the approach to the criticality in our model. We remind the reader that we have chosen the background of a quadratic dilaton $\phi(z)=k z^2$ leading to the scale factor $\zeta(z)$ in \eqref{zetasol}. The temperature as a function of the horizon radius $z_h$ and chemical potential $\mu$ was obtained in \eqref{T} with the functions $C_n(z)$ defined in \eqref{C2}, \eqref{C4} and \eqref{C6}. The entropy as a function of $z_h$ was obtained in \eqref{bhentropy}.

In all results presented in this section the dilaton constant $k$ is set to $k=0.18$ GeV$^2$. We will see later, in section \ref{Sec:Lattice}, that the thermodynamic quantities can be recast in a form that is independent of the value of $k$. In particular, we will see that the value of $k$ can be set by a fit in the limit $\mu \to 0$ to the deconfinement temperature obtained in lattice $SU(N_c)$ gauge theories.  

Below we present our results for the grand canonical potential, entropy density, specific heat, speed of sound, charge density, charge susceptibility and the trace anomaly. We will use those results to arrive at  the $T-\mu$ phase diagram.  We will always distinguish between three possible phases:

\begin{itemize}
    \item Unstable phase : the phase of the non-conformal plasma where the specific heat is negative. This phase is therefore non-physical. 
    \item Metastable phase: the phase of the non-conformal phase where the specific heat is positive but it does not correspond to the ground state (minimum) of the grand canonical potential. This phase is therefore physical but it is not thermodynamically favoured. 
    \item Stable phase: the phase of the non-conformal plasma where the specific heat is positive and also corresponds to the ground state (minimum) of the grand canonical potential. This phase is therefore physical and it is thermodynamically favoured. 
\end{itemize}

In our model it turns out that there will be one unstable phase  and two physical phases that compete with each other. The unstable phase is the non-physical small BH, described in the previous section, where the temperature is an increasing function of the horizon radius (and therefore a decreasing function of the entropy).  The physical phases will correspond to a large BH and a small BH,  where the temperature is a decreasing function of the horizon radius (increasing function of the entropy), as described in the previous section. We will see that the physical large and small BHs will compete with each other and the BH phase where the grand canonical potential is minimum will be thermodynamically favoured (stable). The other BH phase will be in a metastable state. 

In all the figures presented in this section we will represent the unphysical small BH by a red dashed line. The physical large BH will be represented by a solid blue line when it is in a stable state and by a dashed blue line when it is in a metastable state.  Similarly, the physical small BH will be depicted by a green solid line when it is in a stable state and by a green dashed line when it is in a metastable state.

As regards the gravitational constant $\sigma$, one can fix it by the relation \cite{Gursoy:2008za}:
\begin{equation} \label{StefanB}
    \sigma = M_P^3 N_c^2 = \frac{1}{45\,\pi^2} N_c^2  \, ,
\end{equation}
where $M_P$ is the Planck mass and $N_c$ the number of colours of the dual non-Abelian gauge theory. The prescription \eqref{StefanB} for $\sigma$ allows us to reproduce the Stefan-Boltzmann law $P = (\pi^2/45) N_c^2 T^4$ for the non-Abelian plasma in the regime of very high temperatures \cite{Gursoy:2008za}.

\subsection{Grand canonical potential}
The grand canonical potential $\Omega$ was reconstructed in \eqref{OmegaFromT} using the results for the temperature and  entropy. The integration constant $B_{\infty}$ in \eqref{OmegaInfty} was set to zero for consistency whilst the integration constant $A_{\infty}$ can be fixed requiring the vanishing of $\Omega$ at $T=T_{\rm min}$ when $\mu=0$. We find that $A_{\infty} \approx 6.38 \times 10^{-6}$ guarantees the property $\Omega \leq 0$ at $\mu=0$. As explained in the previous section, this also guarantees the property $\Omega \leq 0$ at any finite value of $\mu$.

\begin{figure}[ht]
	\centering
	\includegraphics[scale=0.245]{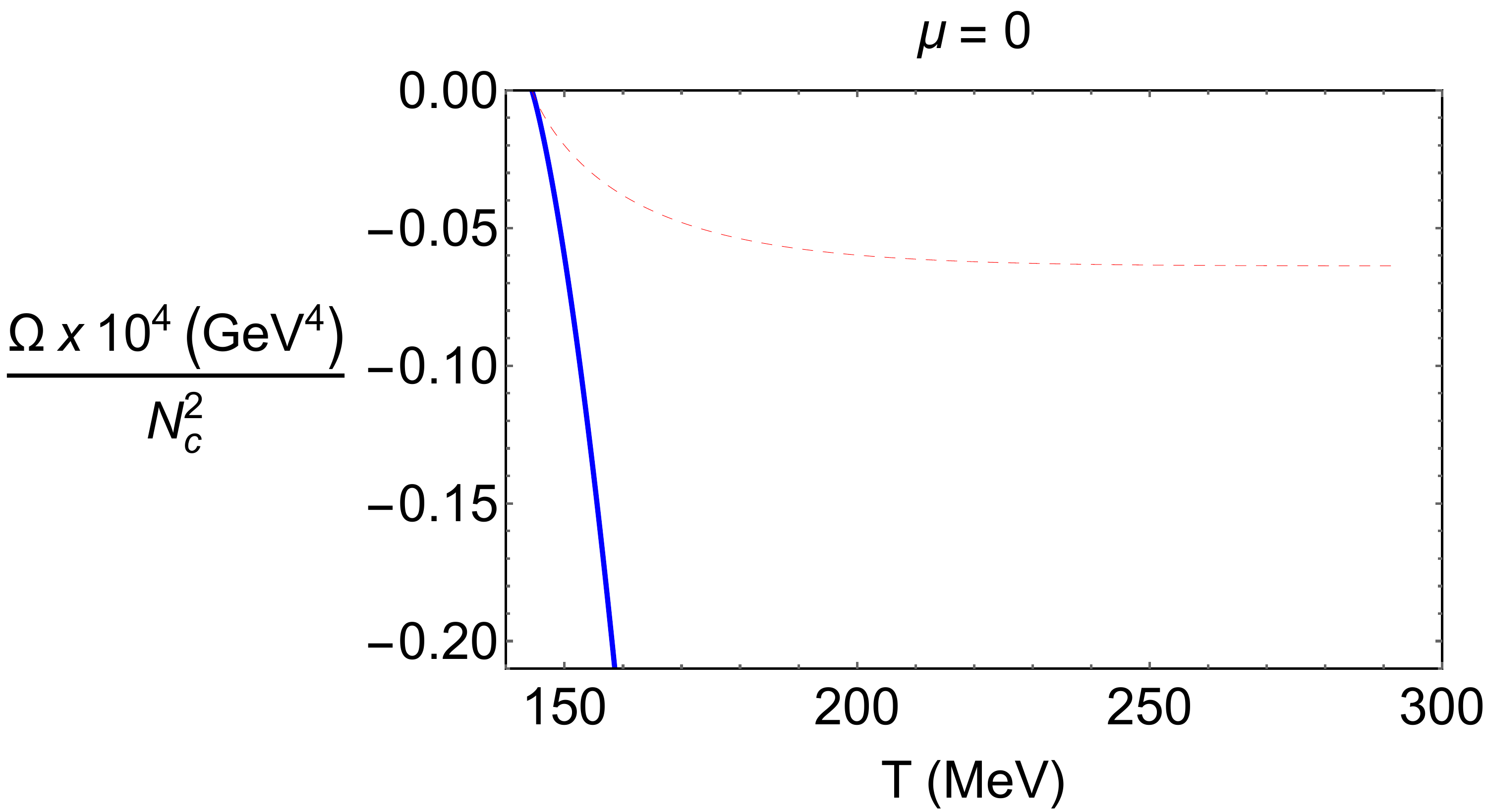}
	\hfill
	\includegraphics[scale=0.245]{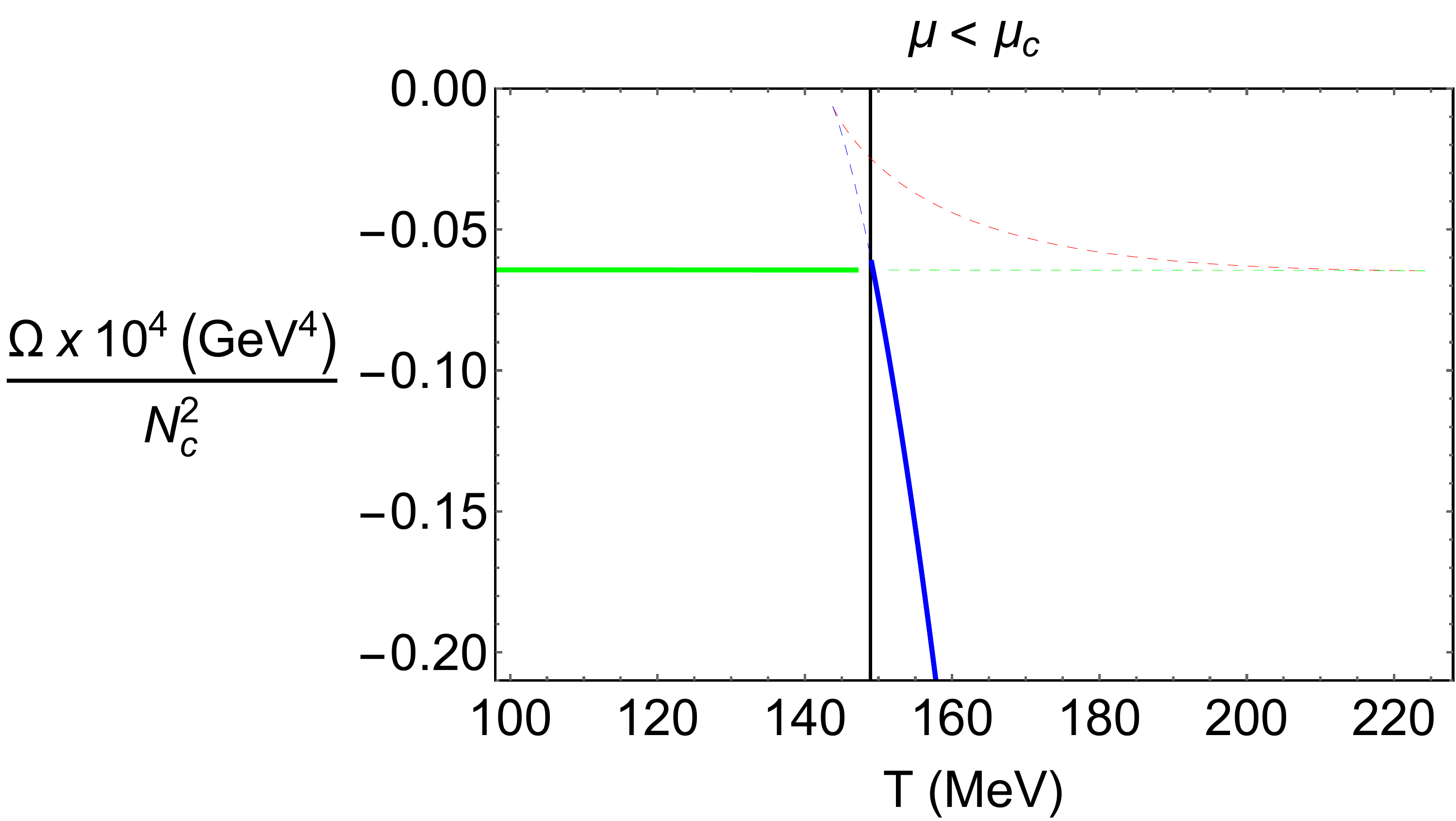}
  \hfill
	\includegraphics[scale=0.245]{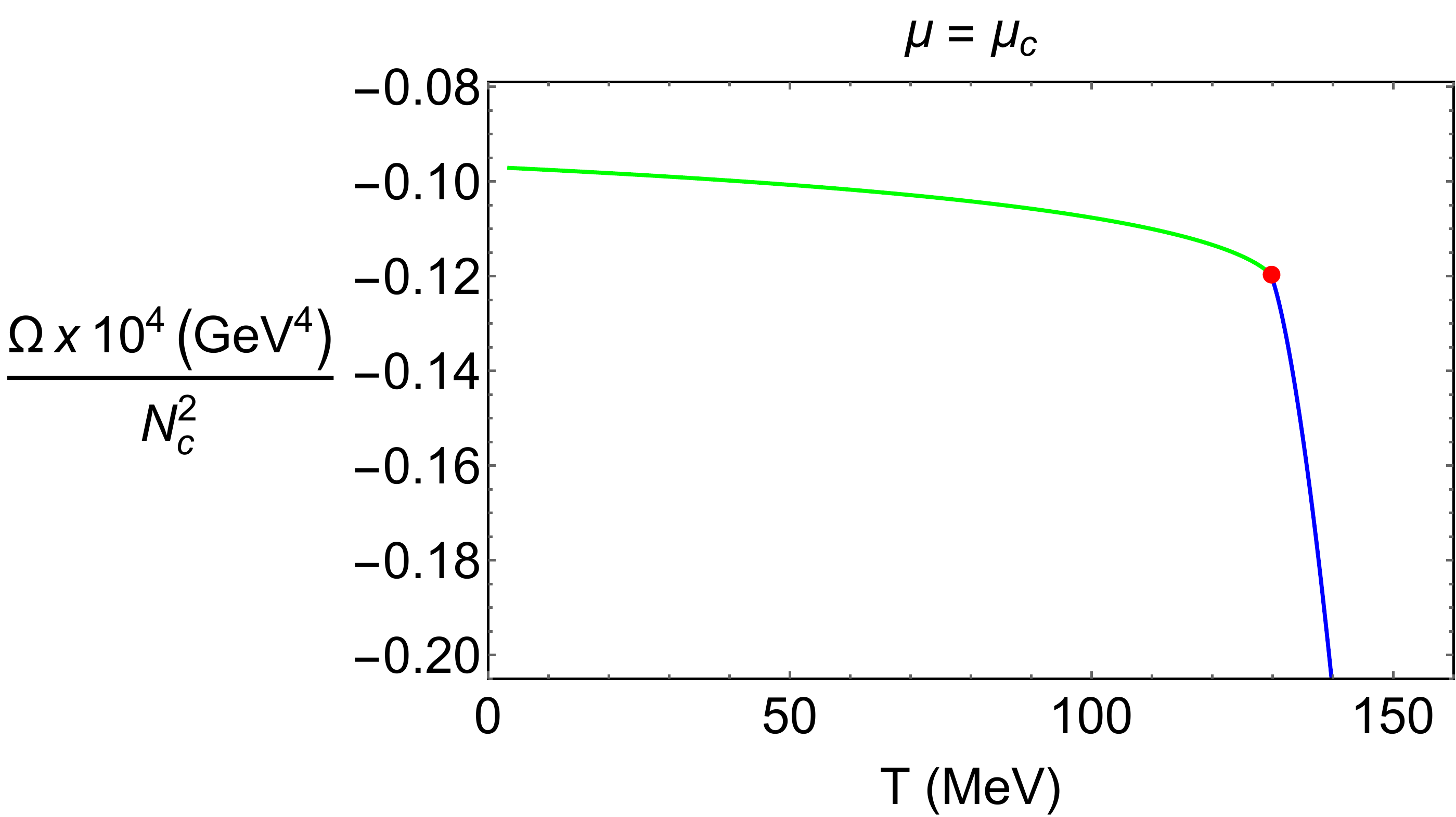}
	\hfill
	\includegraphics[scale=0.245]{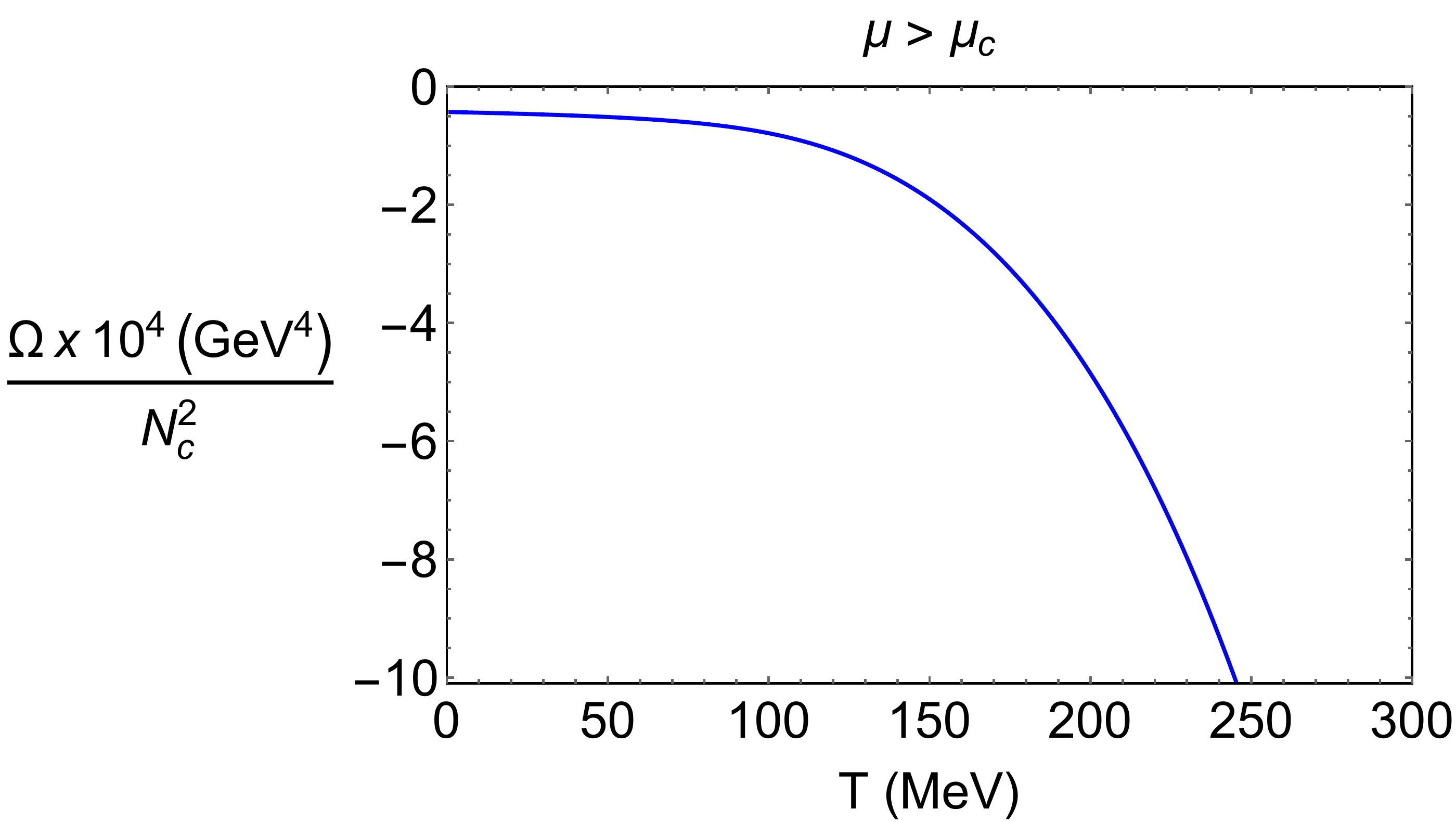}
		\caption{The grand canonical potential $\Omega$ as function of the temperature $T$ for fixed values of the chemical potential: i) $\mu =0$ (upper left panel) ii) $\mu=50$ MeV (upper right panel),  iii) $\mu= \mu^c= 225$ MeV (lower left panel) and $\mu=500$ MeV (lower right panel). 	The red dashed line represents the small non-physical BH, the solid (dashed) blue line represents the large BH in a stable (metastable) state and the solid (dashed) green line represents the small physical BH in a stable (metastable) state.
The vertical solid line marks for $\mu<\mu^c$ the temperature where a first-order phase transition takes place. }
	\label{fig:FreeEnFixedMu}
\end{figure}

Figure \ref{fig:FreeEnFixedMu} shows the behaviour of the grand canonical potential (density) $\Omega$, in units of (GeV$^4$ $N_c^{-2}$), against the temperature $T$, in (MeV), for different values of the chemical potential $\mu$. The first plot (upper left panel) shows the variation of $\Omega$ with respect to $T$ for  $\mu = 0$. In that case we only have the large BH (blue curve) and the unphysical small BH (red curve). When we turn on the chemical potential, i.e.  $\mu =50$ MeV in  Fig. \ref{fig:FreeEnFixedMu} (upper right panel), a new solution appears that corresponds to the physical small BH (green curve). Note the appearance of the swallowtail, which is a main feature in the thermodynamics. The swallowtail indicates a competition between two physical phases, which in our framework are the large BH (blue curve) and the physical small BH (green curve). 
This is a clear signal of a {\it first order transition} and the intersection point between the blue and green curves will correspond to the transition temperature. This will be confirmed later in this section from the analysis of the entropy density. 

There are also two interesting points in the second plot of Fig. \ref{fig:FreeEnFixedMu} (upper right panel); one represents the border between the blue curve and red curve whilst the other represents the border between the red and green curve. These points correspond to the local $T_{\rm min}$ and local $T_{\rm max}$ described in Fig. \ref{fig:Tfinitemuv2} where two secondary transitions take place: i)  transition from the metastable large BH to the unstable small BH (blue-red transition) and ii) transition from the unstable small BH to the metastable small BH  (red-green transition). Later in this section we analyse the entropy density and specific heat and we will conclude that these secondary transitions are {\it second order transitions}.  

When the value of the chemical potential is increased further the swallowtail shrinks. As soon as the chemical potential reaches a critical value, $\mu = \mu^c$ (lower left panel in  Fig. \ref{fig:FreeEnFixedMu}), the swallowtail disappears; the only remaining phases are the large BH (blue curve) and the physical small BH (green curve). The unphysical small BH is now represented by a (red) dot which lies at the border between the physical BHs. For $\mu \gg \mu^c$ (lower right panel in  Fig. \ref{fig:FreeEnFixedMu}) the large BH and small physical small BH have merged smoothly into single curve. We will interpret this as a {\it crossover transition} later in this section. 

As a final remark, when describing phase transitions in the end one is only interested in the stable solutions, that correspond to the ground state (minimum) of the grand canonical potential. These solutions are identified with thick blue and green solid lines in Fig. \ref{fig:FreeEnFixedMu}.

\subsection{Entropy Density}

\begin{figure}[ht]
	\centering
	\includegraphics[scale=0.245]{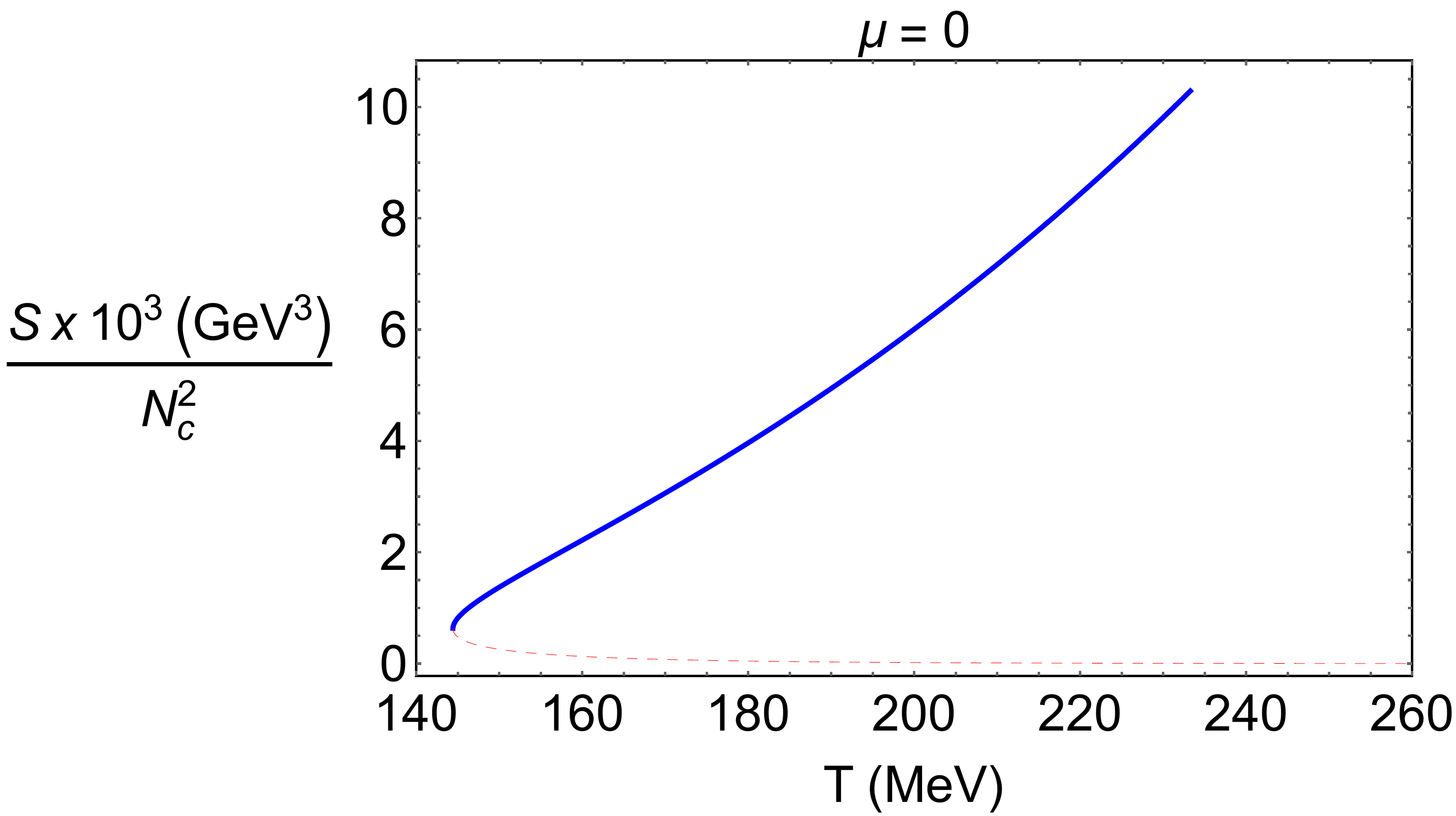}
	\hfill
	\includegraphics[scale=0.245]{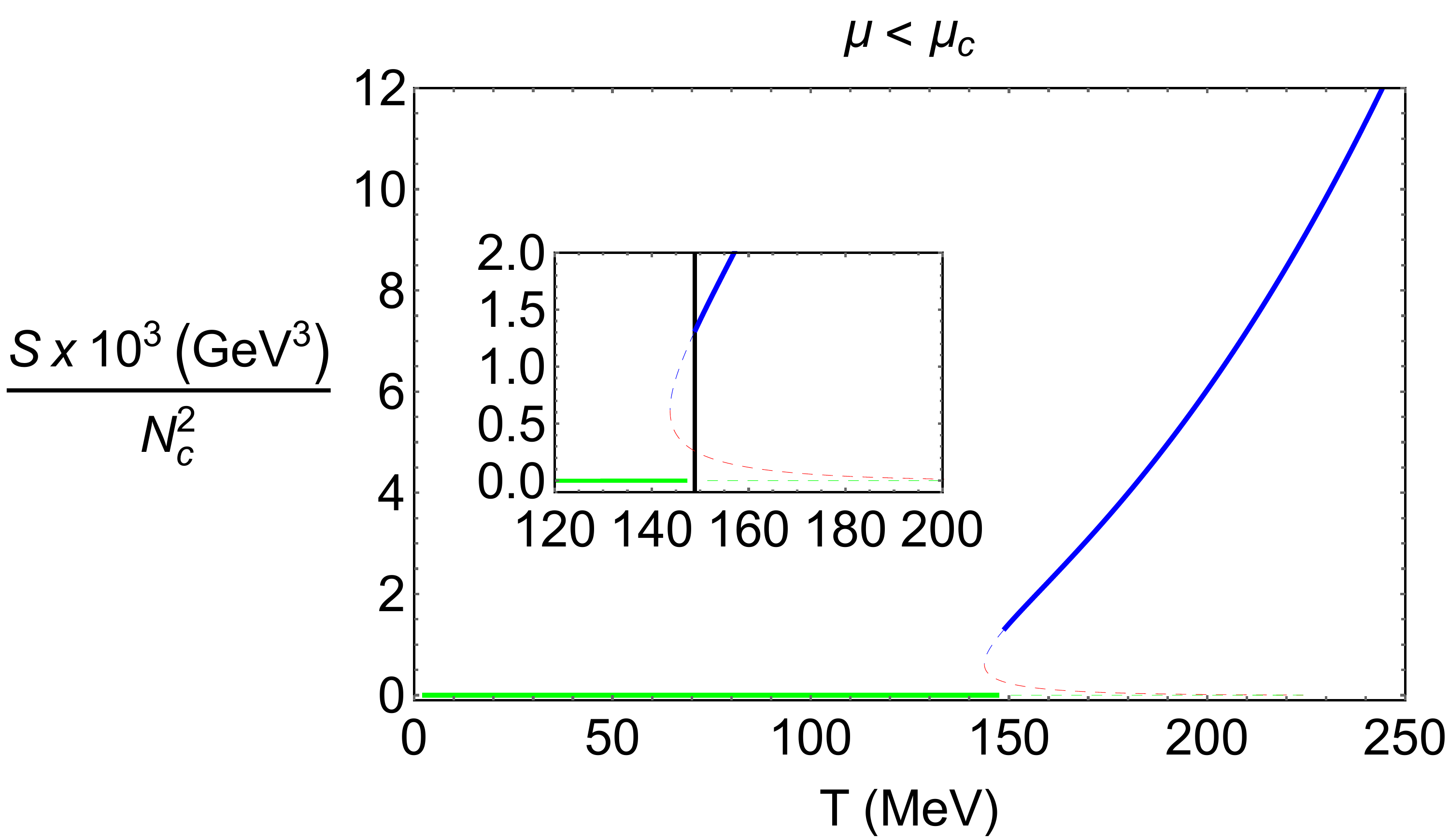}
    \hfill
	\includegraphics[scale=0.245]{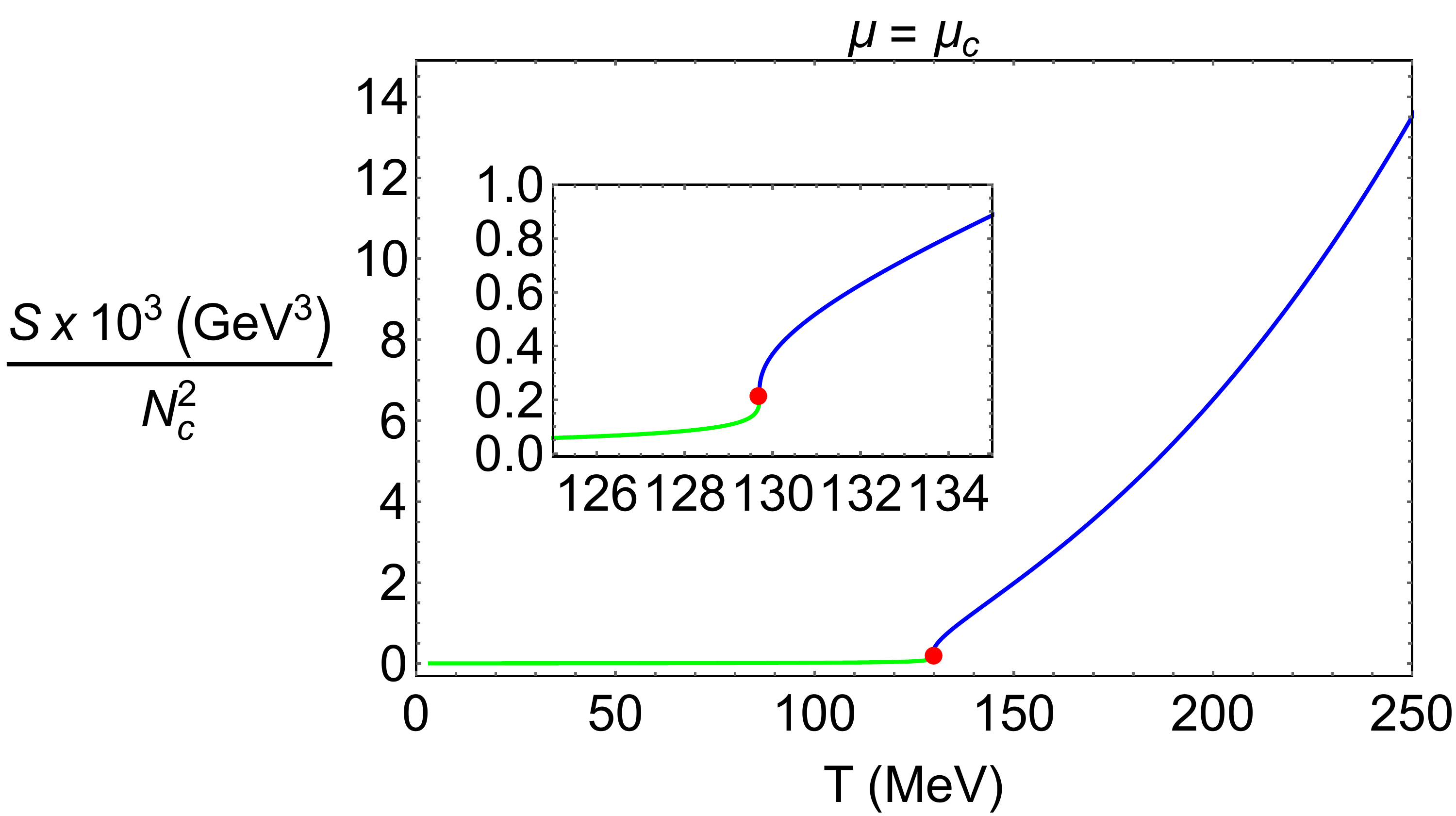}
	\hfill
	\includegraphics[scale=0.245]{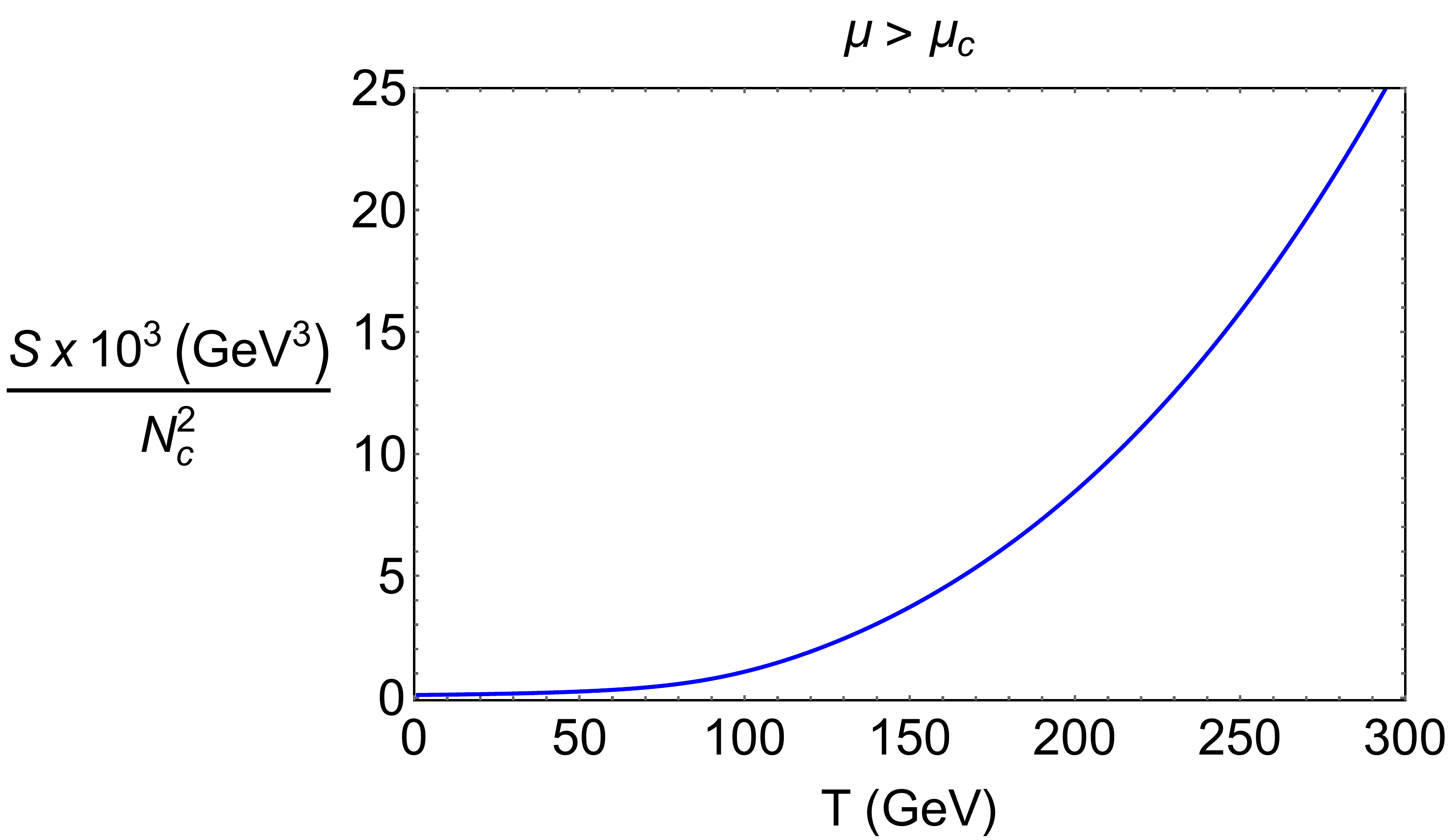}
	\caption{The BH entropy density as a function of the temperature $T$ for fixed values of the chemical potential: i) $\mu =0$ (upper left panel) ii) $\mu=50$ MeV (upper right panel),  iii) $\mu= \mu^c= 225$ MeV (lower left panel) and $\mu=500$ MeV (lower right panel). 	The red dashed line represents the small non-physical BH, the solid (dashed) blue line represents the large BH in a stable (metastable) state and the solid (dashed) green line represents the small physical BH in a stable (metastable) state. The vertical solid line for $\mu<\mu^c$ (inset) marks the temperature where a first-order phase transition takes place.}
	\label{fig:EntropyMuHigh}
\end{figure}

In this subsection we describe the entropy density in our EMD holographic model. We display in Fig. \ref{fig:EntropyMuHigh} the entropy density $S$, in units of (GeV$^3$ N$_c^{-2}$) as function of the temperature $T$, in units of (MeV), for different values of the chemical potential $\mu$. We remind the reader that the entropy was obtained previously in \eqref{bhentropy} as a monotonic function of the horizon radius $z_h$. The non-trivial dependence on $T$, for fixed $\mu$, is obtained considering a parametric plot for the functions $T(z_h,\mu)$ and $S(z_h)$. 

Figure \ref{fig:EntropyMuHigh} displays the following results for the entropy density: i)  At $\mu=0$ (upper left panel), there are only two phases: the physical large BH and the non-physical small BH, represented by the blue and the red curves respectively. The entropy density increases with $T$ for the large BH and decreases with $T$ for the non-physical small BH  ii) As soon as one turns on $\mu$ there are three BH phases, now including a physical small BH (green curve) ($\mu=50$ MeV for the upper right panel). The entropy density for the physical small BH starts at $T=0$ and increases with $T$. iii) Increasing the value of $\mu$ these three phases still coexist but the non-physical small BH shrinks until it completely disappears for $\mu = \mu^c = 225$ MeV (lower left panel). At that value the two physical solutions (blue and green curves) merge iv) For larger values of $\mu$ the physical large BH and small BHs have become a single phase with an entropy that increases with $T$ and starts at $T=0$ ($\mu=500$ MeV for the lower right panel).

As a final remark, when describing phase transitions in the end one is only interested in the stable solutions, that correspond to the ground state (minimum) of the grand canonical potential. These solutions are identified with thick blue and green solid lines in Fig. \ref{fig:EntropyMuHigh}.

\subsection{Specific Heat  and Speed of Sound }

Now we present our results for the specific heat $C_V$ and the speed of sound $c_s$. The specific heat $C_{V}$ is defined by
\begin{equation}
C_{V} = T\left(\dfrac{\partial S}{\partial T} \right)_{\mu} = -T\left(\dfrac{\partial^2 \Omega}{\partial T^2} \right)_{\mu},
\end{equation}
which can also be written as
\begin{equation}\label{CV}
C_V = S\,\dfrac{\partial\ln S}{\partial\ln T}.
\end{equation}

\begin{figure}[htb]
	\centering
	\includegraphics[scale=0.36]{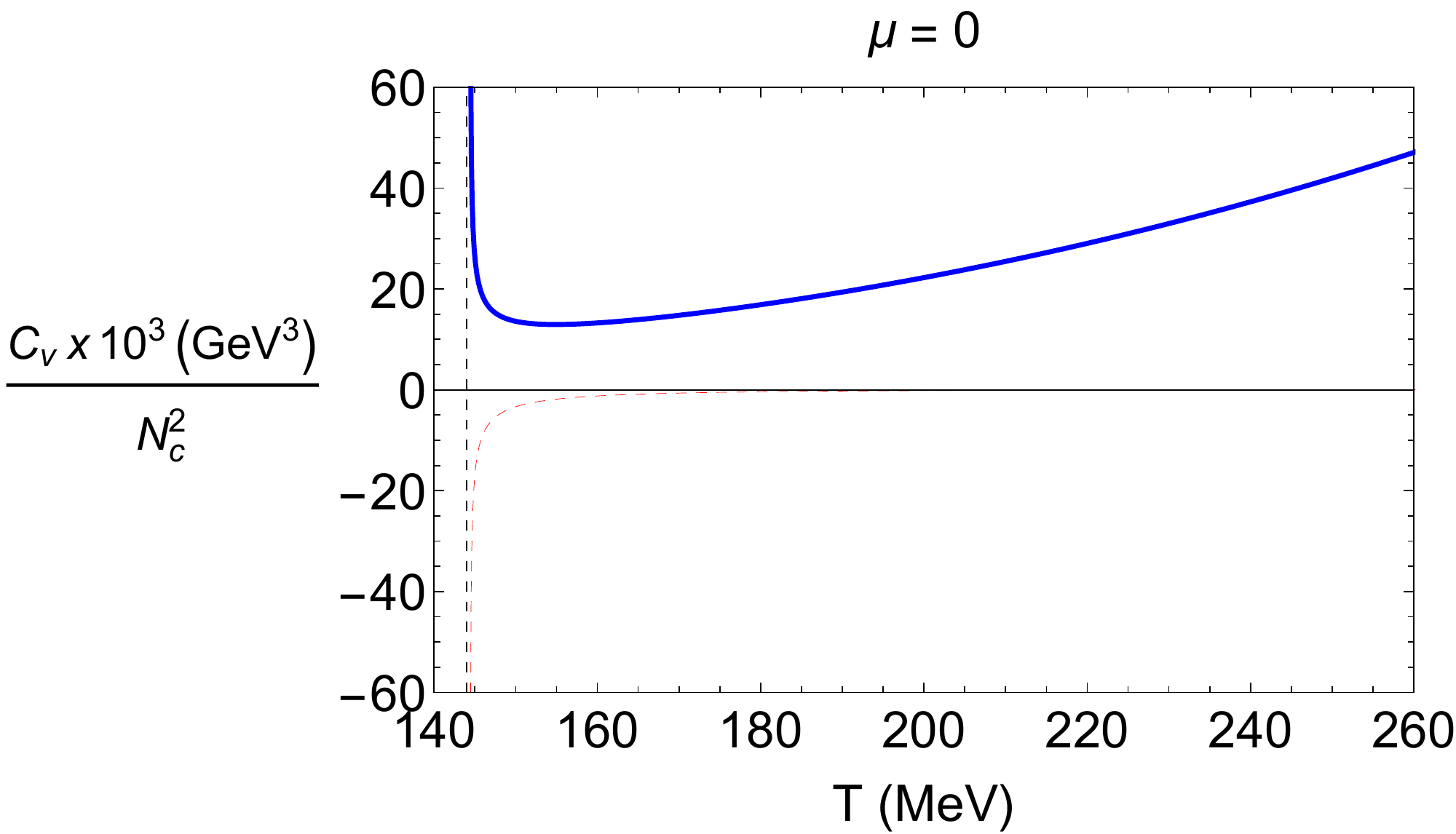}
	\hfill
	\includegraphics[scale=0.36]{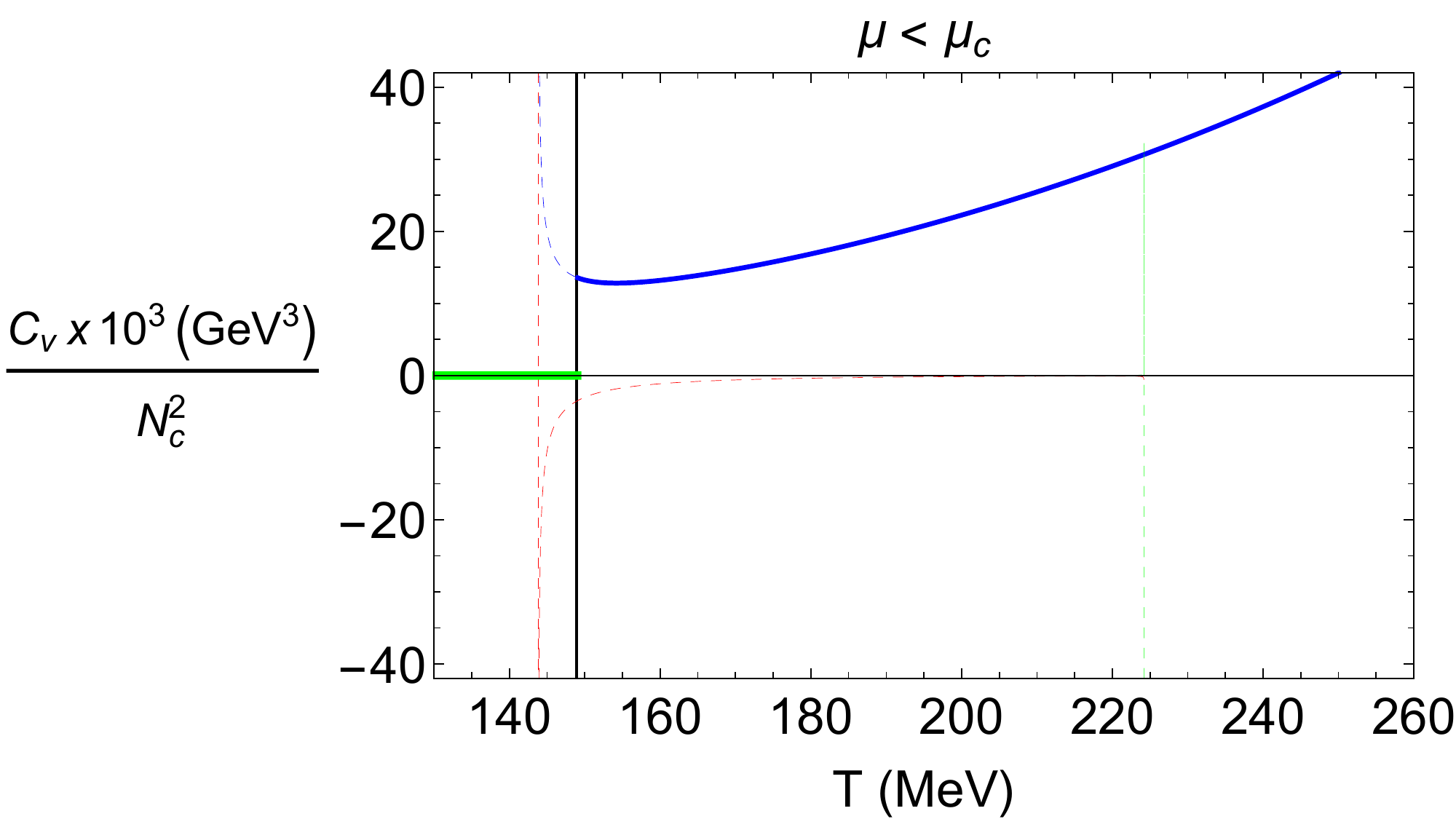}
    \hfill
	\centering
	\includegraphics[scale=0.36]{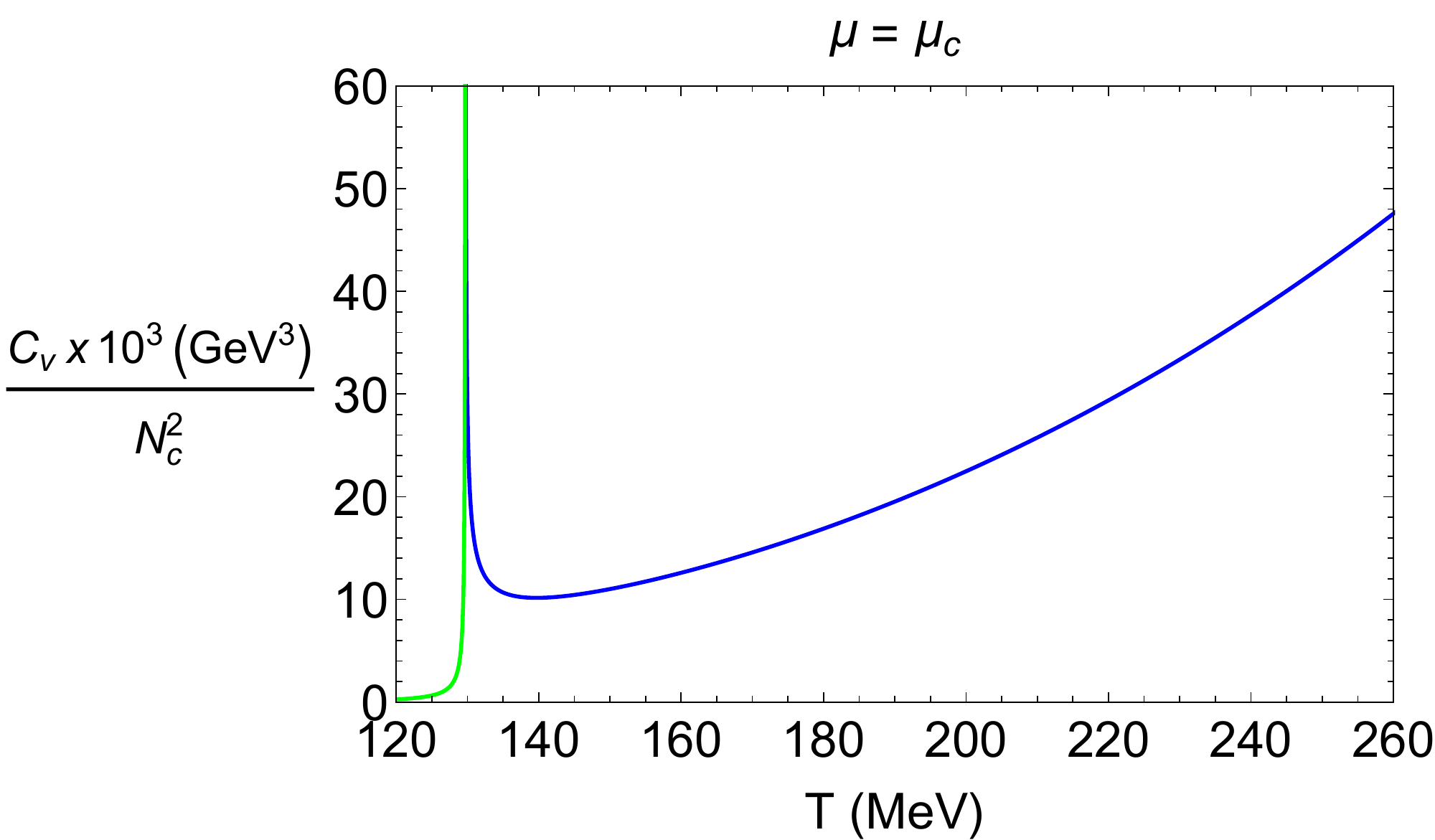}
	\hfill
	\includegraphics[scale=0.36]{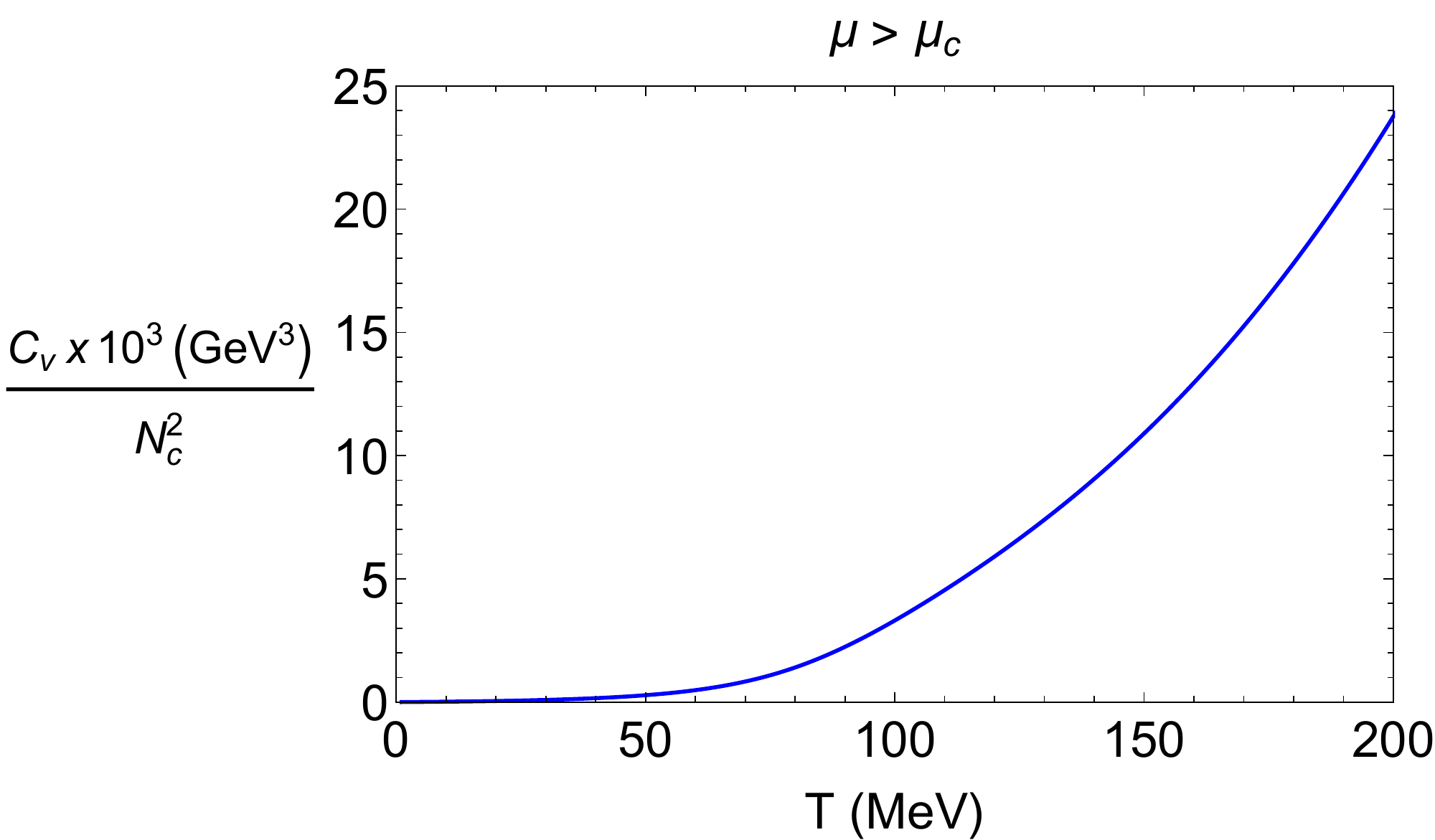}
	\caption{Specific heat $C_{V}$ as a function of the temperature $T$ for fixed values of the chemical potential: i) $\mu =0$ (upper left panel) ii) $\mu=50$ MeV (upper right panel),  iii) $\mu= \mu^c= 225$ MeV (lower left panel) and $\mu=500$ MeV (lower right panel). 	The red dashed line represents the small non-physical BH, the solid (dashed) blue line represents the large BH in a stable (metastable) state and the solid (dashed) green line represents the small physical BH in a stable (metastable) state. The vertical solid line for $\mu<\mu^c$ marks the temperature where a first-order phase transition takes place.}
	\label{fig:cv_all}
\end{figure}

In Fig. \ref{fig:cv_all} we present the behaviour of the specific heat $C_V(T)$, in units of (GeV$^3$ N$_c^{-2}$) as a function of the temperature $T$, in units of MeV, for different values of the chemical potential $\mu$.

For $\mu=0$ (upper left panel) there are only two BH phases, the physical large BH (blue curve) and the non-physical small BH (red curve). The specific heat is positive for the physical large BH  and negative for the non-physical small BH. The latter clearly signalises instability. At the minimum temperature $T_{\rm min}$ the specific heat displays an infinite peak (vertical line in the plot), signifying the transition from the stable large BH to the unstable small BH. 

For small chemical potential, e.g.  $\mu = 50$ MeV (upper right panel), note the appearance of the physical small BH phase (green curve) and the shrinking of the non-physical small BH phase (red curve). This time the specific heat displays two infinite peaks, signifying two transitions: i) From the stable large BH to the unstable small BH (blue to red) and ii) From the unstable small BH to the stable (or metastable) small BH (red to green). 

At the critical chemical potential, $\mu^c = 225$ MeV (lower left panel), one sees the extinction of the non-physical small BH phase and the merging of the physical large and small BHs. There is still a finite peak indicating the transition between these two phases. Note that both phases are characterised by a positive specific heat. 
The distinguished peak fades out as $\mu$ increases  further away from $\mu^c$. In the last plot (lower right panel) we have $\mu = 500$ MeV there is no peak at all and the two physical BHs have already merge into a single continuous curve.
\begin{figure}[ht]
	\centering
	\includegraphics[scale=0.23]{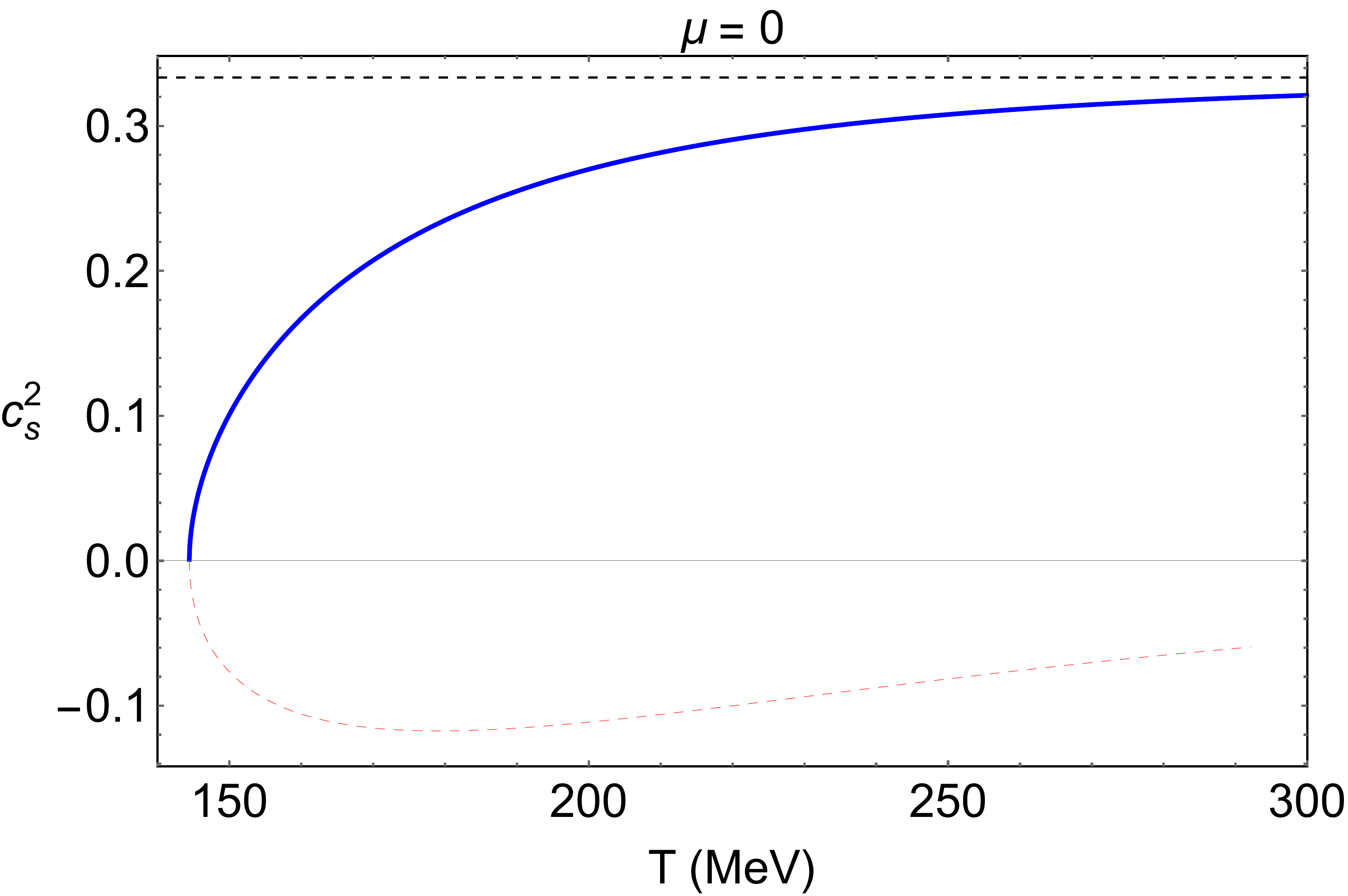}
	\hfill
	\includegraphics[scale=0.23]{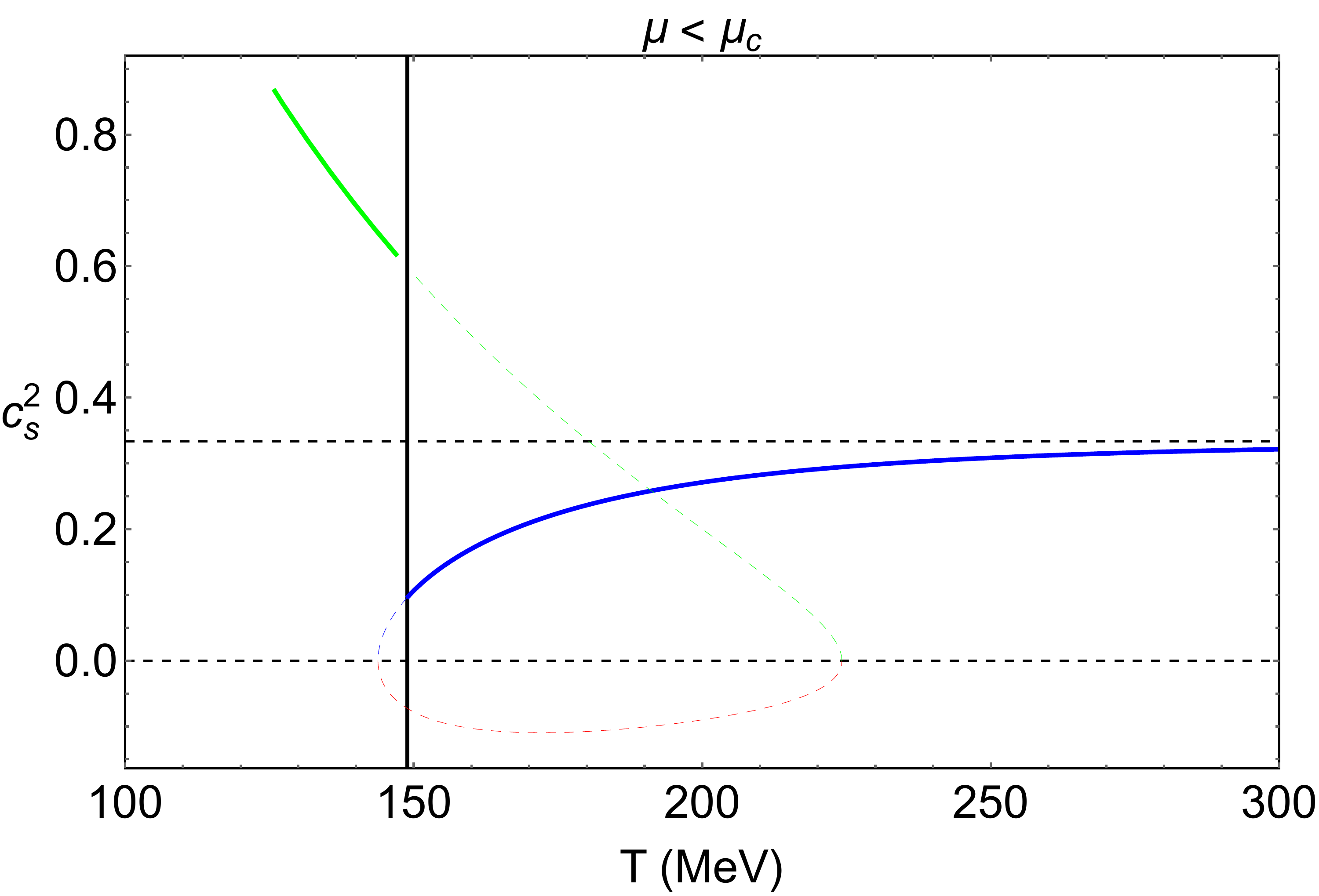}
  \hfill 
	\includegraphics[scale=0.23]{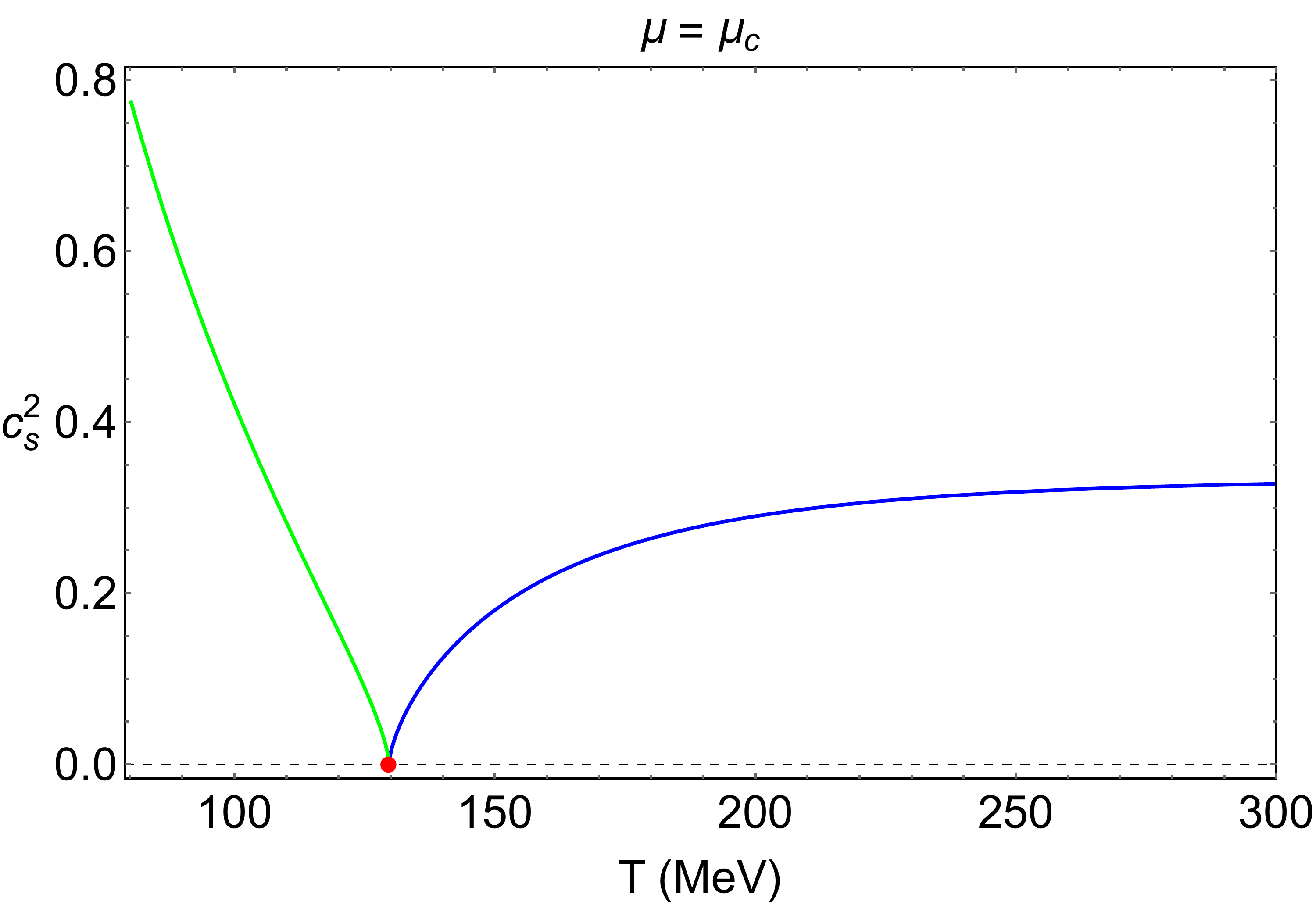}
	\hfill
	\includegraphics[scale=0.23]{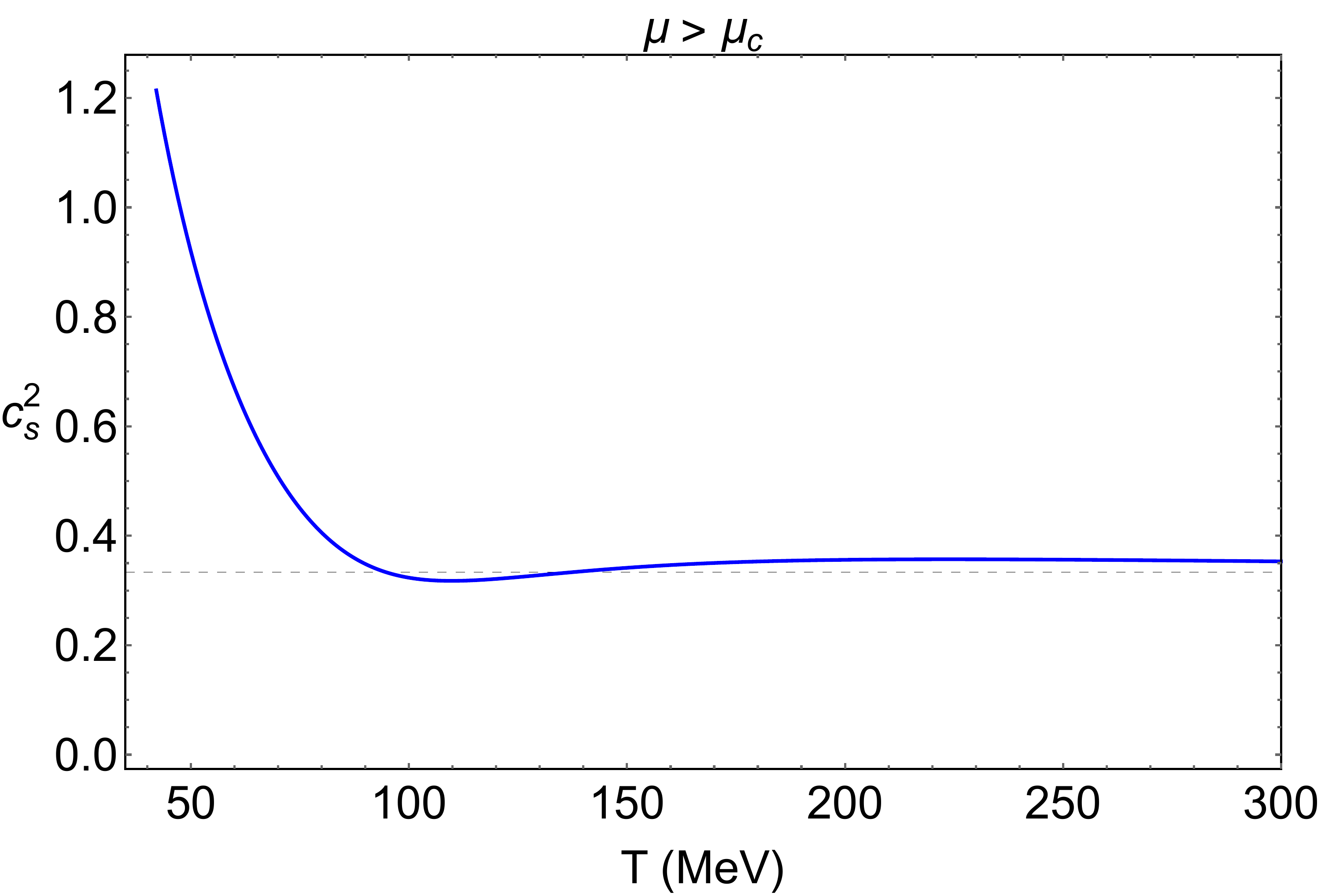}
	\caption{The squared speed of the sound $c^2_{s}$ as a function of the temperature $T$ for fixed values of the chemical potential: i) $\mu =0$ (upper left panel) ii) $\mu=50$ MeV (upper right panel),  iii) $\mu= \mu^c= 225$ MeV (lower left panel) and $\mu=500$ MeV (lower right panel). The red dashed line represents the small non-physical BH, the solid (dashed) blue line represents the large BH in a stable (metastable) state and the solid (dashed) green line represents the small physical BH in a stable (metastable) state. The vertical solid line for $\mu<\mu^c$ marks the temperature where a first-order phase transition takes place.
 }
	\label{fig:cs3}
\end{figure}

The squared speed of sound  $c^2_{s}$ is usually defined by the thermodynamic relation
\begin{equation}
c^2_{s} = \dfrac{S}{C_{V}}.
\end{equation}
Using the result for $ C_{V} $ given in \eqref{CV} one can write the squared speed of sound in terms of the temperature and the entropy:
\begin{equation}
c^2_{s} = \dfrac{\partial\ln T}{\partial\ln S}.
\end{equation}
The figure \ref{fig:cs3} shows the squared speed of sound $c_s^2$ (dimensionless) as a function of the temperature $T$ (MeV) for increasing values of the chemical potential $\mu$, from $\mu= 0$ to $\mu = 500$ MeV. For $\mu=0$ (upper left panel) once again there are only two BHs, the physical large BH (blue curve) and the non-physical small BH (red curve).
For the physical large BH $c_s^2$ is positive and therefore the speed of sound is real. In contrast, for the non-physical small BH $c_s^2$ is negative and the speed of sound becomes imaginary. 
For a small value of the chemical potential, e.g.  $\mu = 50$ MeV (upper right panel), one can already see the appearance of a small physical BH (green curve) and  $c_s^2$ now displays a ``ribbon bow" pattern. Note that $c_s^2$ is positive for the physical large BH (blue) and the physical small BH (green) implying a real value for the speed of sound, as expected. For $\mu = \mu^c = 225$ MeV (lower left panel), the ``ribbon bow" shrinks until it becomes a single point (red dot), and there  remain only the  two physical BH phases (blue and green curves). For $\mu \gg \mu^c$, e.g. $\mu=500$ MeV (lower right panel), the two physical BH phases have already merged into a single continuous curve. It is worth  mentioning that, in all plots, the speed of sound for the physical BH phase approaches the conformal value 1/3 at very high temperatures, 
as expected \footnote{For a more detailed discussion from the holographic perspective see \cite{Hohler:2009tv,Cherman:2009tw,Yang:2017oer} and from a non-holographic approach see \cite{Khaidukov:2018lor,Khaidukov:2018vkv}.}.

As a final remark, when describing phase transitions in the end one is only interested in the stable solutions, that correspond to the ground state (minimum) of the grand canonical potential. These solutions are identified with thick blue and green solid lines in figures \ref{fig:cv_all} and \ref{fig:cs3}.

\subsection{Charge density and charge susceptibility}

In subsection \ref{OmegaReconstruction} we obtained the following formula for the charge density 
\begin{eqnarray} \label{chargedensityformula}
Q(z_h, \mu) = 2 \mu \int_{\infty}^{z_h} S'(z) a(z) b(z) 
\equiv d(z_h) \, \mu \,.  
\end{eqnarray}
We use this formula to evaluate the charge density in our model. 
From the charge density one can obtain the charge susceptibility through the thermodynamic relation 
\begin{eqnarray} \label{SusceptDef}
\chi = \left(\dfrac{\partial Q}{\partial \mu}\right)_{V,\,T} \,. 
\end{eqnarray}
In our model the thermodynamic quantities are functions of the horizon radius $z_h$ and chemical potential $\mu$. Note that the dependence on the temperature $T$ is implicit on $z_h$, so taking the derivative in \eqref{SusceptDef} is not the best way of extracting the charge susceptibility. We describe in appendix \ref{App:Susceptibility} how the charge susceptibility can be obtained for non-conformal plasmas  arising from the EMD holography. 

\begin{figure}[ht]
	\centering
	\includegraphics[scale=0.245]{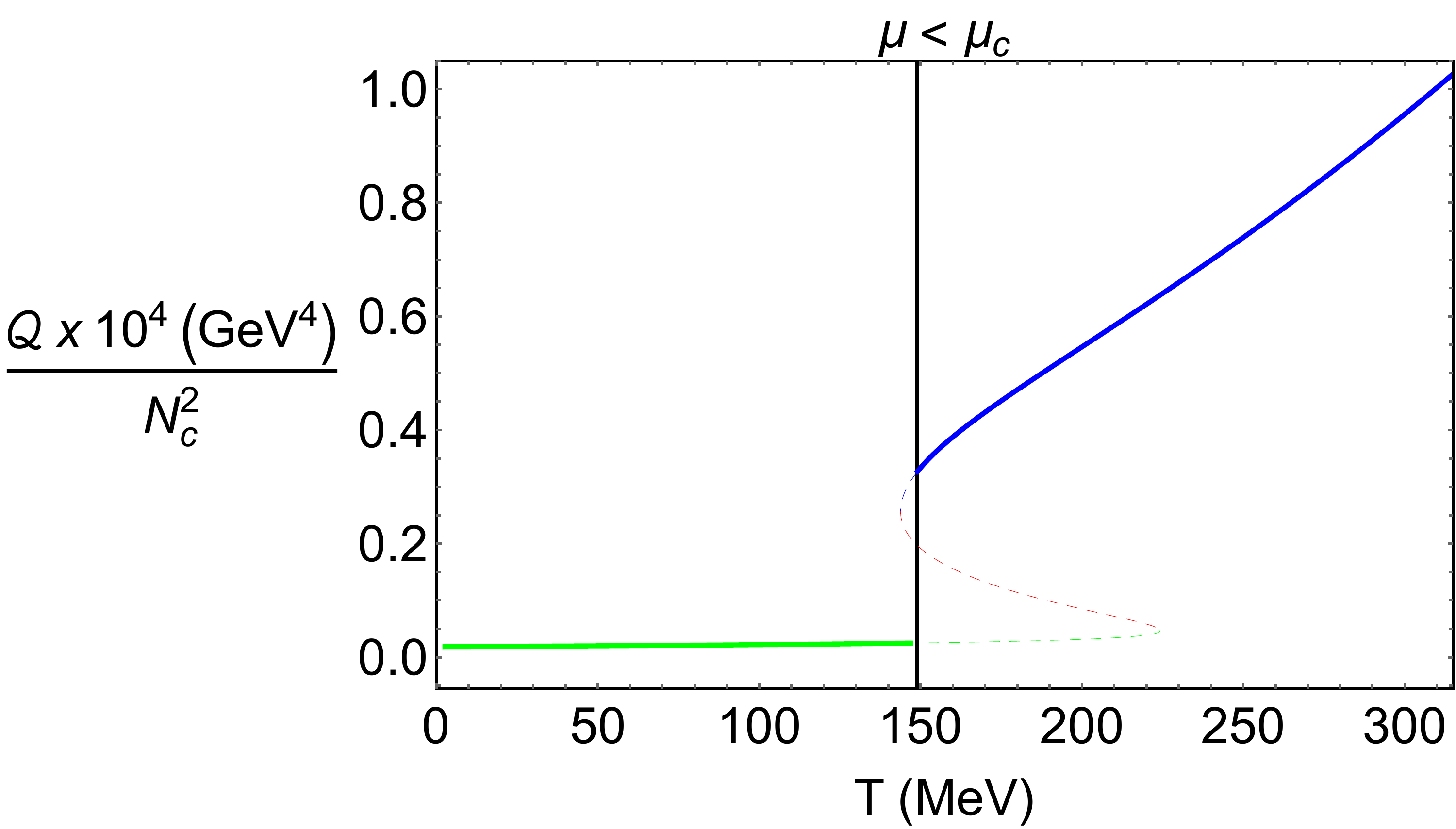}
	\hfill
	\includegraphics[scale=0.24]{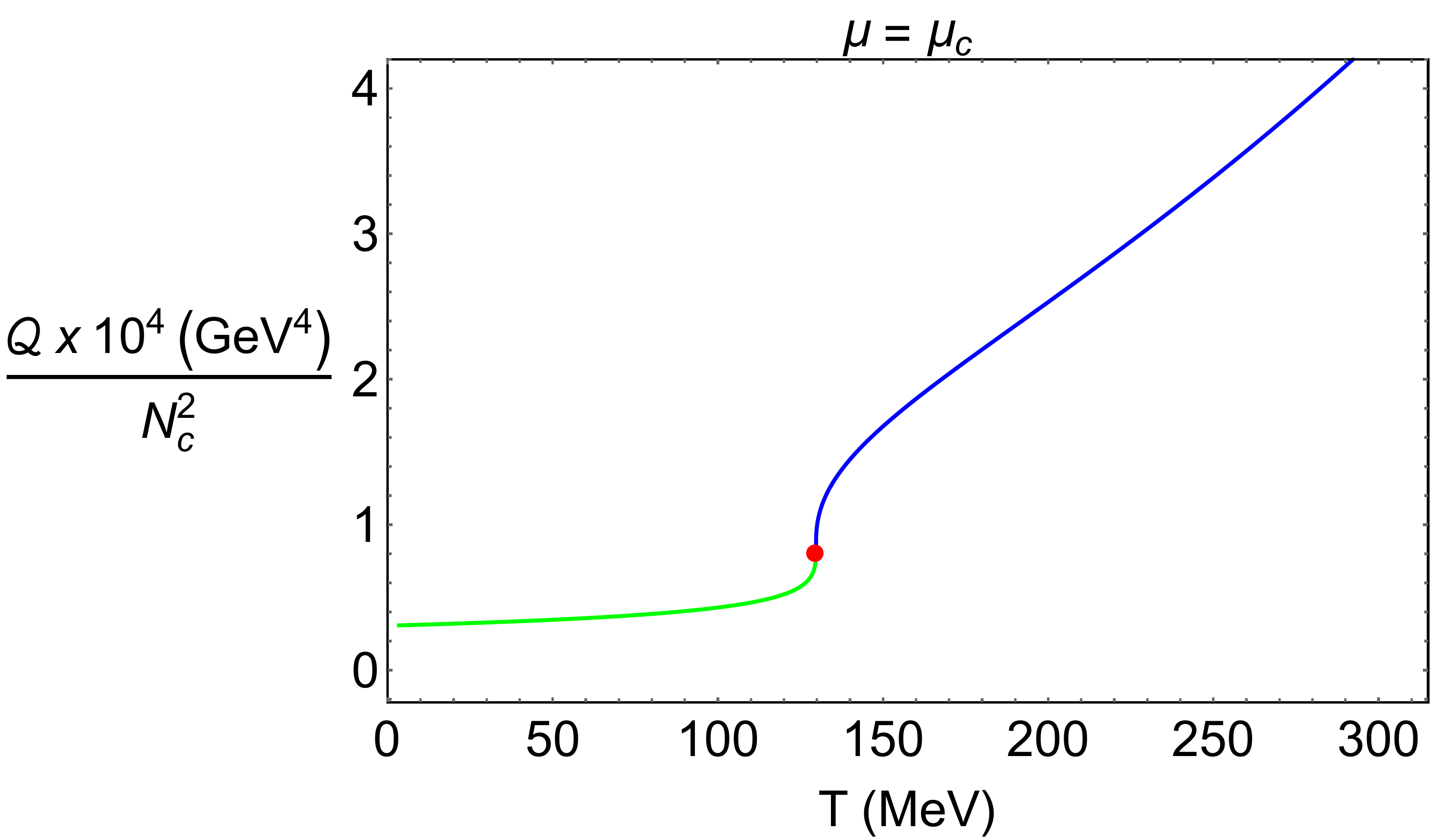}
  \hfill
	\includegraphics[scale=0.25]{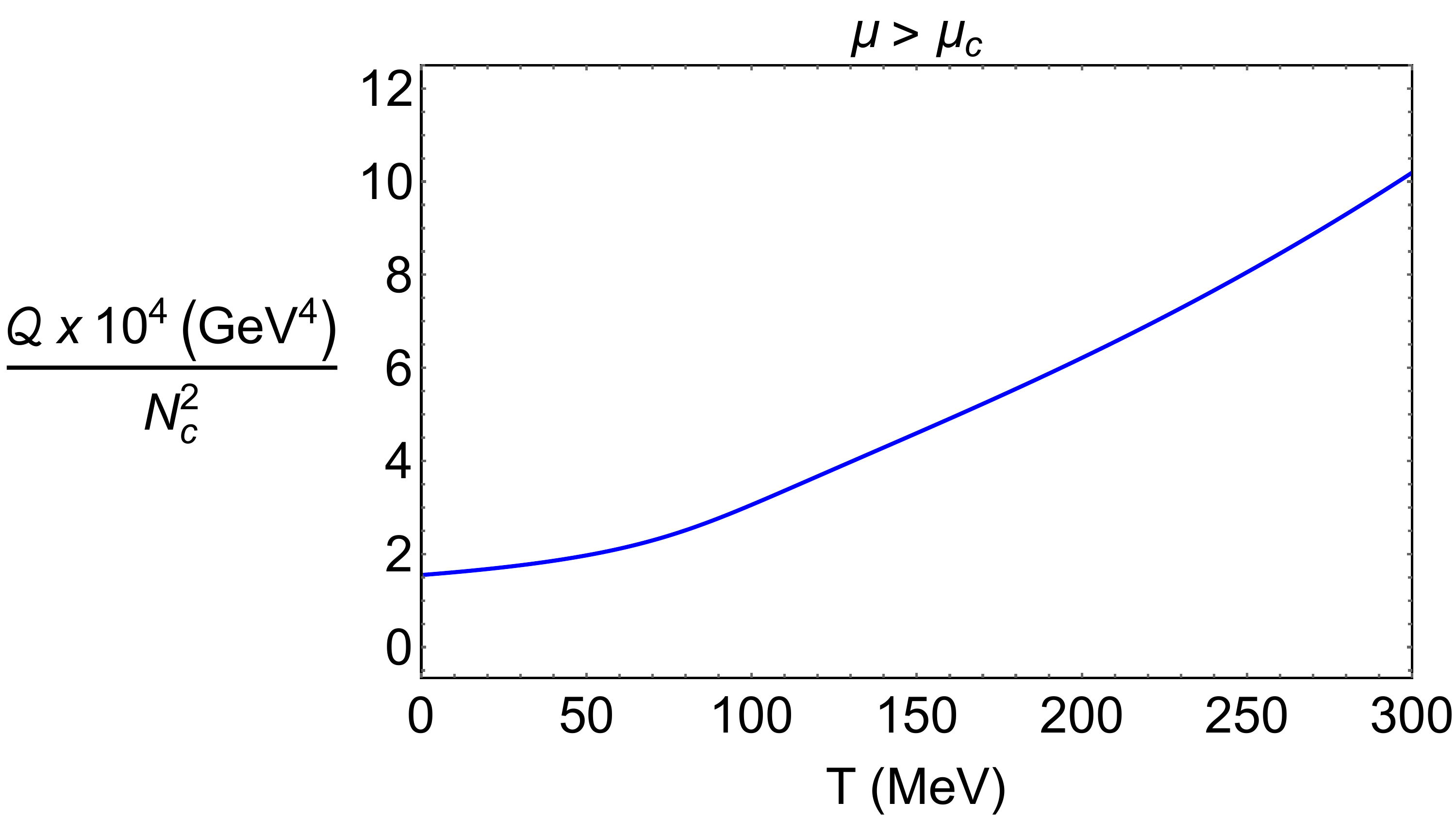}
	\hfill
	\caption{The charge density $Q$ as a function of the temperature $T$ for fixed values of the chemical potential: i) $\mu =50$ MeV (upper left panel) ii) $\mu=\mu^c=225$ MeV (upper right panel) and iii) $\mu=500$ MeV (lower panel). The red dashed line represents the small non-physical BH, the solid (dashed) blue line represents the large BH in a stable (metastable) state and the solid (dashed) green line represents the small physical BH in a stable (metastable) state. The vertical solid line for $\mu<\mu^c$ marks the temperature where a first-order phase transition takes place. }
	\label{fig:Q}
\end{figure}

\begin{figure}[ht]
	\centering
	\includegraphics[scale=0.235]{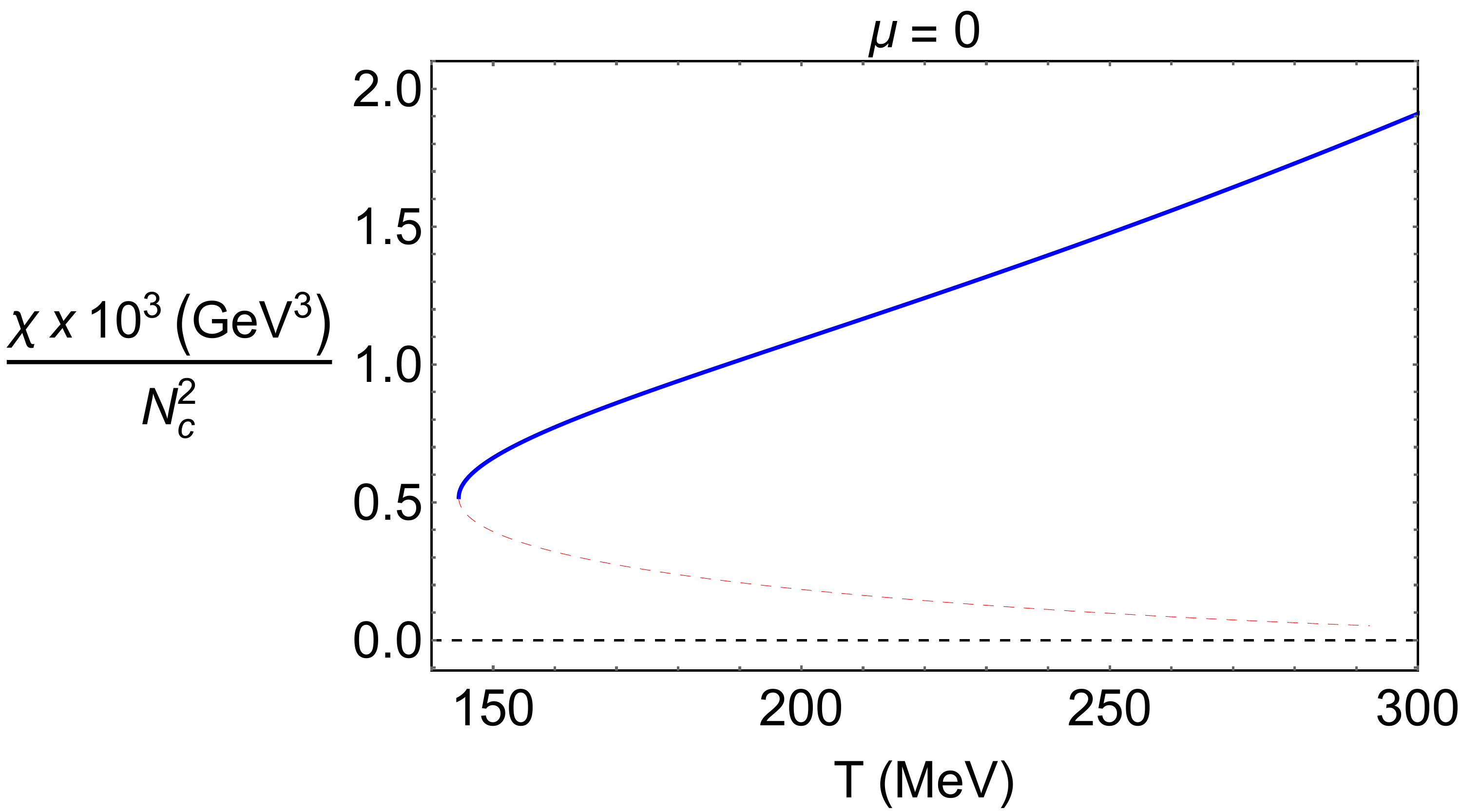}
	\hfill
	\includegraphics[scale=0.235]{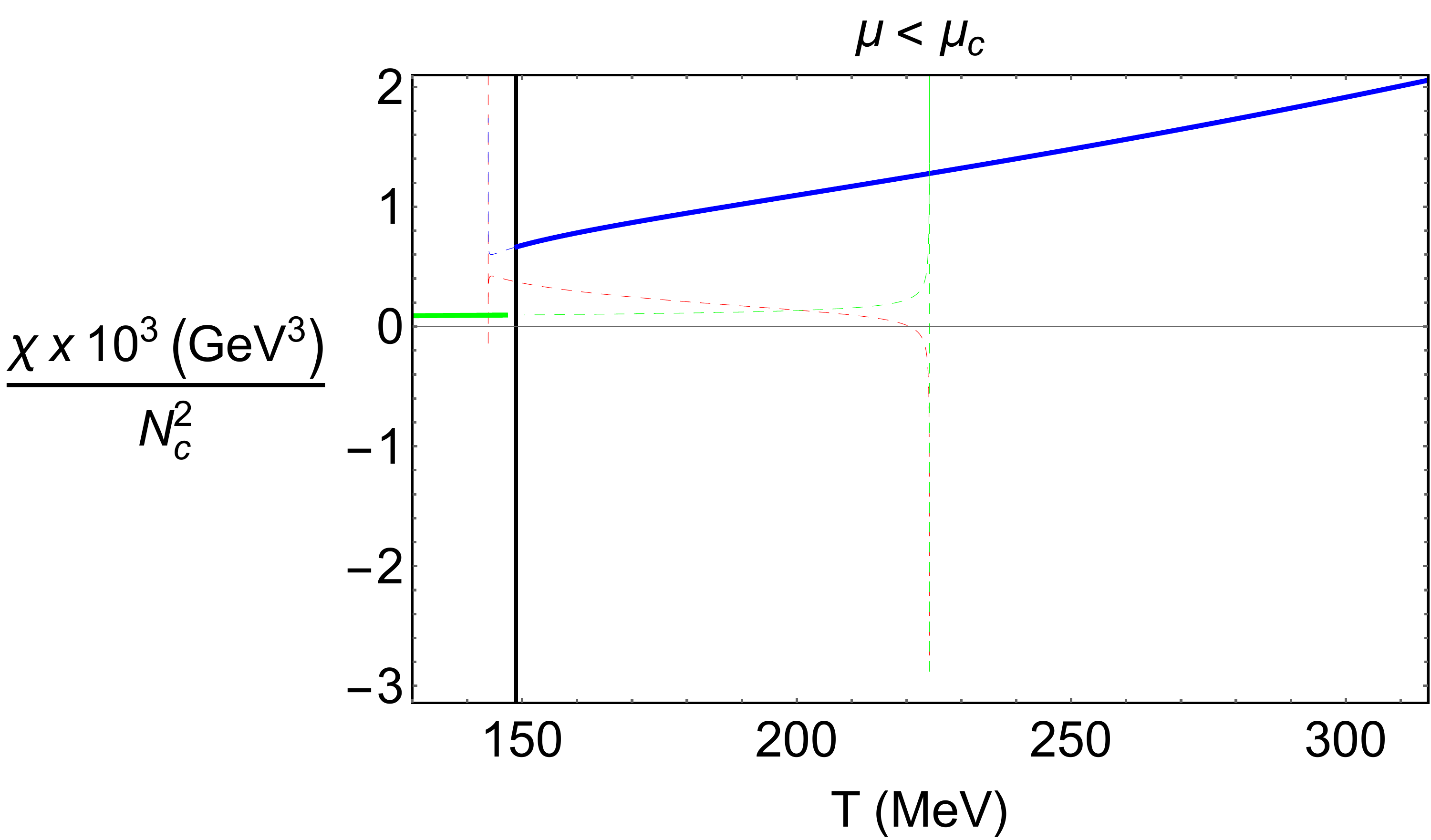}
  \hfill 
	\includegraphics[scale=0.225]{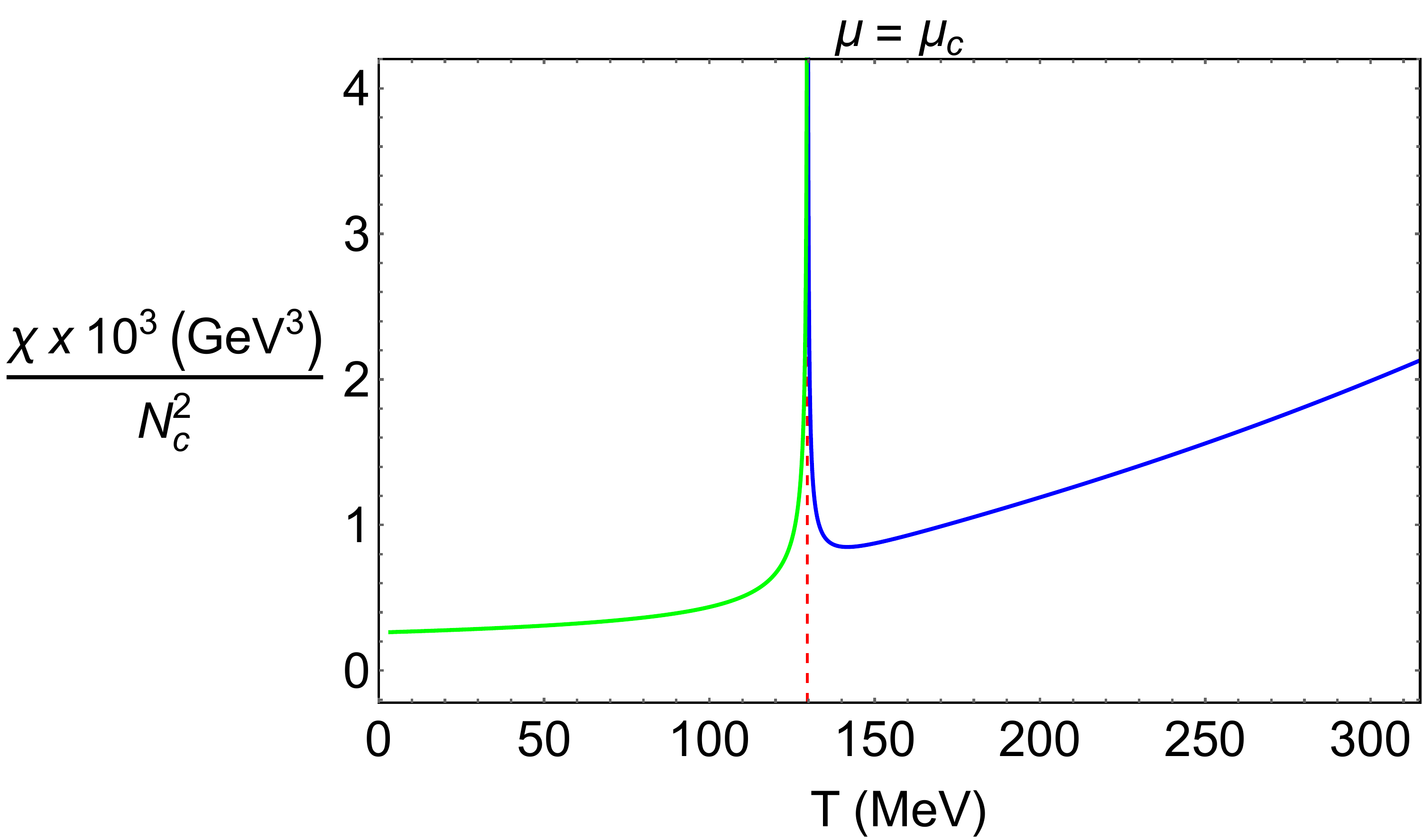}
	\hfill
	\includegraphics[scale=0.245]{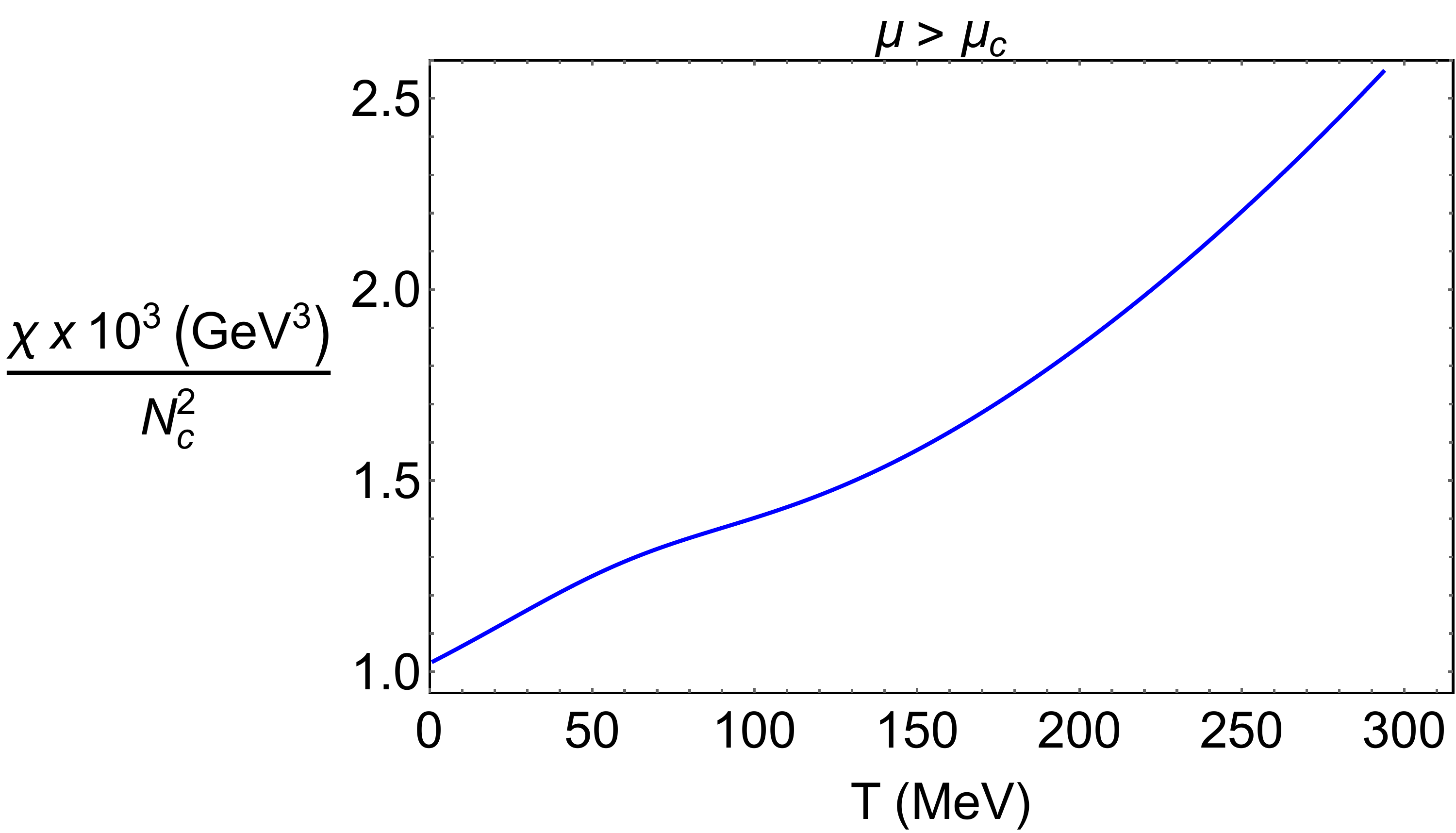}
	\caption{The charge susceptibility $\chi$ as a function of the temperature $T$ for fixed values of the chemical potential: i) $\mu =0$ (upper left panel) ii) $\mu=50$ MeV (upper right panel),  iii) $\mu= \mu^c= 225$ MeV (lower left panel) and $\mu=800$ MeV (lower right panel). The red dashed line represents the small non-physical BH, the solid (dashed) blue line represents the large BH in a stable (metastable) state and the solid (dashed) green line represents the small physical BH in a stable (metastable) state. The vertical solid line for $\mu<\mu^c$ marks the temperature where a first-order phase transition takes place.}
	\label{fig:chi}
\end{figure}

Fig. \ref{fig:Q} presents our numerical results for the charge density for fixed values of the chemical potential. In the following, in Fig. \ref{fig:chi}, the numerical results for the charge susceptibility are presented, using the formula \eqref{chargesuscept}, obtained in appendix \ref{App:Susceptibility}.

In the left panel of Fig. \ref{fig:Q} one can see that the charge density presents the same behaviour as the entropy density (see Fig. \ref{fig:EntropyMuHigh}), showing that in the range $\mu<\mu^c$ we have the three black hole branches, namely the physical small black hole (green curve), the non-physical small black hole (red curve), and the physical large black hole (blue curve). The physical small black hole branch dominates in the low-temperature regime while the physical large black hole branch dominates in the high-temperature regime. In between, at a certain temperature, represented by a vertical dashed line, the system undergoes a first-order phase transition.

On the other hand, at the critical point, $\mu=\mu^c$, in the right panel of Fig. \ref{fig:Q} nothing special happens, but one can see a divergence in the charge susceptibility (lower left panel of Fig. \ref{fig:chi}), characterizing a second-order phase transition.

Beyond the critical point, i.e, in the range $\mu>\mu^c$ (Figs. \ref{fig:Q} (lower middle panel) and \ref{fig:chi} (right lower panel)), we have a analytic crossover, since the grand canonical potential and its related quantities are analytic functions of the temperature.

As a final remark, when describing phase transitions in the end one is only interested in the stable solutions, that correspond to the ground state (minimum) of the grand canonical potential. These solutions are identified with thick blue and green solid lines in figures \ref{fig:Q} and \ref{fig:chi}.

\subsection{Trace Anomaly}

As described in subsection \ref{OmegaReconstruction}, the trace of the energy momentum tensor takes the form 
\begin{equation}
\left\langle T^{a}_{\;\;a}\right\rangle = E-3p = 4\,\Omega + TS + \mu\,Q \,. 
\end{equation}
A non-zero value for $E-3p$ signifies the breaking of conformal symmetry. At $\mu=0$, when the Maxwell term is absent, the presence of the dilaton deforms the AdS space leading to a non-zero value for $E-3p$. In our model, the breaking of conformal symmetry at $\mu=0$ will be associated with the dilaton parameter $k$; namely the trace anomaly will be proportional to $k^2$. In the absence of the dilaton term, i.e. $k=0$, we recover the Einstein-Maxwell action with the solution given by the charged Reissner-N\"ordstrom AdS$_5$ black hole and $E-3p$ vanishes, see e.g. \cite{Hartnoll:2009sz}. It turns out that the simultaneous presence of  the dilaton and the Maxwell term lead to a non-vanishing value for $E-3p$ at any point in the $T-\mu$ phase diagram excepting one special point: the critical end point (CEP).  At the CEP the trace anomaly $E-3p$ vanishes and we therefore expect the restoration of conformal symmetry and the presence of a non-trivial CFT. 

\begin{figure}[ht]
	\centering
		\includegraphics[scale=0.24]{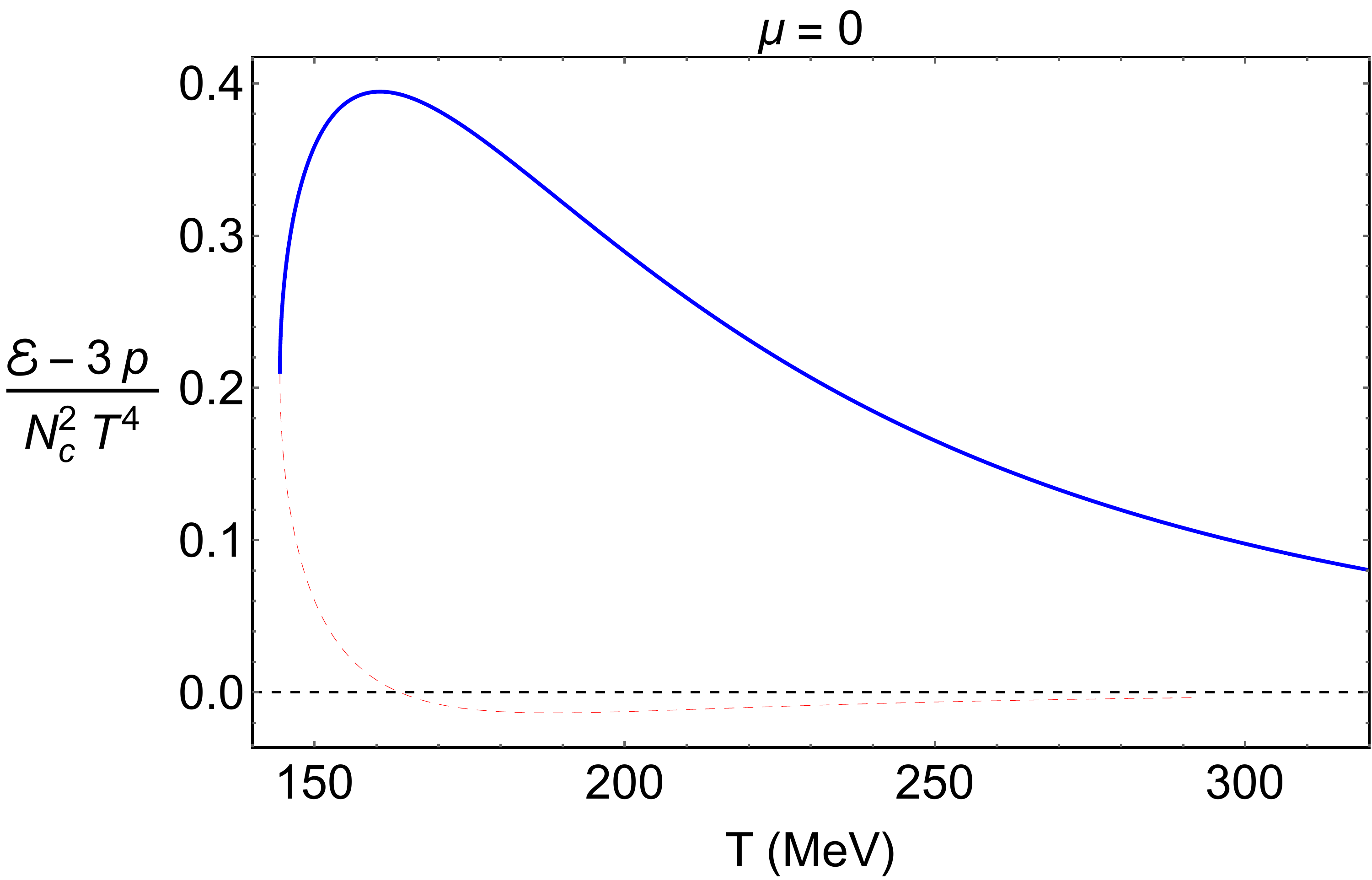}
	\hfill
	\includegraphics[scale=0.24]{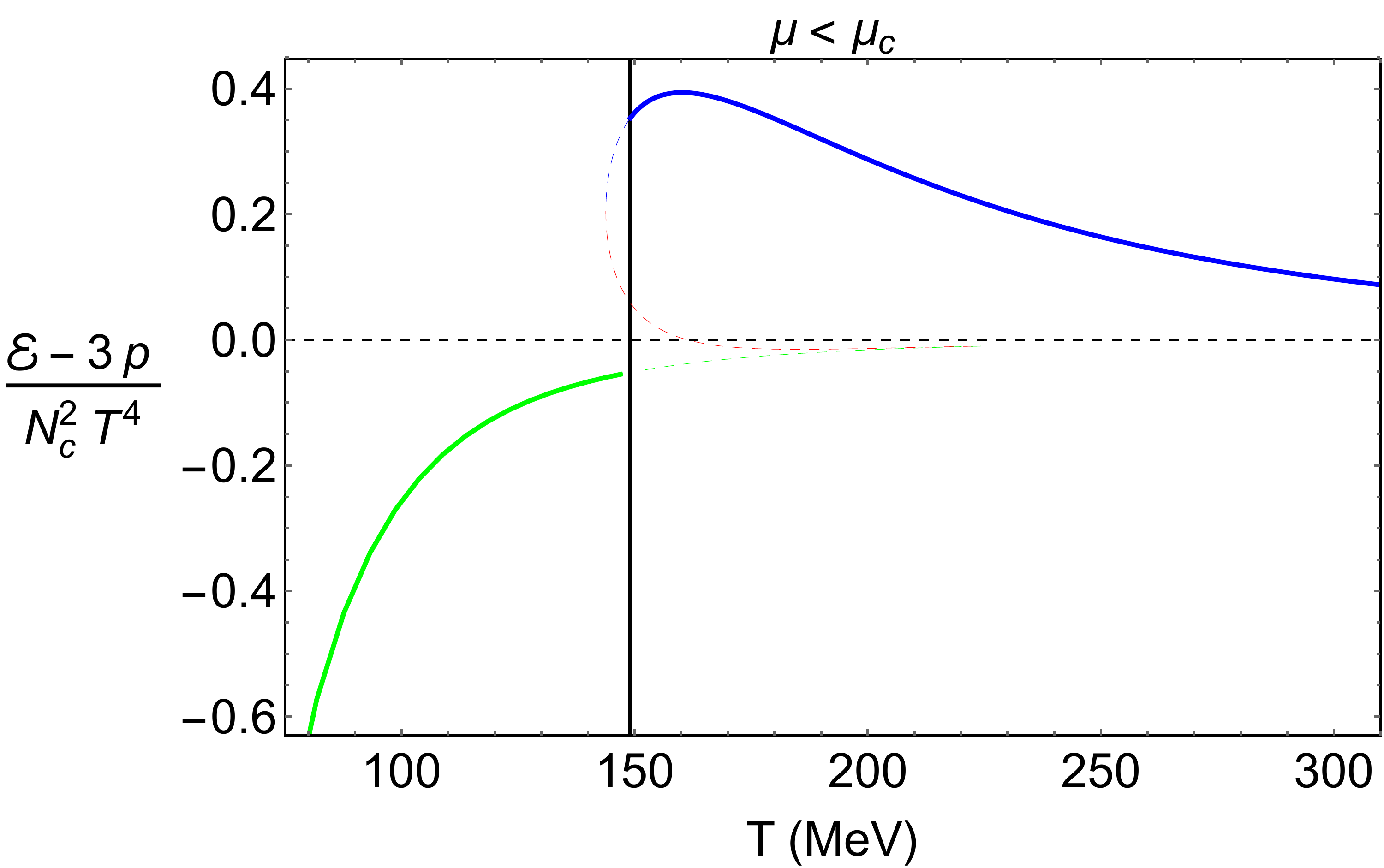}
 \hfill
	\includegraphics[scale=0.24]{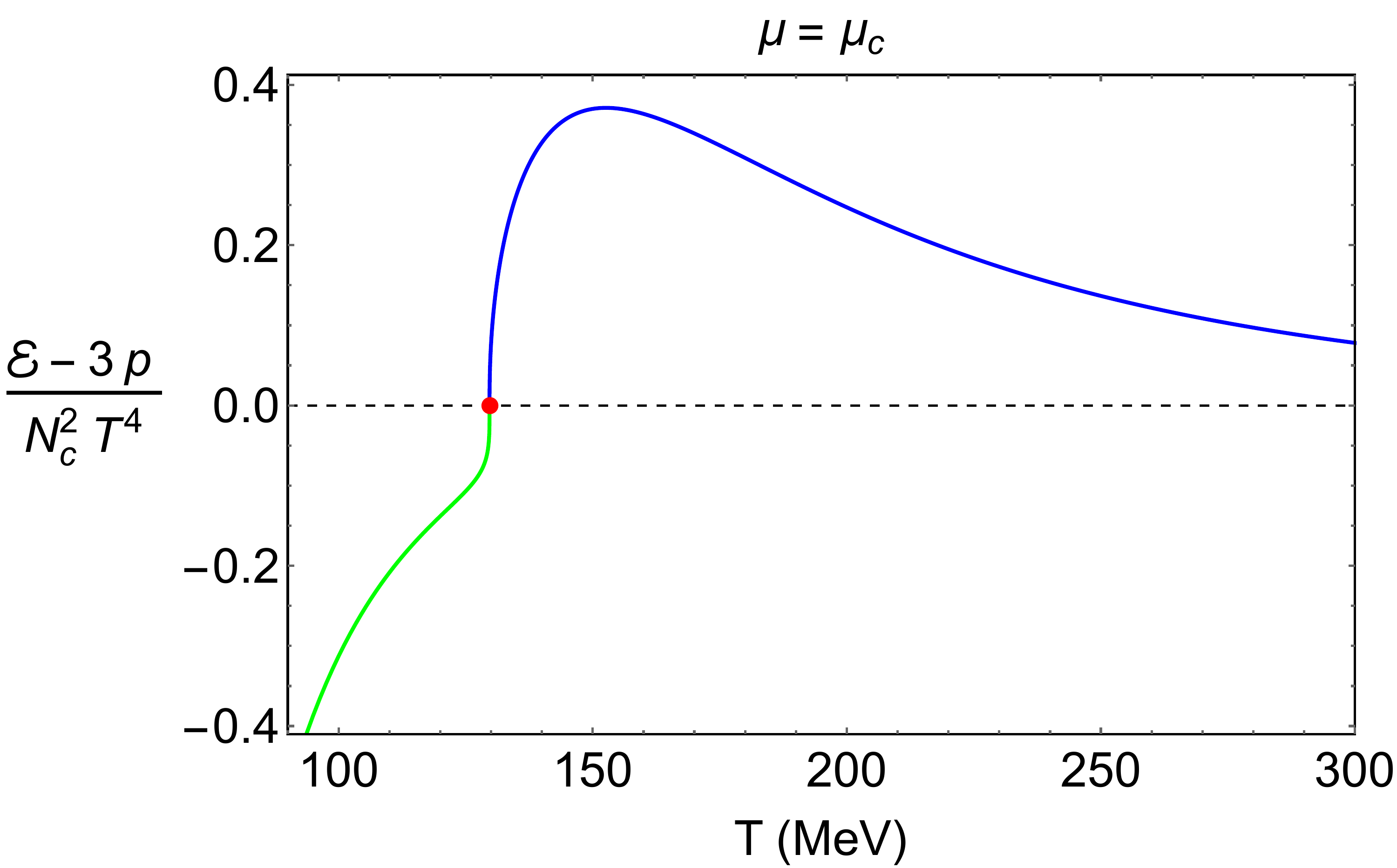}
	\hfill
	\includegraphics[scale=0.24]{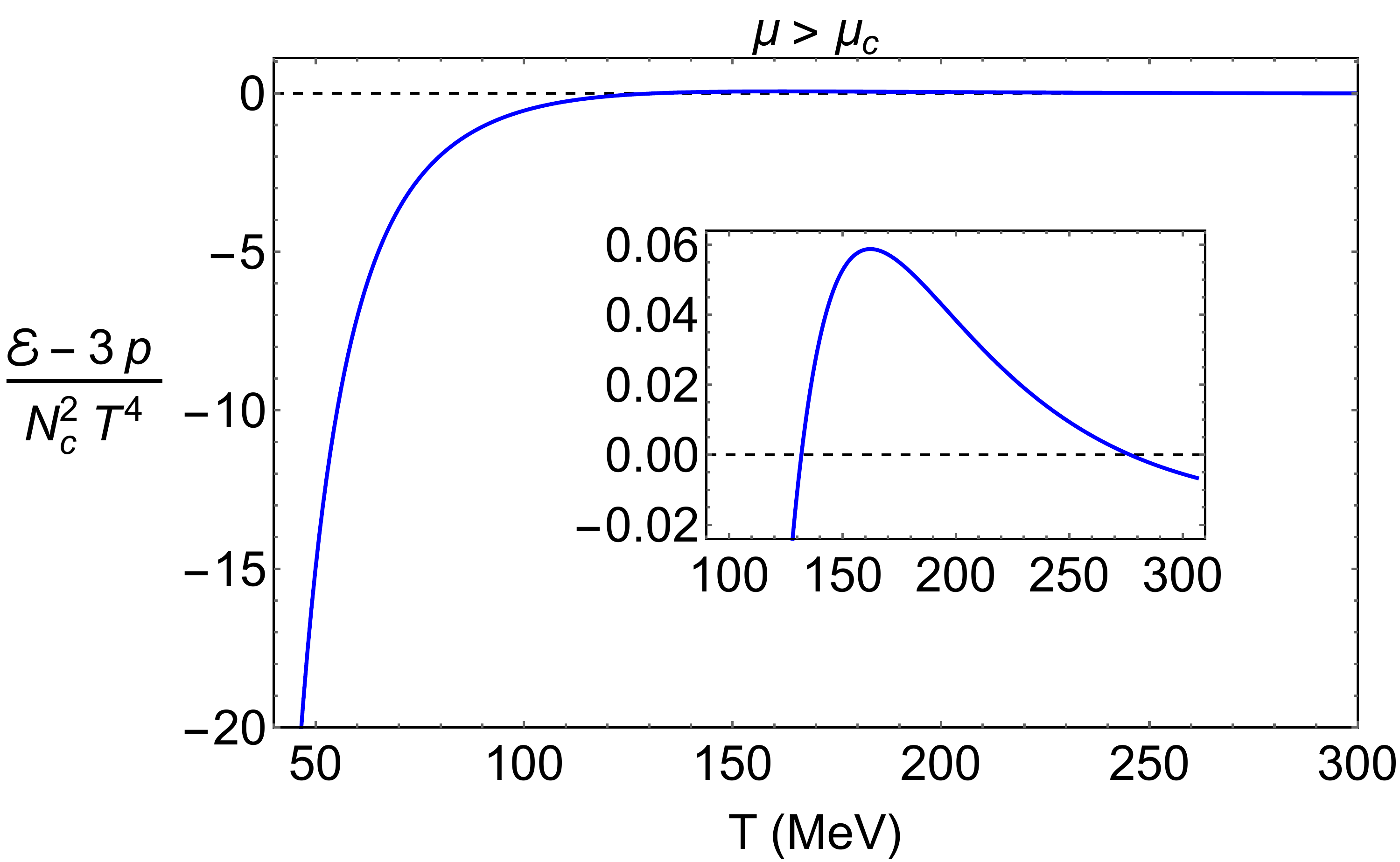}
	\caption{The trace anomaly $E - 3p$ as a function of the temperature $T$ for fixed values of the chemical potential: i) $\mu =0$ (upper left panel) ii) $\mu=50$ MeV (upper right panel),  iii) $\mu= \mu^c= 225$ MeV (lower left panel) and $\mu=500$ MeV (lower right panel). The red dashed line represents the small non-physical BH, the solid (dashed) blue line represents the large BH in a stable (metastable) state and the solid (dashed) green line represents the small physical BH in a stable (metastable) state. The vertical solid line for $\mu<\mu^c$ marks the temperature where a first-order phase transition takes place.}
	\label{fig:tranomaly}
\end{figure}

In figure \ref{fig:tranomaly} we present our numerical results for the trace anomaly $E- 3p$. Interestingly, the region associated with the unstable small BH phase (red curves) always crosses the axis where $E-3p=0$. See e.g. the plot for $\mu=0$ (upper left panel) or the plot for $\mu=50$ MeV (upper right panel).  In general, the region associated with the unstable small BH is non-physical but at the critical regime $(\mu^c, T^c)$ (lower left panel) the unstable BH phase shrinks to a point (red dot) and it it becomes the limit of the physical BH phases (blue and green curves). At that critical point the trace anomaly $E-3p$ vanishes and therefore conformal symmetry is restored. We conclude that we are reaching a non-trivial conformal field theory (CFT) at the critical point. Later in this work we will evaluate the thermodynamics near the critical point and extract the critical exponents in order to learn more about this non-trivial CFT.  For sufficiently large values of $\mu>\mu^c$, e.g. $\mu=500$ MeV (lower right panel), the physical BHs have already merged into a single continuous curve.  

As a final remark, when describing phase transitions in the end one is only interested in the stable solutions, that correspond to the ground state (minimum) of the grand canonical potential. These solutions are identified with thick blue and green solid lines in Fig.\ref{fig:tranomaly}.

\subsection{The $T-\mu$ Phase Diagram}

We finish this section presenting the most important result in this work, namely the $T-\mu$ Phase Diagram. 

\begin{figure}[ht]
	\centering
	\includegraphics[scale=0.245]{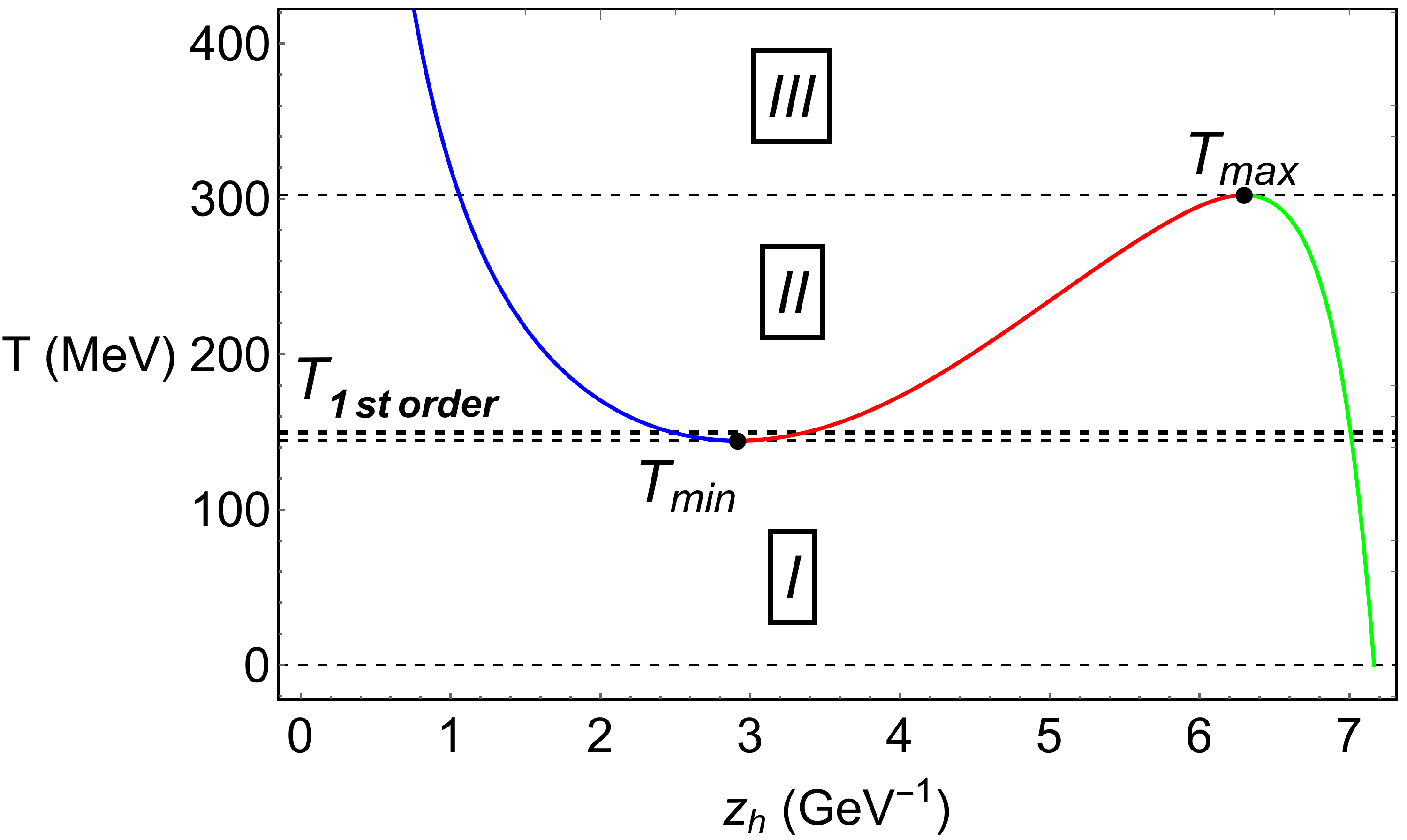}
	\hfill
	\includegraphics[scale=0.245]{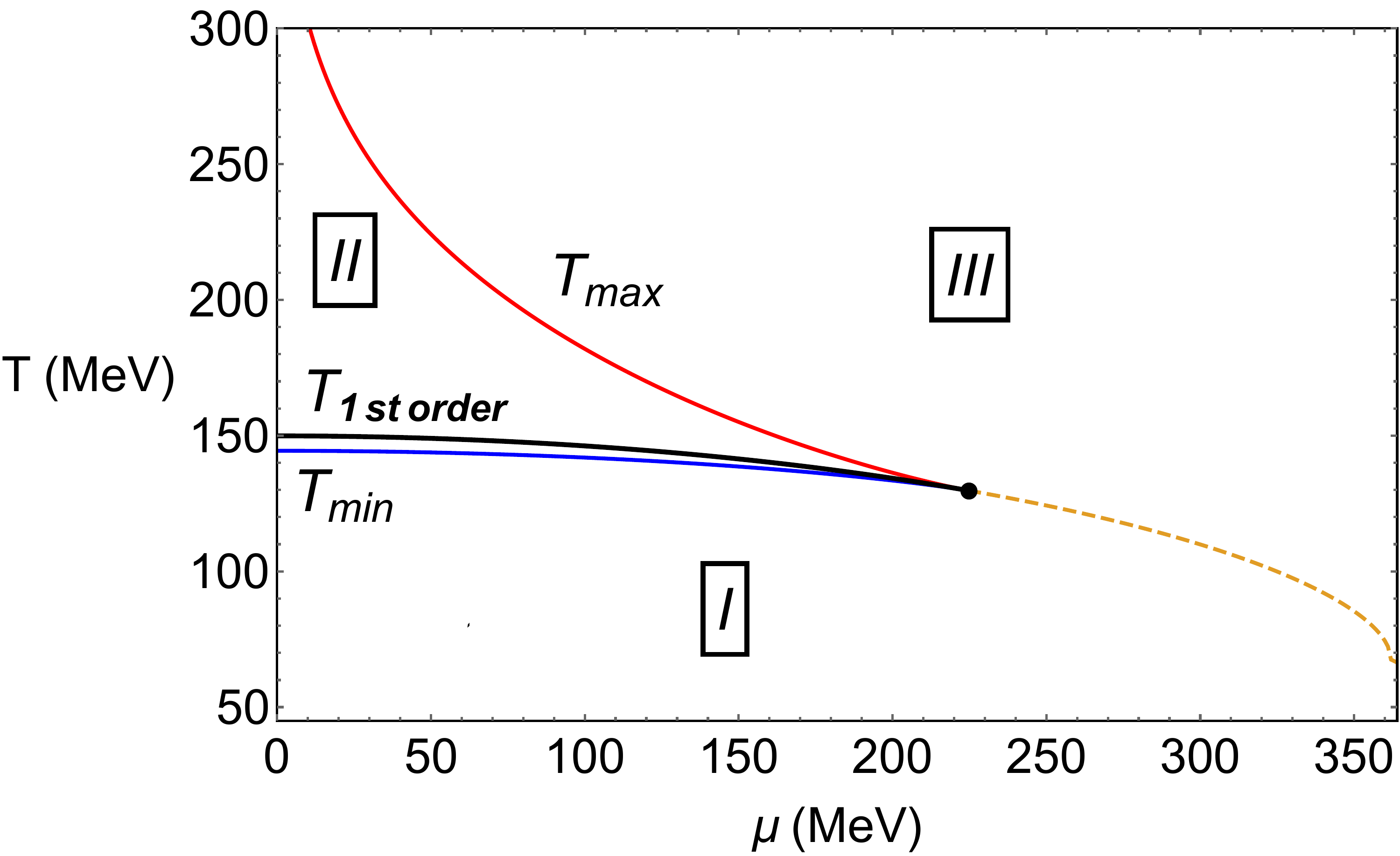}
	\caption{Left panel: Description of the three temperature regimes: I) $T<T_{\rm min}$, II) $T_{\rm min} < T < T_{\rm max}$ and III)
	$ T > T_{\rm max}$ in the plot of the temperature $T$ as a function of the horizon radius $z_h$. The blue, red and green curves represent the large BH, small non-physical BH and small physical BH respectively. Right panel: Description of the temperature regimes I), II) and III) this time in the $T- \mu$ plane.  } 
	\label{fig:phasediag}
\end{figure}

But first we remind the reader that in figure \ref{fig:Tfinitemuv2}  we presented a plot of the BH temperature $T$ as a function of the horizon radius $z_h$ for $\mu<\mu^c$ (upper right panel). In the left panel of figure \ref{fig:phasediag} we present the same plot but this time distinguishing three regions: 
\begin{itemize}
    \item Region I: $0<T<T_{\rm min}$ ,
    \item Region II: $T_{\rm min} < T < T_{\rm max}$,
    \item Region III: $T> T_{\rm max} $.
\end{itemize}

\begin{figure}[ht]
	\centering
	\includegraphics[scale=0.3]{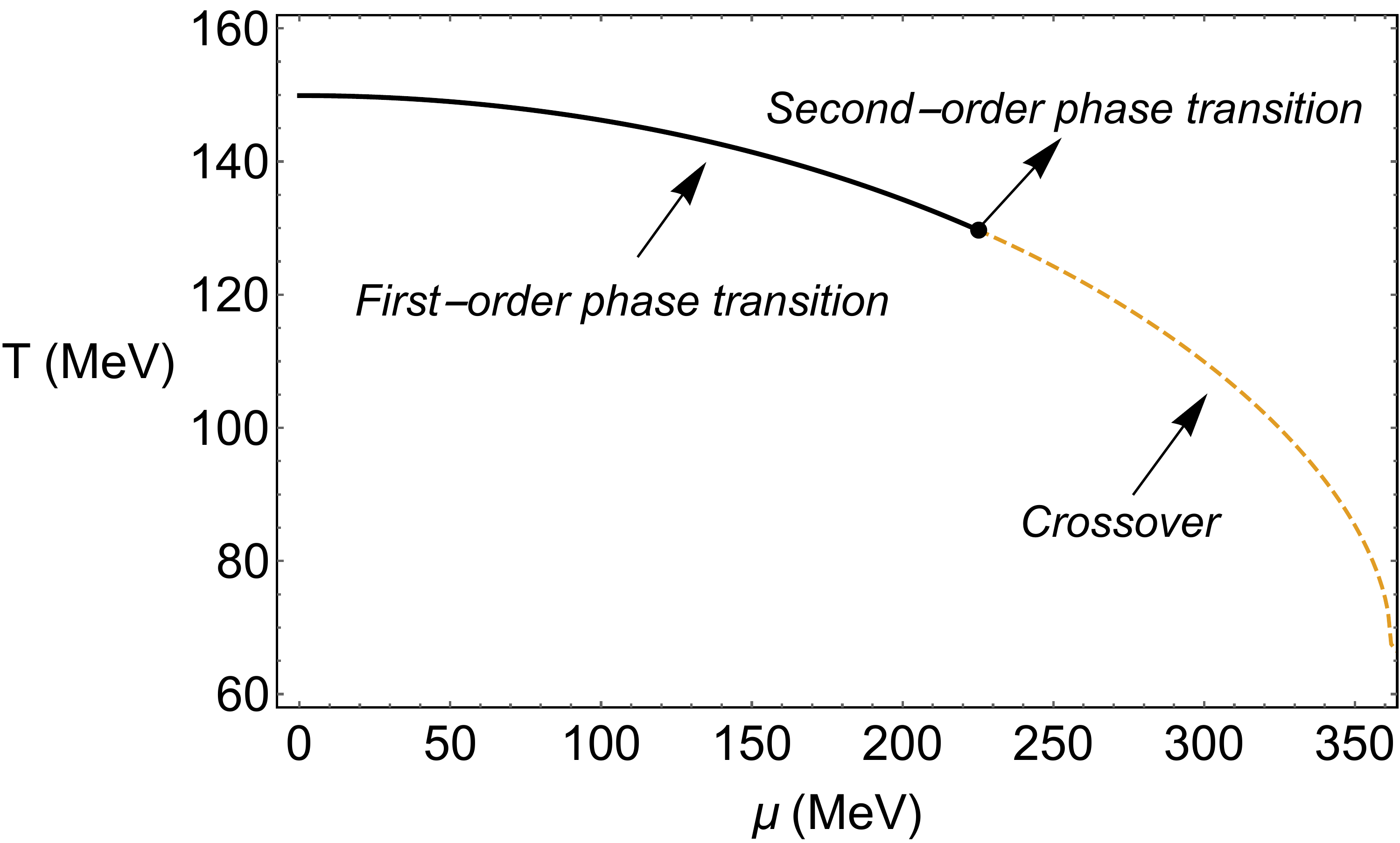}
	\caption{The final $T-\mu$ phase diagram, obtained within our EMD holographic model, presenting a critical line and the critical point (black disk) which has the following coordinates $(\mu^c,T^c)\simeq (225, 130)$ MeV.}
	\label{fig:phasediag1}
\end{figure}

In region I one can see only the physical small BH (green curve). In  region II we have a coexistence of three BHs phases, the physical large BH (blue curve), the non-physical small BH (red curve) and the physical small BH (green curve). In region III, the only possible phase is the physical large BH (blue curve). The dashed horizontal line represents the critical temperature $T_{\rm 1st \, order}$, where a first-order transition takes place between the physical BH phases (green and blue curves). This critical temperature lies  above the minimum temperature $T_{\rm{min}}$ and below the maximum temperature $T_{\rm{max}}$. The right panel of Fig. \ref{fig:phasediag} presents the phase diagram $T - \mu$ for the black hole phases, displaying also the regions I, II and III. The main feature of the phase diagram is the fist-order transition represented by the black line. This is the transition between the physical small BH (low temperature) and the physical large BH (high temperature).  The other interesting feature in the phase diagram is the presence of two other transitions taking place at $T_{\rm min}$ and $T_{\rm max}$, represented by a blue and a red curve respectively. These are second-order transitions between the physical BHs and the non-physical BH:  $T_{\rm min}$ ($T_{\rm max}$) corresponds to the transition between the physical large (small) BH and the non-physical small BH. Note that the black, blue, and red curved lines converge to a single point. This is the critical end point (CEP) in the $T-\mu$ phase diagram, with coordinates $T^c, \mu^c$,  where the first-order transition becomes second-order. For $\mu > \mu^c$ the unstable small BH disappeared and there is a crossover transition between the physical small BH and large BH. 

We present in Fig.  \ref{fig:phasediag1} our final $T-\mu$ phase diagram, obtained within our EMD holographic model. The black line, black dot and orange dashed line represent the first-order transition, the critical point and the crossover transition respectively. The thermodynamic analysis performed in this section confirm our previous result for the critical point. Namely, the critical point in our holographic model is located at
\begin{equation}
(\mu^c,T^c)\simeq (225, 130) \; \mathrm{MeV}.
\end{equation}

\section{Thermodynamics near the Critical Point} 
\label{Sec:CriticalThermod}

In this section we will study the behaviour of some thermodynamic quantities near the critical point found in this work. We will provide useful expansions for the temperature and the grand canonical potential near the critical point.  These expansions will be useful for calculating other thermodynamic quantities near the critical point and find the critical exponents. Moreover, they shed some light in the connection between the criticality found in this work and the criticality usually described  catastrophe theories $A_3$. This connection was originally proposed in  \cite{Chamblin:1999tk,Chamblin:1999hg} and here we provide a more concrete realisation. 

To simplify the presentation we will omit units in this section, but they can easily be obtained via dimensional analysis.

\subsection{Temperature near the critical point}

 The temperature is a function of two variables, the horizon radius $z_h$ and the chemical potential $\mu$. The explicit form, found in subsection \ref{subsec:Temperature}, is conveniently written as
\begin{equation} \label{Tempv2}
T(z_h , \mu) = a(z_h) -  \mu^2 a(z_h) b(z_h) \,.     
\end{equation}
The critical point $(z_h^c, \mu^c)=(3.466,0.225)$ was obtained imposing the conditions $T_{10}(z_h^c,\mu^c)=0$, and $T_{20}(z_h^c , \mu^c)=0$, where we have introduced the notation
\begin{equation}
T_{mn}(z_h, \mu) = \frac{\partial^{m+n}}{\partial z_h^m \partial \mu^n} T(z_h, \mu) \,  .      
\end{equation}
To investigate the physics near the critical point we consider the Taylor expansion
\begin{align} \label{TaylorExpT}
T(z_h^c + \delta z_h , \mu^c + \delta \mu) &= T^{\, c} + T_{01}^{\, c} \delta \mu + \frac12 T_{02}^{\, c} \, \delta \mu^2  + T_{11}^{\, c} \delta z_h \delta \mu + \frac12 T_{21}^c \delta z_h^2 \delta \mu \nonumber \\ 
&+ \frac12 T_{12}^c \delta z_h  \delta \mu^2 
+ \frac16 T_{30}^{\, c} \delta z_h^3 + \dots \, ,
\end{align}
where $T^{\, c} = T(z_h^c , \mu^c) = 0.1297$ and $T_{mn}^{\, c} = T_{mn}(z_h^c , \mu^c)$. Note the absence of the pure $\delta z$ and $\delta z^2$ terms because the coefficients $T_{10}^{\, c}$ and $T_{20}^{\, c}$ vanish.  Any term containing powers $\delta \mu^3$ or higher are also absent because the coefficients $T_{mn}^c$ vanish when $n>2$.

\begin{figure}[ht]
	\centering
	\includegraphics[scale = 0.4]{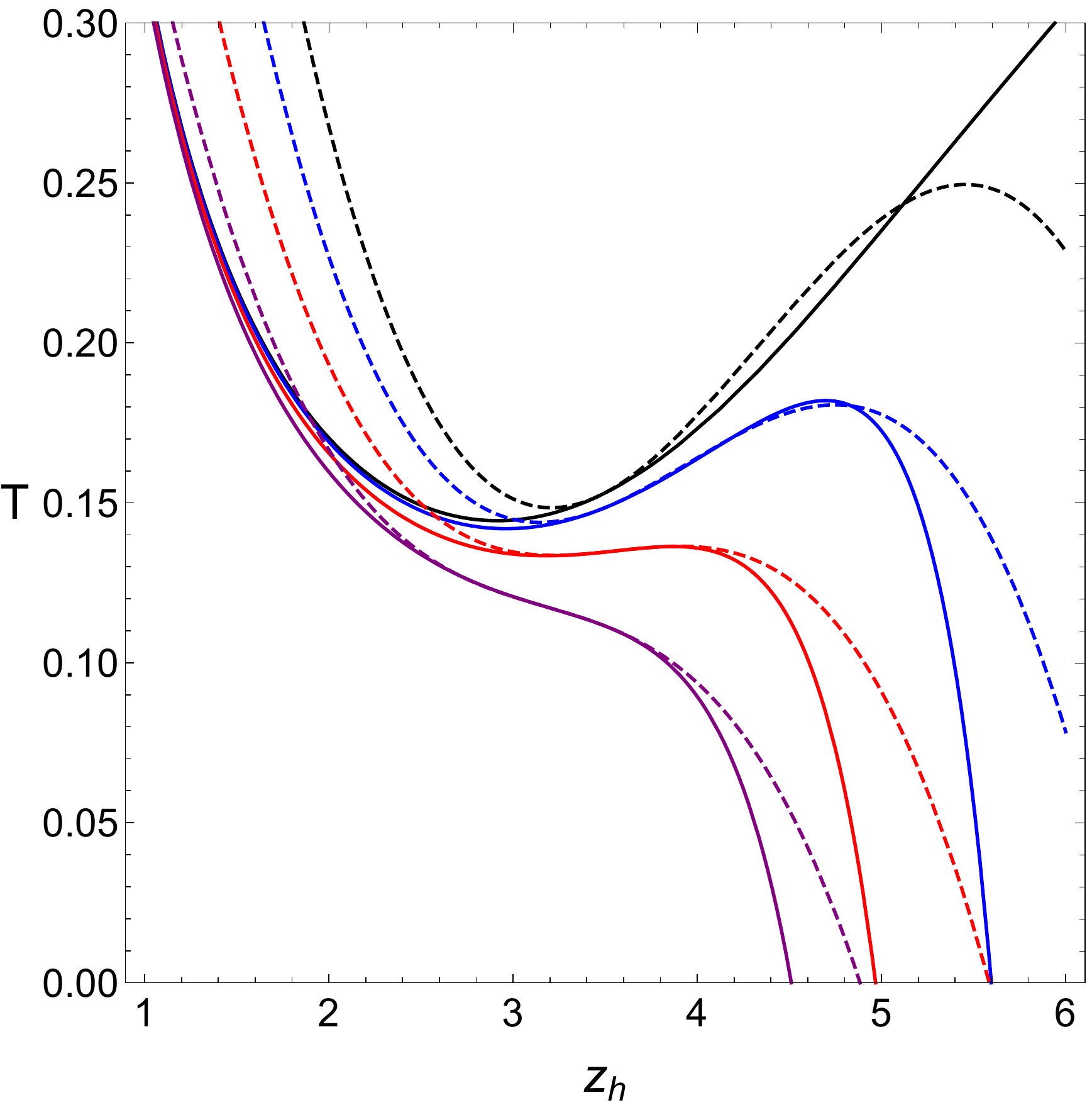}
	\caption{Temperature as a function of the horizon radius $z_h$ obtained from the analytic approximation in \eqref{TempNearTc} and compared with the full numerical result. The dashed (solid) curves represent the analytic (numerical) results. The chemical potential varies from $\mu=0$ (black) to $\mu=0.3$ (purple) in steps of $0.1$.}
	\label{fig:CubicExp}
\end{figure}

It will be sufficient to truncate the series expansion in \eqref{TaylorExpT} at cubic order. From \eqref{Tempv2} we evaluate the other coefficients and find
\begin{align}
T_{01}^{\, c} &\approx -0.1996 \quad , \quad 
T_{02}^{\, c} \approx -0.8872 \quad , \quad 
T_{11}^{\, c} \approx -0.2438 \, , \nonumber \\
T_{21}^{\, c} &\approx -0.4113 \quad , \quad  
T_{12}^{\, c} \approx -1.084 \quad , \quad T_{30}^{\, c} \approx -0.107 \,. 
\end{align}
From these results we are able to approximate the temperature difference $\delta T = T - T^{\, c}$ by a polynomial of cubic order in $\delta z$, i.e.
\begin{align} \label{TempNearTc}
\delta T &\approx  - \delta \mu (0.1996 + 0.4436  \delta \mu) 
- (0.2438 +0.5418 \delta \mu ) \, \delta \mu \delta z \nonumber \\
&  - 0.2057 \, \delta \mu \delta z^2  -0.0178 \, \delta z^3 \,.  
\end{align}
Alternatively, we can fix $\delta T$ and $\delta \mu$ and find $\delta z$ solving the cubic equation
\begin{align}
 & 0.0178 \, \delta z^3 + 0.2057 \, \delta \mu \delta z^2 
+ (0.2438 +0.5418 \delta \mu ) \, \delta \mu \delta z \nonumber \\
&+ \delta \mu (0.1996 + 0.4436  \delta \mu)  + \delta T = 0 \,.  
\end{align}
The three roots $z_h^1 < z_h^2 < z_h^3$  that solve the cubic equation at finite $0<\mu<\mu^c$ correspond to the stable (or metastable) large BH, unstable small BH and stable (or metastable) small BH respectively. In Fig. \ref{fig:CubicExp} we compare the temperature $T(z_h, \mu)$ obtained from the cubic approximation in \eqref{TempNearTc} against the full numerical result obtained in subsection \ref{subsec:Temperature}. As expected, the analytic expansion \eqref{TempNearTc} provides a good approximation for the temperature near the critical point, i.e. when $\mu$ and $z_h$ are near the critical values $\mu^c$ and $z_h^c$ and therefore $T$ is near $T^c$. 

The fact that the temperature $T$ can be approximated near the critical point by a cubic polynomial in $z_h$ strongly suggests a possible  connection between  the critical behaviour of non-conformal plasmas arising from EMD equations and the $A_3$ catastrophic theory \cite{catastrofe1, catastrofe2, catastrofe3}.  This connection was originally suggested by the authors in \cite{Chamblin:1999tk,Johnson:2013dka} for the case of RN AdS$_{n+1}$ BHs in global coordinates, which are solutions with $S^{n-1}$ symmetry  arising from the pure Einstein-Maxwell theory.

\subsection{Grand canonical potential near the critical point}

We want to describe the grand canonical potential $\Omega$ near the critical point. The grand canonical potential is also a function of two variables:  the horizon radius $z_h$ and the chemical potential. As it was done for the temperature, we introduce a simplified notation for the derivatives in $z_h$ and $\mu$ :
\begin{equation} 
\Omega_{mn}(z_h, \mu) = \frac{\partial^{m+n}}{\partial z_h^m \partial \mu^n} \Omega(z_h, \mu) \,  .      
\end{equation}
In subsection \ref{OmegaReconstruction} we provided a method for reconstructing the grand canonical potential $\Omega$. We found that the potential takes the form
\begin{eqnarray} \label{OmegaSol}
\Omega(z_h, \mu) = - A(z_h) - \mu^2 B(z_h) \, , 
\end{eqnarray}
with the functions $A(z_h)$ and $B(z_h)$ obtained from the integrals in \eqref{ABsols}. Interestingly, it turns out that the derivatives $\Omega_{10}(z_h^c,\mu^c)$ and $\Omega_{20}(z_h^c , \mu^c)$ vanish at the critical point $(z_h^c, \mu^c)=(3.466,0.225)$ .

\begin{figure}[ht]
	\centering
	\includegraphics[scale = 0.4]{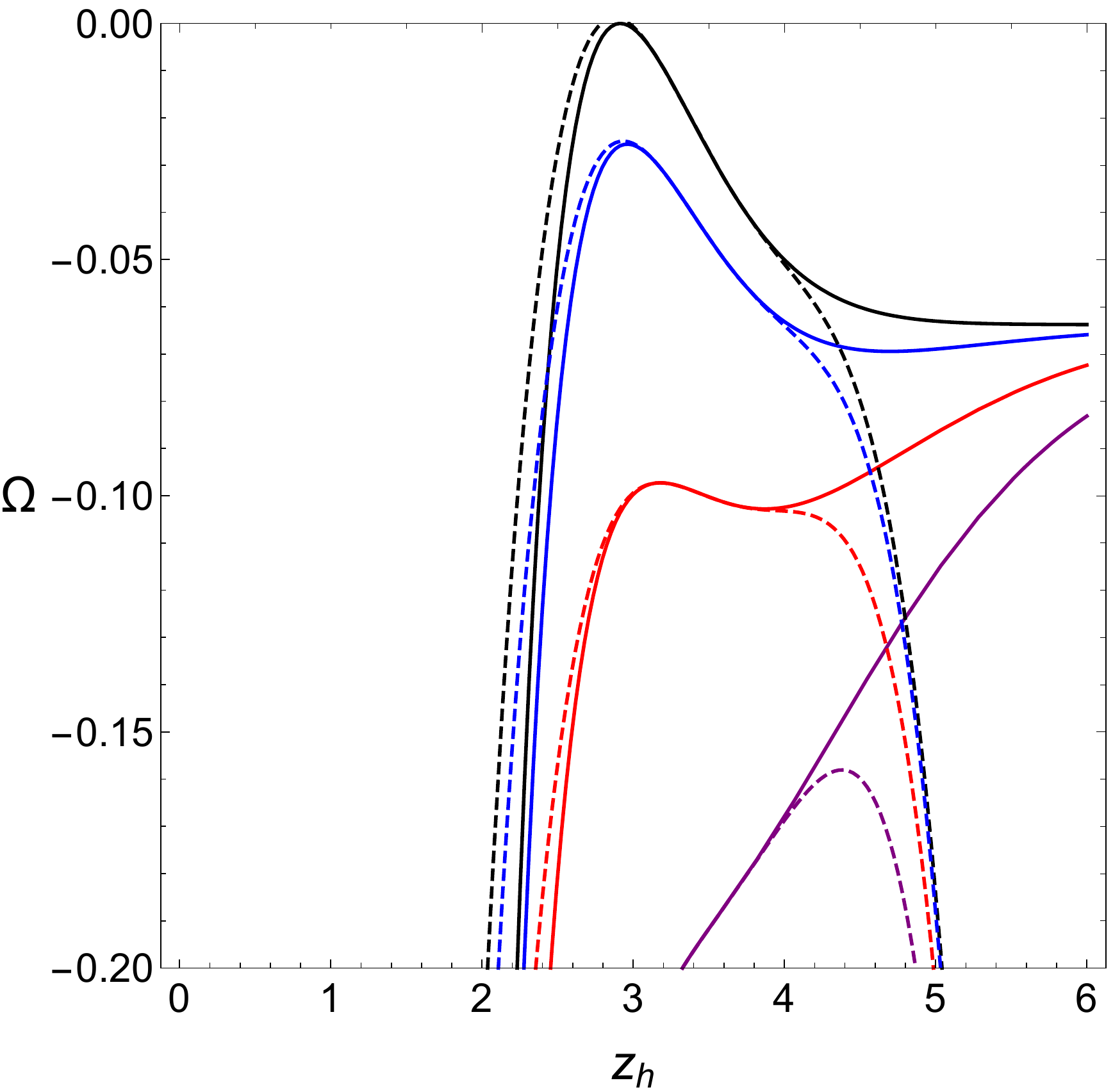}
	\caption{Grand canonical potential $\Omega$, as a function of the horizon radius $z_h$, obtained from the analytic approximation in \eqref{OmegaNearTc} and compared with the full numerical result. The dashed (solid) curves represent the analytic (numerical) results. The chemical potential varies from $\mu=0$ (black) to $\mu=0.3$ (purple) in steps of $0.1$.}
	\label{fig:QuarticExp}
\end{figure}

To investigate the physics near the critical point we consider this time the Taylor expansion
\begin{align} \label{TaylorExpOmega}
\Omega(z_h^c + \delta z_h , \mu^c + \delta \mu) &= \Omega^{\, c} + \Omega_{01}^{\, c} \delta \mu + \frac12 \Omega_{02}^{\, c} \, \delta \mu^2  + \Omega_{11}^{\, c} \delta z_h \delta \mu + \frac12 \Omega_{21}^c \delta z_h^2 \delta \mu \nonumber \\ 
&+ \frac12 \Omega_{12}^c \delta z_h  \delta \mu^2 
+ \frac16 \Omega_{30}^{\, c} \delta z_h^3 + \frac14 \Omega_{22}^c \delta z_h^2 \delta \mu^2  \nonumber \\
&+ \frac16 \Omega_{31}^c \delta z_h^3 \delta \mu + 
\frac{1}{24} \Omega_{40}^c \delta z_h^4 + \dots \, ,
\end{align}
where $\Omega^{\, c} = \Omega(z_h^c , \mu^c) \approx -1.197 \times 10^{-5}$ and $\Omega_{mn}^{\, c} = \Omega_{mn}(z_h^c , \mu^c)$. Note the absence of pure  $\delta z$ and $\delta z^2$ terms because the coefficients $\Omega_{10}^{\, c}$ and $\Omega_{20}^{\, c}$ vanish. Any term containing powers $\delta \mu^3$ or higher are also absent because the coefficients $\Omega_{mn}^c$ vanish when $n>2$.  This time a reasonable approximation for the grand canonical potential can only be found in the regime $z_h < z_h^c$  truncating the series expansion \eqref{TaylorExpOmega} at quartic order. From \eqref{OmegaSol} we evaluate the coefficients $\Omega_{mn}^c$ and find 
\begin{align}
\Omega_{01}^{\, c} &\approx -0.8406 \times 10^{-4} \quad , \quad 
\Omega_{02}^{\, c} \approx -3.736 \times 10^{-4}\quad , \quad 
\Omega_{11}^{\, c} \approx 0.5256 \times 10^{-4}\, , \nonumber \\
\Omega_{21}^{\, c} &\approx -0.2131 \times 10^{-4}\quad , \quad  
\Omega_{12}^{\, c} \approx 2.336 \times 10^{-4}\quad , \quad 
\Omega_{30}^{\, c} \approx 0.234 \times 10^{-4}\,, \nonumber \\
\Omega_{22}^{\, c} &\approx -0.947 \times 10^{-4}\quad , \quad 
\Omega_{31}^{\, c} \approx 0.0403 \times 10^{-4}\quad , \quad 
\Omega_{40}^{\, c} \approx -1.004 \times 10^{-4}\,.
\end{align}
From these results we are able to approximate  $\delta \Omega = \Omega - \Omega^{\, c}$ by a polynomial of quartic order in $\delta z$, i.e.
\begin{align} \label{OmegaNearTc}
10^4 \, \delta \Omega &\approx - \delta \mu (0.8407 + 1.868  \delta \mu) + (0.5256 +1.168 \delta \mu ) \, \delta \mu \delta z   \nonumber \\
&- (0.1065+0.2367 \delta \mu ) \, \delta \mu \delta z^2 + (0.039 + 0.0067 \delta \mu )  \, \delta z^3 - 0.0418 \, \delta z^4  
 \,.  
\end{align}
In figure \ref{fig:QuarticExp} we compare the grand canonical potential $\Omega(z_h, \mu)$ obtained from the quartic approximation in \eqref{OmegaNearTc} against the full numerical result obtained in section \ref{Sec:Thermodynamics}. The analytic expansion \eqref{OmegaNearTc} provides a reasonable approximation for the potential in the regime $z_h < z_h^c$ corresponding to the stable (or metastable) large BH and the unstable small BH. It seems that polynomial expansions like  \eqref{OmegaNearTc} are not sufficient to describe the regime $z_h > z_h^c$ associated with the stable (or metastable) small BH. This is related to the rapid decrease in the curves of Fig. \ref{fig:QuarticExp} at large $z_h$ associated with the exponential decrease of the entropy. 

\subsection{Critical exponents}

Having found the temperature and grand canonical potential near the critical point we can investigate the behaviour of other thermodynamic quantities near the critical point. In this subsection we will analyse the specific heat and the charge susceptibility near the critical point and find the corresponding critical exponents. 

In this analysis it is convenient to work with the dimensionless quantities
\begin{eqnarray} 
\xi &\equiv&  \frac{z_h - z_h^c}{z_h^c} \quad , \quad 
u \equiv  \frac{\mu - \mu^c}{\mu^c} \quad , \quad 
t \equiv  \frac{T-T^c}{T^c} \quad , \quad \cr
\omega &\equiv&  \frac{\Omega - \Omega^c}{\Omega^c} \quad , \quad
s \equiv \frac{S-S^c}{S^c} \quad , \quad 
q \equiv \frac{Q-Q^c}{Q} \, . \label{dimlessqts}
\end{eqnarray}
Armed with this notation the temperature expansion in \eqref{TempNearTc} takes the form
\begin{equation} \label{TempNearTcv2}
t(\xi,u) \approx    (\# + \#  u) u   
+ (\# + \# u ) \, u \, \xi+ \# u \, \xi^2  + \# \xi^3 \,,  
\end{equation}
where the symbol $\#$ represent non-zero numerical coefficients whose values are not important for this discussion. The potential expansion in \eqref{OmegaNearTc} becomes 
\begin{eqnarray} \label{OmegaNearTcv2}
\omega(\xi,u) &\approx&   (\# + \# u) u   + (\# + \# u ) \, u \, \xi 
 - (\# + \# u ) \, u \xi^2 \cr 
 &+&
 (\# + \# u )  \, \xi^3 +  \# \xi^4 \, .
 \end{eqnarray}
Since the entropy, obtained in \eqref{bhentropy} is a monotonically decreasing function of $z_h$, near the critical value $z_h^c$ the dimensionless entropy difference $s$, defined in \eqref{dimlessqts}, is simply expanded as 
\begin{eqnarray} \label{entropynearTc}
s(\xi) = \# \xi + \# \xi^2 + \dots \, . 
\end{eqnarray}
Using the expansions \eqref{TempNearTcv2}, \eqref{OmegaNearTcv2} and \eqref{entropynearTc} we can find the expansions for other thermodynamic quantities near the critical point. We will describe below how we can use these expansions to find the critical exponents associated with the specific heat and the charge susceptibility.

We found that the charge density $Q$ can be written as 
\begin{eqnarray}
Q(z) = d(z_h) \mu \, ,
\end{eqnarray}
with $d(z_h)$ a decreasing function of $z_h$. Then near the critical point the dimensionless charge density difference $q(\xi, u)$, defined in \eqref{dimlessqts}, can be expanded as 
\begin{eqnarray} \label{chargenearTc}
q(\xi,u) = \# \xi + \# u + \# \xi^2 + \# u \xi + \dots \, . 
\end{eqnarray}

\subsubsection{Specific heat}

We want to analyse the specific heat, defined by 
\begin{eqnarray} \label{specificheat}
C_V = S \frac{ \partial \ln S}{ \partial \ln T} \, ,
\end{eqnarray}
near the critical point. We found in the previous section that at the critical point the specific heat diverges. In order to find how the divergence scales with the temperature difference $t$ or chemical potential difference $u$ we first rewrite \eqref{specificheat} as 
\begin{eqnarray} \label{specificheatv2}
C_V^{-1} = S^{-1} \frac{ \partial \ln T}{ \partial \ln S} = T^{-1} \frac{\partial T}{\partial S} \,. 
\end{eqnarray}
In terms of the dimensionless temperature and entropy differences $s$ and $t$, eq. \eqref{specificheatv2} takes the form
\begin{eqnarray}  \label{specificheatv3}
C_V^{-1} = \# (1+t)^{-1} \frac{dt}{ds}
= \# (1+t)^{-1} \left ( \frac{ds}{d \xi} \right )^{-1} \frac{\partial t}{\partial \xi} \,. 
\end{eqnarray}
We are particularly interested in the vertical line $\mu=\mu^c$ in the $T - \mu$ phase diagram. This corresponds to fixing the dimensionless variable $u$ to zero. Plugging the expansions in \eqref{entropynearTc} and \eqref{chargenearTc}  into eq.  \eqref{specificheatv3} we find 
\begin{eqnarray}
C_V^{-1} \vert_{u=0} = \# \xi^2 + \dots 
= \# t^{2/3} + \dots \, . 
\end{eqnarray}
This means that for $\mu=\mu^c$ the specific heat diverges near $T=T^c$ as 
\begin{eqnarray} \label{criticalCV}
C_V (T, \mu^c)  \, \propto (T - T^c)^{-\alpha} \, ,
\end{eqnarray}
with $\alpha =2/3$. 
The power scaling \eqref{criticalCV}  for the specific heat near the critical point was obtained from expansions near the critical point that are not particular to this model and should be valid in a general class of holographic QCD models constructed  from EMD theory.  The critical exponent $\alpha=2/3$ is therefore universal for this class of holographic models. The same critical exponent for the specific heat was also found  in \cite{Chamblin:1999tk,Johnson:2013dka} for the case of a RN AdS BH in global coordinates.

\subsubsection{Charge susceptibility}

The charge susceptibility is obtained in appendix \ref{App:Susceptibility}. Here we just recast the main result \eqref{chargesuscept} as 
\begin{eqnarray} \label{chargesusceptv2}
\chi =  \left ( \frac{\partial T}{\partial z_h} \right )^{-1} \Big [ \frac{\partial Q}{\partial \mu} \frac{\partial T}{\partial z_h} - \frac{\partial Q}{\partial z_h} \frac{\partial T}{\partial \mu} \Big ] \,. 
\end{eqnarray}
Taking the inverse of \eqref{chargesusceptv2} and writing the RHS in  terms of dimensionless thermodynamic differences we obtain
\begin{eqnarray}
\chi^{-1} = \# \frac{\partial t}{\partial \xi} \Big [
\# \frac{\partial q}{\partial u} \frac{\partial t}{\partial \xi}
+ \# \frac{\partial q}{\partial \xi} \frac{\partial t}{\partial u}
\Big ]^{-1} \,. 
\end{eqnarray}
Again, we are particularly interested in the case $\mu=\mu^c$ corresponding to a vertical line in the $T-\mu$ phase diagram. Using the expansions in \eqref{TempNearTcv2} and \eqref{chargenearTc} and setting the dimensionless difference $u$ to zero we obtain
\begin{eqnarray}
\chi^{-1} = \# \xi^2 + \dots = \# t^{2/3} + \dots \,. 
\end{eqnarray}
Then at $\mu=\mu^c$ the charge susceptibility diverges near the critical point $T=T^c$ according to the scaling 
\begin{eqnarray}
\chi (T, \mu^c) \, \propto  (T - T^c)^{-\gamma} \, , 
\end{eqnarray}
with $\gamma=2/3$.  Interestingly, the critical exponent found for the charge susceptibility is equal to the one found for the specific heat. We remark that these results were obtained from expansions for the thermodynamic quantities that should be valid for a general class of non-conformal plasmas arising from EMD holography.   We conclude that the critical exponents $\alpha=2/3$ for the specific heat and $\gamma=2/3$ for the charge susceptibility are universal in this class of holographic models.

\section{Comparing our model to lattice $SU(N_c)$ gauge theories in the limit $\mu \to 0$}
\label{Sec:Lattice}

In this work we have described the phase diagram of a non-conformal plasma using a holographic model based on Einstein-Maxwell-dilaton theory.  We have obtained the corresponding $T-\mu$ phase diagram, which is characterised by a critical line at low $\mu$ ending on a critical point at $(T^c,\mu^c)$. 

The critical line at low $\mu$ corresponds to a first order transition for the non-conformal plasma. In particular, in the limit $\mu \to 0$ the first order transition takes place at a temperature $T_c  \approx 0.149 \, {\rm GeV}$  for $k=0.18 \, {\rm GeV}^2$, as shown in Fig. \ref{fig:phasediag1}. In fact, what really matters in our model is the dimensionless temperature
\begin{equation}
\frac{T_{c}}{\sqrt{k}}  =  0.354    \,. \label{dimlessTc}
\end{equation}
Our model only depends on one parameter, namely the dimensionful constant $k$ appearing on the quadratic ansatz for the dilaton field $\phi(z) = k z^2$. For concreteness we chose the value $k=0.18 {\rm GeV}^2$ that leads to a mass for the $\rho$ meson close to the experimental value $m_{\rho} = 0.775 \, {\rm GeV}$ \cite{Li:2013oda,Chelabi:2015cwn}.  We remark, however, that the result in \eqref{dimlessTc} is independent of the choice of $k$ because it is a dimensionless quantity. 

Our results for the phase diagram strongly suggest an interpretation of EMD holography in terms of the thermodynamics of large $N_c$ QCD, dominated by the dynamics of pure $SU(N_c)$ Yang-Mills theory. A similar conclusion was found in a recent work \cite{Afonin:2018era}. 
In fact, the first order transition found in this paper in the limit $\mu \to 0$ can be interpreted in terms of the deconfinement transition found in $SU(N_c)$ Yang-Mills theories in the limit of large $N_c$. The order parameter for this transition is the entropy that, in the limit $\mu \to 0$ jumps from zero to a finite value at the critical temperature $T_c$ \footnote{The transition is quite similar to a Hawking-Page transition between a thermal and a black hole solution, as described, for example, in \cite{Herzog:2006ra,BallonBayona:2007vp,Gursoy:2008za}.}.

In $SU(N_c)$ lattice  gauge theories the critical temperature for the deconfinement transition can be described by the empirical formula \cite{Lucini:2012wq}
\begin{equation}
T_c/\sqrt{\sigma} = 0.5949 (17)  + 0.458 (18) /N_c^2  \, ,
\end{equation}
where $\sigma$ is the string tension. Choosing  the phenomenological value for the string tension $\sqrt{\sigma} = 0.44 \, {\rm GeV}$
\cite{Lucini:2013qja} one finds  $T_c \approx 0.262 \,{\rm GeV}$.
We can reproduce this result in our model by fixing the parameter $k$ as follows
\begin{equation}
k = \left ( \frac{T_c}{0.354} \right )^2 \approx 0.548 \, {\rm GeV}^2 \,.
\end{equation}
It is very interesting to compare the thermodynamic properties of our model in the limit $\mu \to 0$ against the results obtained in $SU(N_c)$ lattice  gauge theories. Below we present a quantitative comparison for the pressure, trace anomaly and latent heat in our model against the results obtained in \cite{Panero:2009tv} for different values of $N_c$.  We will compare only dimensionless quantities since they are independent of the choice of $k$. 

\subsection{Pressure}

The pressure is related to the grand canonical potential by the  relation $p= - \Omega$. We will be interested in the dimensionless ratio
\begin{equation}
\frac{3 p}{N_c^2 T^4} = -  \frac{3 \Omega}{N_c^2 T^4}  \,.  \label{dimlessp}
\end{equation}
As described in section \ref{Sec:Results}, the pressure was normalised to recover the Stefan-Boltzmann result for a non-Abelian plasma in the limit of very high temperatures. Particularly, the dimensionless ratio in \eqref{dimlessp} reaches the value $3 \pi^2/(45)$ in that limit. 

\begin{figure}[ht]
	\centering
		\includegraphics[scale=0.6]{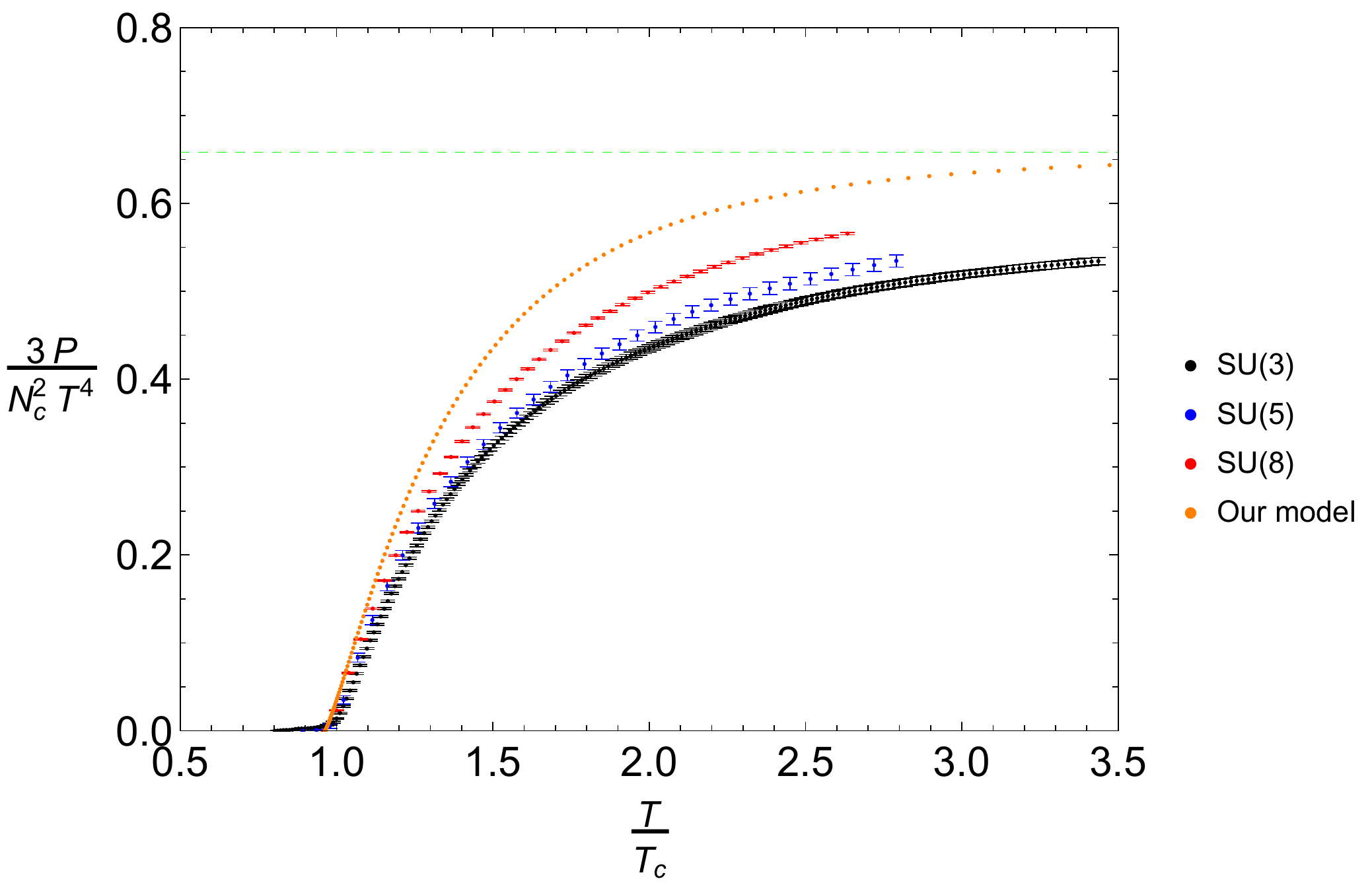}
	\caption{The dimensionless ratio $3 P/(N_c^2 T^4)$, associated with the pressure, as a function of the dimensionless temperature $T/T_c$ in the limit $\mu \to 0$ compared against the $SU(N_c)$ lattice results obtained in \cite{Panero:2009tv}. The thin dashed line represents the Stefan-Boltzmann limit $3 \pi^2/(45)$. }
		\label{fig:PvsT}
\end{figure}

Using our formulas \eqref{OmegaFromT} for the grand canonical potential 
and \eqref{T} for the temperature we can evaluate the dimensionless ratio
\eqref{dimlessp} at any value of $\mu$ and $T$. In Fig. \ref{fig:PvsT} we compare our results in the limit $\mu \to 0$ against the results obtained in \cite{Panero:2009tv} for lattice $SU(N_c)$ Yang-Mills theories. The figure displays the dimensionless ratio in \eqref{dimlessp} as a function of the dimensionless temperature $T/T_c$ where $T_c$ is the deconfinement temperature (at $\mu=0$). The black, blue and red dots with error bars represent the lattice results for $N_c=3$, $N_c=5$ and $N_c=8$ respectively. The orange dots represent our results and the green dashed line represent the Stefan-Boltzmann limit. From the figure we conclude that our results for the pressure are consistent with the lattice results in the limit of large $N_c$.

\subsection{Trace anomaly}

\begin{figure}[ht]
	\centering
		\includegraphics[scale=0.7]{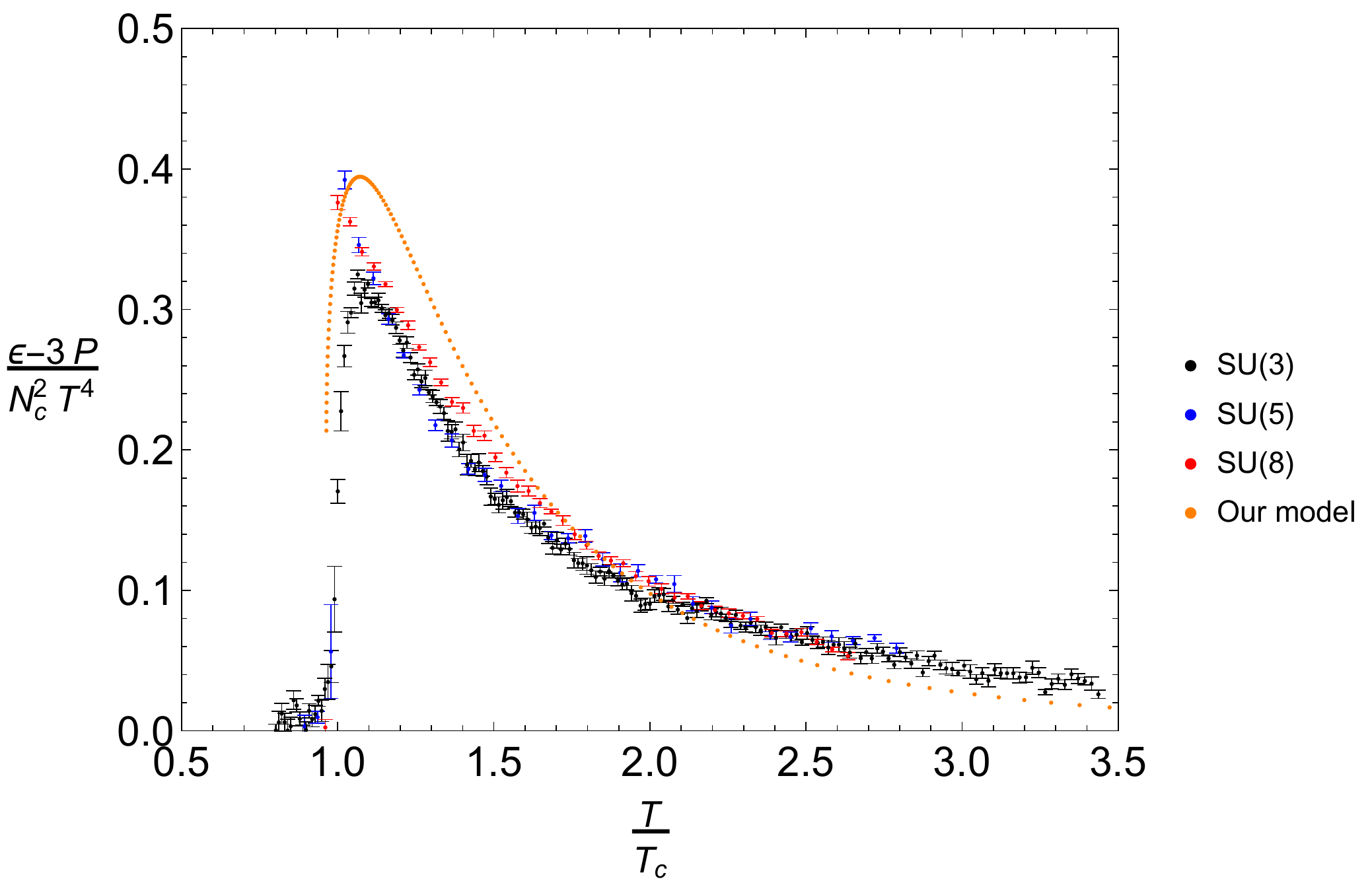}
	\caption{The dimensionless ratio $(E-3p)/(N_c^2 T^4)$, associated with the trace anomaly, as a function of the dimensionless temperature $T/T_c$ in the limit $\mu \to 0$ compared against the $SU(N_c)$ lattice results obtained in \cite{Panero:2009tv}.}
		\label{fig:TracevsT}
\end{figure}

In QCD, the breaking of conformal symmetry is described by the trace anomaly, which is the VEV of the trace of the energy-momentum tensor. At finite $\mu$ and $T$ it takes the form 
\begin{equation}
\langle T^{a}_{\, \, a} \rangle = E - 3 p  = 4 \Omega + T S + \mu Q  \,.
\end{equation}
In lattice QCD, this quantity is also known as the interaction measure, denoted by $\Delta$.
We are interested in the dimensionless ratio 
\begin{equation}
\tilde \Delta \equiv \frac{ \Delta}{ N_c^2 T^4} = \frac{E - 3p}{N_c^2 T^4} \,.     
\label{dimlessTrace}
\end{equation}
In Fig. \ref{fig:TracevsT} we compare our results for the trace anomaly against the results obtained in \cite{Panero:2009tv} for lattice $SU(N_c)$ Yang-Mills theories. The figure displays the dimensionless ratio in \eqref{dimlessTrace} as a function of the dimensionless temperature $T/T_c$ where $T_c$ is the deconfinement temperature (at $\mu=0$). Again, the black, blue and red dots with error bars represent the lattice results for $N_c=3$, $N_c=5$ and $N_c=8$ respectively whilst the orange dots represent our results. Interestingly, the dimensionless trace anomaly  in our model behaves very similarly to the corresponding quantity in lattice $SU(N_c)$ Yang-Mills theories. Particularly, both quantities display a peak near the deconfinement temperature $T_c$ and decrease quickly for temperatures lower than $T_c$. We note, however, a small discrepancy in the regime of high temperatures. We suspect that this is related to the fact that in our model we considered a dilaton field that is always quadratic in the radial coordinate. In more realistic holographic models for QCD, this ansatz is slightly modified in order to account for the conformal dimension of the gluon condensate, see e.g. \cite{Ballon-Bayona:2017sxa}. As a final comment regarding Fig. \ref{fig:TracevsT}, notice that the trace anomaly is always non-negative when $\mu \to 0$. However, we have shown in section \ref{Sec:Results} that in our model  the trace anomaly at finite $\mu$ suffers a transition to negative values at low temperatures , see Fig. \ref{fig:tranomaly}. In particular, the critical point in the $T-\mu$ phase diagram corresponds to the case where the trace anomaly vanishes. 

\subsection{Latent heat}

Lastly, we estimate the latent heat for the deconfinement transition in the limit $\mu \to 0$ and compare our estimate against the lattice result found in \cite{Panero:2009tv} for $SU(N_c)$ Yang-Mills theories.
The latent heat is defined by 
\begin{equation}
L_h \equiv T_c  \Delta S(T_c) \, ,    
\end{equation}
where $\Delta S$ is the entropy jump at the deconfinement temperature 
$T_c$. We are interested in the dimensionless ratio
\begin{equation}
\tilde L_h \equiv \frac{L_h}{N_c^2 T_c^4}   \,. 
\end{equation}
In the limit $\mu \to 0$ we find in our model
\begin{equation}
\tilde L_h  \approx (0.788)^4 \,.     
\end{equation}
This can be compared to the lattice estimate
\begin{equation}
\tilde L_h \approx (0.759 \pm 19)^4 \, ,    
\end{equation}
found in \cite{Panero:2009tv} in the limit of large $N_c$. We conclude that in our model the latent heat associated with the deconfinement transition in the limit $\mu \to 0$ almost agree with the value obtained in lattice $SU(N_C)$ Yang-Mills theories; although our result is very close to the lattice result it does not fall within the error interval.

\section{Conclusions and Discussions}
\label{Sec:Conclusions}


In this work we have analytically constructed asymptotically AdS charged black brane solutions from five dimensional EMD theory and used the gauge/gravity duality to map these solutions to four dimensional non-conformal plasmas at finite temperature and density. For the dilaton field we considered a quadratic profile  consistent with the confinement criterion in the IR. We have studied the thermodynamics in the grand canonical ensemble (fixed $T$ and $\mu$) and the critical behaviour of those  non-conformal plasmas.  We have also shown, through the behaviour of several thermodynamic quantities, that this system displays a variety of phenomena and a rich phase structure such as first-, second-order and continuous phase transitions depending on the value of the chemical potential $\mu$. These properties are intimately connected with dual phase transitions between different branches of charged BHs. 

At $\mu=0$ we found one large and one small BH phases, with the large BH branch in a stable phase (positive specific heat) and the small BH branch always unstable (negative specific heat). For $0<\mu<\mu^c$ we showed that besides the large BH branch, which dominates in the high temperature regime,  there are two small BH branches. one of them is stable or metastable (low temperature regime) whilst the other is always unstable. We described how the competition between the stable (or metastable) small and large BHs leads to a first-order phase transition from the low temperature regime to the high temperature regime. The first-order transition occurs until we reach the critical point at $\mu=\mu^c$,  where it becomes second-order and we find a power-law behaviour for the specific heat and the charge susceptibility as $T\to T^c$. The corresponding critical exponents were also computed and we  found $\alpha=\gamma=\frac{2}{3}$, with $\alpha$ ($\gamma$)  being the critical exponent associated with the specific heat (charge susceptibility). 

It is well-known that any quantum field theory that displays a critical point in the phase diagram should enjoy scale invariance at that point \cite{DiFrancesco:1997nk}. This can be seen by considering correlation functions near the critical point. The correlation functions are characterised by a length scale known as the correlation length $\xi$, which is the  maximum length at which the degrees of freedom affect each other. From dimensional arguments  the inverse of the correlation length can be interpreted as the characteristic mass scale of the system. A universal result in the theory of phase transitions is that as we approach the critical point the correlation length diverges. Then  the characteristic mass scale of the system vanishes which in turn implies that the system is scale invariant.
The scale invariance indicates the presence of a conformal field theory, usually associated with a non-trivial fixed point in the renormalisation group  \cite{Weinberg:1996kr}.

In this work we have confirmed the restoration of conformal symmetry at the critical point by evaluating the thermal trace anomaly $E- 3p$  and showing that it indeed vanishes at the critical point, as shown in Fig.\ref{fig:tranomaly}. For larger values of chemical potential, i.e., for $\mu>\mu^c$, we showed that the phase transition becomes  continuous with the thermodynamic observables displaying a smooth behaviour as we vary the temperature. These results were summarized in Fig.\ref{fig:phasediag1}, where we presented our final phase diagram in the $T-\mu$ plane. 

The thermodynamic quantities presented in section \ref{Sec:Results} and \ref{Sec:CriticalThermod} were obtained for a fixed value of the model parameter $k$, namely $k=0.18 \, {\rm GeV}^2$ motivated by the meson spectroscopy. We remark, however, that dimensionless ratios of thermodynamic quantities are independent of the choice of $k$. In section \ref{Sec:Lattice} we defined dimensionless ratios for the pressure, trace anomaly and latent heat (dividing by a suitable power of the temperature). In the limit $\mu \to 0$ we compared our results with the lattice results of $SU(N_c)$ Yang-Mills theories and found that our model is consistent with $SU(N_c)$ Yang-Mills theories in the large $N_c$ limit. We showed in section \ref{Sec:Lattice} that the critical temperature for the deconfinement transition in lattice $SU(N_c)$ Yang-Mills theories can be used to fix $k$, namely for $T_c \approx 0.262 \, {\rm GeV}$ we find $k=0.548 \, {\rm GeV}^2$.

It is noteworthy that the phase transitions of non-conformal plasmas found in this work showed some similarities with the van der Waals-Maxwell liquid-gas transition. We found a very similar phase diagram structure, consisting of a critical line, associated with a first-order phase transition, ending in a critical point from which we have a continuous phase transition. We elaborated on the analogy between these systems mapping the temperature and horizon radius of charge BHs  to the pressure and volume in the van der Waals model. We also provided some connections and analogies with the $A_3$ catastrophic theory from the thermodynamic analysis near the critical point. It is fascinating to observe that two very distinct systems, a gravitational one and a liquid-gas one, share the same qualitative features and behaviours.

Finally, we have developed  a consistent and systematic method for reconstructing the thermodynamic potential from the entropy density and temperature by using the first law of thermodynamics, and which does not require performing any holographic renormalisation procedure. It would be interesting to extend the analysis done in this paper through the holographic renormalisation lens \cite{Papadimitriou:2011qb,Elvang:2016tzz} in order to show the equivalence between the two approaches. It would also be interesting to extend the Maxwell term in the EMD action considered in this work to the nonlinear DBI or tachyon-DBI action to make contact with other holographic approaches, such as VQCD \cite{Alho:2013hsa}. We leave these problems for future work.


\section*{Acknowledgements}
The authors would like to acknowledge Luis Mamani for useful conversations and Marco Panero for useful correspondence and for sharing his results with us.  The work of A.B-B.  is partially funded by Conselho Nacional de Desenvolvimento Científico e Tecnológico (CNPq), grants No. 306528/2018-5 and No. 434523/2018-6. DMR is supported by  Conselho Nacional de Desenvolvimento Científico e Tecnológico (CNPq) under Grant No. 152447/2019-9. H.B.-F. is partially supported by Coordenação de Aperfeiçoamento de Pessoal de Nível Superior (CAPES),  and Conselho Nacional de Desenvolvimento Científico e Tecnológico (CNPq) under Grant No. 311079/2019-9. 

\begin{appendix}



\section{Thermodynamics of AdS/RN BH: Low and  high temperature expansions}
\label{App:RNAdS}


When one turns off the scalar field, by setting $k=0$, one should recover the results for the AdS/RN solution. Indeed, for $k=0$, we have that the temperature T \eqref{T} and entropy S \eqref{bhentropy} given by
\begin{equation}
T = \frac{6-\mu^2\,z_h^2}{6\pi\,z_h}; \quad S = \frac{4\pi\sigma}{z_h^3},
\end{equation}
where $\sigma = M_{P}^{3}\,N^{2}_{c}\,V_{3} $. Here, we are using the units from \cite{Gursoy:2008za} in which
\begin{equation}
(M_{P}\,l)^{3} = \frac{1}{45\,\pi^2},
\end{equation}
where $l$ is the AdS radius. From now on we will not write explicitly the factor of $N_c^2$. 

One can express the horizon position $z_h$ in terms of the temperature $T$ and the chemical potential $\mu$ as
\begin{equation}\label{zhfunctionTmu}
z_{h} = \frac{\sqrt{6\,\mu^2 + 9\, \pi^2\,T^2} -3\,\pi\,  T}{\mu^2}.
\end{equation}
Therefore,
the entropy density along with its low- and high-temperature expansions, respectively, can be expressed as functions of ($T,\,\mu$) as
\begin{eqnarray}
s(T,\mu) &=& \frac{4\, \mu ^6}{45\, \pi \left(\sqrt{6\, \mu ^2+9\, \pi ^2
\,T^2}-3\, \pi\,  T\right)^3},
\\*
s(T,\mu) &=& \frac{\sqrt{\frac{2}{3}}\, \mu ^3}{135 \,\pi }+\frac{\mu ^2\,T}{45}+O\left(T^2\right),
\\*
s(T,\mu) &=& \frac{4\, \pi ^2\, T^3}{45}+\frac{2\, \mu ^2
\,T}{45}+O\left(\frac{1}{T}\right).
\end{eqnarray}
The specific heat can be obtained as
\begin{equation}
C_V = T\,\frac{\partial S}{\partial T}.
\end{equation}
Its expansion, for $\mu>>T$ and $\mu<<T$ are given by, respectively
\begin{eqnarray}
C_V &=& \frac{\mu ^2\, T}{45}+O\left(T^2\right),
\\
C_V &=& \frac{4\, \pi ^2\, T^3}{15}+\frac{2\,\mu ^2\,T}{45} + O\left(\frac{1}{T^2}\right).
\end{eqnarray}
The speed of sound, in turn, can be obtained as
\begin{equation}
c^2_s = \frac{S}{C_V}.
\end{equation}
Its expansion, for $\mu>>T$ and $\mu<<T$ are given by, respectively
\begin{eqnarray}
c^2_s &=& \frac{\sqrt{\frac{2}{3}}\, \mu }{3\, \pi\, T}+O\left(T\right),
\\
c^2_s &=& \frac 13 + O\left(\frac 1T\right).
\end{eqnarray}

The grand canonical potential $\Omega (T,\mu)$ can be expressed as
\begin{equation}
\Omega (T,\mu) = -\frac{\mu^6\, \left(3\,\mu^2 + 6\,\pi^2\,T^2 -2\,\pi\,T\sqrt{6\,\mu^2+9\,\pi^2\,T^2}\right)}{45\,\pi^2\,\left(\sqrt{6\,\mu^2 + 9\,\pi^2\,T^2} - 3\,\pi\,T\right)^4},
\end{equation}
where we used \eqref{OmegaAdSRN} and \eqref{zhfunctionTmu}.
Its expansions for $\mu>>T$ and $\mu<<T$ are given by, respectively
\begin{eqnarray}
\Omega(T,\mu) &=& -\frac{\mu^4}{540 \,\pi^2} + O\left(T\right) , 
\\*
\Omega(T,\mu) &=&  -\frac{\pi^2\,T^4}{45}-\frac{\mu^2\,T^2}{45}+O\left(\frac{1}{T}\right).
\end{eqnarray}

From \eqref{chargedensityformula}, one can obtain the charge density as a function of $z_h$ and using \eqref{zhfunctionTmu} we obtain it as function of $(T,\mu)$, which reads 
\begin{equation}
Q = \frac{2\,\mu^5}{45\,\pi^2\,\left(\sqrt{6\,\mu^2+9\,\pi^2\,T^2}-3\,\pi\,T\right)^2}.
\end{equation}
Its expansion for $\mu>>T$ and $\mu<<T$ are given by, respectively
\begin{eqnarray}
Q &=& \frac{\mu^3}{135\,\pi^2}+O\left(T\right),\label{chargeexp1}
\\*
Q &=& \frac{2\,\mu\,T^2}{45}+O\left(\frac{1}{T}\right).\label{chargeexp2}
\end{eqnarray}
The charge susceptibility $\chi$ can be easily obtained from the above expansions by deriving w.r.t $\mu$ keeping $T$ constant.

Finally, the trace anomaly is
\begin{equation}
\left\langle T^{a}_{\;\;a}\right\rangle = 4\Omega + Ts + \mu\,Q.
\end{equation}
Using the above formulas for $s(\mu,T)$, $\Omega(\mu,T)$ and $Q(\mu,T)$ one can verify that the trace anomaly for the AdS/RN black brane vanishes
\begin{equation}
\left\langle T^{a}_{\;\;a}\right\rangle = 0.
\end{equation}

\section{Alternative method for reconstructing the grand canonical potential}
\label{App:AltMethod}

We start with the following ansatz for the grand canonical potential
\begin{eqnarray}  \label{OmegaAnsatzApp}
\Omega(z_h, \mu) = - A(z_h) - \mu^2 B(z_h)   \,. 
\end{eqnarray}
Now we define the auxiliary potential
\begin{align} \label{ABDErel}
\tilde \Omega &\equiv \Omega + T S     \nonumber \\
&= - A(z_h) + S(z_h) a(z_h) - \mu^2 \Big [B(z_h) + S(z_h) a(z_h) b(z_h) \Big ] \nonumber \\
&\equiv D(z_h) - \mu^2 E(z_h) \,. 
\end{align}
The differential of $\tilde \Omega$ takes the form
\begin{eqnarray}
d \tilde \Omega = \Big [ D'(z_h) -  \mu^2 E'(z_h) \Big ] d z_h - 2 \mu \, E(z_h) d \mu \,. 
\end{eqnarray}
We identify this differential with
\begin{align} 
d \tilde \Omega &=  T dS - Q d \mu \nonumber \\
&= S'(z_h) T(z_h, \mu) dz_h - Q(z_h, \mu) d \mu \nonumber \\
&= \Big [  S'(z_h) a(z_h)  - \mu^2 S'(z_h) a(z_h) b(z_h)\Big ] dz_h - Q(z_h, \mu) d \mu \,. 
\end{align}
Therefore, we find the differential equations
\begin{align} \label{DEeqs}
D'(z_h) &= S'(z_h) a(z_h)   \, , \nonumber \\
E'(z_h) &= S'(z_h) a(z_h) b(z_h) \, , 
\end{align}
as well as the relation
\begin{eqnarray}
2 \mu \, E(z_h) = Q(z_h, \mu) \,. 
\end{eqnarray}
Solving  the differential equations \eqref{DEeqs} we find $D(z_h)$ and $E(z_h)$ and from \eqref{ABDErel} we obtain the functions $A(z_h)$ and $B(z_h)$ so that we reconstruct the grand canonical potential in \eqref{OmegaAnsatzApp}. The procedure described in this appendix  is equivalent to the one described in subsection \ref{OmegaReconstruction}. The advantage of this alternative procedure is that we arrive at simpler thermodynamic relations. 

\section{Extracting the charge susceptibility}
\label{App:Susceptibility}

Here we use the notation
\begin{eqnarray}
X_{mn} = \frac{ \partial^{m+n} X}{ \partial z_h^m \partial \mu^n} \, 
\end{eqnarray}
for any thermodynamic quantity $X$. 

In subsection \ref{OmegaReconstruction} we found the charge density can be written as 
\begin{eqnarray}
Q(z_h, \mu) &=& 2 \mu \int_{\infty}^{z_h} S'(z) a(z) b(z) 
\equiv d(z_h) \, \mu \,. 
\end{eqnarray}
This expression was obtained from the thermodynamic relations involving the derivatives of the temperature and grand canonical potential.

Since the charge density $Q$ is a function of $z_h$ and $\mu$, the differential takes the form
\begin{eqnarray} \label{ChargeDiff}
d Q = Q_{10} dz_h + Q_{01} d \mu \, .
\end{eqnarray}
Considering the charge density as a function of $T$ and $\mu$ we  get instead
\begin{eqnarray} \label{ChargeDiffv2}
d Q &=& \Sigma \, dT  + \chi d \mu  \cr
&=& \Sigma T_{10} \, dz_h + \Big ( \Sigma T_{01} + \chi \Big ) d \mu \,.  
\end{eqnarray}
Identifying the differentials in \eqref{ChargeDiff} and \eqref{ChargeDiffv2} we find the relations
\begin{align}
Q_{10} &= \Sigma \, T_{10}  \nonumber \\
Q_{01} &= \Sigma T_{01} + \chi \, . 
\end{align}
From these relations we extract the charge susceptibility
\begin{eqnarray} \label{chargesuscept}
\chi = Q_{01} - \frac{Q_{10} T_{01}}{T_{10}} \,. 
\end{eqnarray}

\end{appendix}

\end{document}